\documentclass[12pt]{article}
\usepackage{jheppub}
\usepackage{tikz}
\usepackage{tikz-cd}
\usetikzlibrary{matrix,calc}
\usetikzlibrary{decorations.markings,shapes.geometric,shapes.misc}
\usetikzlibrary{decorations.pathreplacing,positioning}
\usetikzlibrary{arrows}
\usepackage{color}
\usepackage{xcolor}
\usepackage{graphicx}

\DeclareMathOperator{\Tr}{Tr}

\renewcommand{\Im}{\operatorname{Im}}

\title{Kondo line defects and affine Gaudin models}

\author[a]{Davide Gaiotto,}
\author[a]{Ji Hoon Lee,}
\author[b]{Beno\^{\i}t Vicedo,}
\author[a]{and Jingxiang Wu}

\affiliation[a]{Perimeter Institute for Theoretical Physics, Waterloo, Ontario, Canada N2L 2Y5}
\affiliation[b]{Department of Mathematics, University of York, York YO10 5DD, United Kingdom}

\abstract{We describe the relation between integrable Kondo problems in products of chiral $SU(2)$ WZW models and affine $SU(2)$ Gaudin models. We propose a full ODE/IM solution of the spectral problem for these models.}

\begin{document}
\maketitle
\section{Introduction and Summary}
The purpose of this paper is to explore the connections between three topics: 
\begin{enumerate}
\item Integrable Kondo defects in products of chiral $SU(2)$ WZW models $\prod_i \mathfrak{\widehat {sl}}(2)_{k_i}$ \cite{Kondo:1964nea,Wilson:1974mb,Andrei:1980fv,Andrei:1982cr,tsvelick1985exact,PhysRevLett.52.364,doi:10.1080/00018738300101581,PhysRevB.46.10812,Cardy:1989ir,Saleur:1998hq,Affleck:1990by,Affleck:1990iv,Affleck:1991yq,Affleck:1995ge,Fendley:1995kj,Saleur:2000gp,Bachas:2004sy,Nakagawa_2018,KondoGLW}. 
These are families of mutually commuting line defects parameterized by a conformal symmetry-breaking scale $e^\theta \equiv \lambda^{-1}$. 
\item Bethe equations for an affine $SU(2)$ Gaudin model \cite{Frenkel:2003qx, Feigin:2007mr, Feigin:2017gcv}. The Kondo lines can be identified with a renormalized version of the quantum transfer matrices for the affine Gaudin model. The corresponding Bethe equations (and Bethe vectors) should thus control the spectrum of the transfer matrices.
\item Solutions of the affine $\mathfrak{\widehat {sl}}(2)$ Bethe equations can be used to produce $PSU(2)$ {\it $\lambda$-opers with singularities of trivial monodromy} \cite{Bazhanov:1998wj, Frenkel:2003qx, Fioravanti:2004cz, Feigin:2007mr, Bazhanov:2013cua, Frenkel:2016gxg, Masoero:2018rel}. We identify the spectrum of the transfer matrices with the Stokes data of the $\lambda$-opers. This provides a complete ODE/IM correspondence for integrable Kondo problems. 
\end{enumerate}

We will begin by a quick review of the affine Gaudin model in Section \ref{sec:affine}. We then define $\lambda$-opers with singularities of trivial monodromy and derive the affine Bethe equations in Section \ref{sec: opers}. We analyze the Stokes data at large and small $\lambda$ in Sections \ref{sec:UVexpansion} and \ref{sec:IRexpansion}, respectively, and compare it with direct calculations for the Kondo defects. 

In the large $\lambda$ regime, the Stokes data are obtained with the help of (exact) WKB analysis \cite{aoki2005virtual,AIF_1993__43_1_163_0, VOROS1983, silverstone1985jwkb, kawai2005algebraic, takei2017wkb,iwaki2014exact,10.1007/3-540-09532-2_85, voros1982spectre,GMN:2009hg,Gaiotto:2012rg}. The examples under study have some unusual features that require us to generalize the standard WKB analysis in order to evaluate the complete collection of the Stokes data. In order not to clutter the main body, we will only quote the results in Section \ref{sec:IRexpansion} while leaving the detailed review of the WKB analysis and our generalizations in the Appendix \ref{app:sec:wkb}.

We will also see that the construction automatically includes integrable defects in coset models $\frac{\prod_i \mathfrak{\widehat {g}}_{k_i}}{ \mathfrak{\widehat {g}}_{\sum_i k_i}}$. We discuss briefly some alternative semiclassical limits in Section \ref{sec:semi}.

Although we focus on $SU(2)$ examples in the main body of the paper, we expect the results to extend to general affine ADE Lie algebras and will comment on that in Section \ref{sec: sl3}. 

We conclude the paper with a list of open questions in Section \ref{sec:open}.
\section{Affine Gaudin models, classical and quantum} \label{sec:affine}

The affine Gaudin model, first studied in \cite{Feigin:2007mr}, is a somewhat conjectural integrable system which quantizes the classical affine Gaudin model, in a manner analogous to 
the relation between the classical and quantum KdV integrable systems \cite{Feigin:1991qm, Feigin:1993sb, Bazhanov:1994ft}\footnote{We will actually find that the quantum KdV integrable system can be recovered in a 
certain decoupling limit from the affine Gaudin model, adjusting parameters in such a way that the total Kac-Moody currents decouple and leave behind a coset model.}.

The classical affine Gaudin model is defined by a collection of Poisson-commuting 
Hamiltonians built from classical currents ${\cal J}^a_i$ with Kac-Moody Poisson brackets. The latter are organized into a Lax matrix
\begin{equation} \label{Lax matrix}
\varphi(z) {\cal L}^a(z;\sigma) = \sum_{i=1}^N \frac{{\cal J}^a_i(\sigma)}{z-z_i}
\end{equation}
where the $z_i$ are couplings and we use the auxiliary 1-form
\begin{equation} \label{twist function}
\varphi(z) dz = \bigg( 1 + \frac12 \sum_{i=1}^N \frac{k_i}{z-z_i} \bigg) dz
\end{equation}
where $k_i$ are the levels for the currents ${\cal J}^a_i$.

The Lax matrix is used to define both the Poisson-commuting transfer matrices
\begin{equation}
{\cal T}[z] = \textup{Tr} \, \mathcal{P}\exp \bigg( \oint {\cal L}^a(z;\sigma) t_a d \sigma \bigg),
\end{equation}
and families of local Hamiltonian densities ${\cal H}_u^{(n)}(\sigma)$ labelled by exponents $n$ of the affine Kac-Moody algebra and zeros $\zeta_u$ of the twist function \eqref{twist function}. In classical types, the ${\cal H}_u^{(n)}$ are given by specific homogeneous polynomials \cite{Evans:1999mj, Evans:2000qx, Evans:2001sz, Lacroix:2017isl} in
\begin{equation}
\mathrm{Tr} \big( \varphi(z) {\cal L}^a(z;\sigma) t_a \big)^r \big|_{z = \zeta_u},
\end{equation}
of total degree $n+1$ in the currents ${\cal J}^a_i$.

One of our main proposals is that the correct quantization of the affine Gaudin model involves Kondo line defects defined by coupling a 
spin to a collection of quantum Kac-Moody currents $J_i^a$.  These Kondo lines are defined in the UV in the same manner as the classical transfer matrices:
\begin{equation} \label{quantum transfer}
T[z] = \textup{Tr} \, \mathcal{P}\exp \bigg( \oint \sum_i g_i[z] J_i^a(\sigma) t_a d \sigma \bigg).
\end{equation}
The couplings $g_i[z]$ need to be renormalized and acquire a scale dependence. It was conjectured in \cite{KondoGLW} that the RG flow
factors through a flow of the spectral parameter, so that the spectral parameter may be identified with the (complexified) renormalization scale $e^\theta$ via dimensional transmutation.
In other words, the RG flow defines a one-parameter family of commuting line defects. The specific functional form of the couplings $g_i[z]$ along the commuting family 
depends on the chosen renormalization scheme. 

The RG flow is physically rich and depends sensitively on the spin of the auxiliary $\mathfrak{sl}_2$ generators $t^a$ and on the relative UV couplings. The endpoints is some IR-free line defect whose nature 
can be predicted with the help of the ODE/IM correspondence. For special choices of spin and couplings, the endpoint is an irrelevant deformation of a single identity line defect. 
Such a deformation must take the form 

\begin{equation}
	\textup{Tr} \, \mathcal{P}\exp \bigg( \oint \sum_n e^{-  n \theta} {\cal O}_u^{(n)}(\sigma) d \sigma \bigg)
\end{equation}
for some collection of bulk quasi-primaries $\mathcal{O}_u^{(n)}(\sigma)$ of dimension $n+1$.

Now we denote with $u$ the choice of UV line defect flowing to the identity line.
We will see that this generalizes naturally the choice of a zero $\zeta_u$ for $\varphi(z)$ above. A special property of such deformations by bulk chiral currents is that 
there exists a renormalization scheme where the path-ordered exponential becomes effectively integration along separate contours. Therefore the path-ordered exponential is essentially Abelian, and reduces to the exponential of the zero-modes of the ${\cal O}_u^{(n)}(\sigma)$.

These IR effective line defects commute with the transfer matrices by construction and thus the zero-modes of ${\cal O}_u^{(n)}$ can be identified with the quantum version of the classical Hamiltonians ${\cal H}_u^{(n)}$. 
With the help of a WKB analysis of the ODE/IM solution, we will match the vevs of the zero-modes of ${\cal O}_u^{(n)}$ on eigenstates with the conjectural eigenvalues of the 
quantum version of the ${\cal H}_u^{(n)}$ proposed in \cite{Lacroix:2018fhf}.

\section{Opers, $\lambda$-opers and affine opers} \label{sec: opers}
In this section, for simplicity, we specialize to the case of $\mathfrak{sl}_2$. The generalisation to $\mathfrak{sl}_3$ will be discussed in Section \ref{sec: sl3}.
The main objective of this section is to introduce the family of Schr\"odinger operators which provides the conjectural full ODE/IM solution of the Kondo defects
spectrum problem:
\begin{equation}
\partial_x^2 - \lambda^{-2} P(x) - t(x),
\end{equation}
where $P(x) = e^{2 \alpha x} \prod_{a=1}^N (x-z_a)^{k_a}$ and $t(x)$ is an auxiliary meromorphic classical stress tensor which will be determined by a solution of the Bethe equations. 

We will motivate some of our definitions in analogy to the well-known correspondence between the (non-affine) Gaudin model and 
opers with singularities of trivial monodromy \cite{Babujian:1993ts,Feigin:1994in,Frenkel:2004qy}. The latter is one of the most basic manifestations of the Geometric Langlands correspondence and can be investigated with the help of supersymmetric gauge theory \cite{Kapustin:2006pk,Gaiotto:2011nm}. It would be very nice to give a similar derivation of the ODE/IM proposal based on gauge theory or string theory
constructions.

\subsection{$\mathfrak{sl}_2$ opers}
An \emph{$\mathfrak{sl}_2$ oper} is a complexified Schr\"odinger operator 
\begin{equation} \label{Schro oper}
\partial_x^2 - t(x)
\end{equation}
with a natural transformation law under a change of coordinate
\begin{equation}
\partial_x^2 - t(x) = ( \partial_x \tilde x)^{\frac32} \Big(\partial_{\tilde x}^2 - \tilde t(\tilde x) \Big) ( \partial_x \tilde x)^{\frac12}.
\end{equation}
This implies that $t(x)$ transforms as a classical stress tensor
\begin{equation}
t(x) = (\partial_x \tilde x )^2 \tilde t(\tilde x) + \frac{3}{4} \left( \frac{\partial_x^2 \tilde x}{\partial_x \tilde x} \right)^2 - \frac{1}{2} \frac{\partial_x^3 \tilde x}{\partial_x \tilde x}.
\end{equation}
We will always consider $\mathfrak{sl}_2$ opers for which $t(x)$ is a rational function and only allow coordinate transformations which preserve this property.

We will typically denote a solution/flat section of the Schr\"odinger equation as $\psi(x)$:
\begin{equation}
\partial_x^2 \psi(x) = t(x) \psi(x)
\end{equation}
and the (constant) Wronskian of two solutions as
\begin{equation}
(\psi,\psi') = \psi(x) \partial_x \psi'(x) - \psi'(x) \partial_x \psi(x).
\end{equation}

The data of an $\mathfrak{sl}_2$ oper \eqref{Schro oper} is equivalent to that of a flat connection 
\begin{equation} \label{canonical form}
\partial_x + \begin{pmatrix}0 & t(x) \cr 1 & 0\end{pmatrix}.
\end{equation}
More generally, see for instance \cite{FrenkelLanglandsBook}, an $\mathfrak{sl}_2$ oper can be described as a flat connection of the form
\begin{equation} \label{oper sl2}
\partial_x + \begin{pmatrix}a(x) & b(x) \cr 1 & - a(x) \end{pmatrix}
\end{equation}
where $a(x)$ and $b(x)$ are rational functions, modulo gauge transformations by unipotent upper-triangular matrices whose entries are rational functions. We can fix the gauge invariance completely by bringing \eqref{oper sl2} to its unique \emph{canonical form} \eqref{canonical form} with stress tensor given by
\begin{equation} \label{stress tensor sl2}
t(x) = b(x) + \partial_x a(x) + a(x)^2.
\end{equation}

\subsection{$\mathfrak{sl}_2$ $\lambda$-opers}

An \emph{$\mathfrak{sl}_2$ $\lambda$-oper}, or simply \emph{$\lambda$-oper}, is a complexified Schr\"odinger operator with standard dependence on 
a quantization parameter $\hbar$, here denoted as $\lambda$, namely
\begin{equation}
\partial_x^2 - \frac{P(x)}{\lambda^2} - t(x).
\end{equation}
The coordinate transformations act in the same way as for an $\mathfrak{sl}_2$ oper, so that $t(x)$ is a classical stress tensor and $P(x)$ is a quadratic differential. We will always work with $\lambda$-opers for which $t(x)$ is a rational function on $\mathbb{C}$ but allow $P(x)$ to be a more general analytic function, typically with branch points or an essential singularity at infinity, but whose logarithmic derivative is a rational function.

The data of a $\lambda$-oper can be encoded in a flat connection of the form 
\begin{equation} \label{lambda-oper connection}
\partial_x + \begin{pmatrix} 0& P(x)\lambda^{-1}  + t(x) \lambda \cr  \lambda^{-1} & 0 \end{pmatrix}.
\end{equation}
More generally, an $\mathfrak{sl}_2$ $\lambda$-oper is a flat connection
\begin{equation} \label{lambda-oper general}
\partial_x + \begin{pmatrix} a(x) & P(x) \lambda^{-1}  + b(x) \lambda \cr \lambda^{-1} & -a(x) \end{pmatrix},
\end{equation}
where $a(x)$ and $b(x)$ are rational functions, modulo gauge transformations by upper-triangular matrices of the form
\begin{equation} \label{lambda-oper gauge tr}
\begin{pmatrix} 1 & v(x) \lambda \cr 0 & 1 \end{pmatrix}.
\end{equation}
for some rational function $v(x)$.
Every $\lambda$-oper admits a unique \emph{canonical form} as in \eqref{lambda-oper connection} with stress tensor as in \eqref{stress tensor sl2}.

Of course, we could equally describe a $\lambda$-oper using a flat connection of the form 
\begin{equation} \label{dual lambda-oper connection}
\partial_x + \begin{pmatrix} 0& \lambda^{-1} \cr P(x)\lambda^{-1}  + t(x) \lambda & 0 \end{pmatrix}.
\end{equation}
This leads to another (equivalent) way of describing $\lambda$-opers, namely as a flat connection
\begin{equation} \label{dual lambda-oper general}
\partial_x + \begin{pmatrix} a(x) & \lambda^{-1} \cr P(x) \lambda^{-1}  + b(x) \lambda & -a(x) \end{pmatrix}
\end{equation}
modulo gauge transformations by lower-triangular matrices of the form
\begin{equation} \label{dual lambda-oper gauge tr}
\begin{pmatrix} 1 & 0 \cr v(x) \lambda & 1 \end{pmatrix}.
\end{equation}

\subsection{Miura opers and singularities of trivial monodromy} 

A \emph{Miura $\mathfrak{sl}_2$ oper}, or simply \emph{Miura oper} for short, is a connection of the form \eqref{oper sl2} with $b(x) = 0$, namely
\begin{equation} \label{Miura connection}
\partial_x + \begin{pmatrix}a(x) & 0 \cr 1 & - a(x) \end{pmatrix}.
\end{equation}
Its equivalence class modulo gauge transformations by unipotent upper-triangular matrices defines an oper, with stress tensor $t(x) = a(x)^2 + \partial_x a(x)$, which we refer to as the oper underlying \eqref{Miura connection}. The corresponding Schr\"odinger operator \eqref{Schro oper} factorises as $(\partial_x + a(x)) (\partial_x - a(x))$ and has an obvious solution $\psi(x) = e^{\int a(x) dx}$ which is an eigenline of the monodromy around each singularity of $t(x)$.

It is useful to allow the Miura oper to have apparent singularities where the monodromy eigenline $\psi(x)$ has a simple zero
but where $t(x)$ is regular. At such an apparent singularity, $a(x)$ behaves as 
\begin{equation}
a(x) = \frac{1}{x-w} + O(x-w).
\end{equation}

Another important type of singularity is one of the form
\begin{equation}
a(x) = -  \frac{l}{x-z} + O(1)
\end{equation}
for a non-negative half integer $l$. Then 
\begin{equation}
t(x) =  \frac{l(l+1)}{(x-z)^2}+ \cdots
\end{equation}
and the Schr\"odinger operator \eqref{Schro oper} has a local solution with $\pm 1$ monodromy around $z$, of the form
\begin{equation}
\psi(x) \sim (x-z)^{-l} + \cdots
\end{equation}
The Miura condition then gives a second local solution with $\pm 1$ monodromy around $z$, of the form
\begin{equation}
\psi'(x) \sim (x-z)^{l+1} + \cdots
\end{equation}
It follows that the monodromy of any flat section around $z$ must be  $\pm 1$. Therefore $z$ is a regular singularity of trivial monodromy for $t(x)$.
One can also see this by noting that the Miura oper \eqref{Miura connection} is gauge equivalent to the connection
\begin{equation}
\partial_x + \begin{pmatrix} r(x) & 0 \cr (x-z)^{2l} & - r(x) \end{pmatrix}
\end{equation}
where $r(x) = a(x) + \frac{l}{x-z}$, which is manifestly regular at $z$ for non-negative $l$.

A quick discussion of the term ``trivial monodromy'' is in order here. If $l$ is allowed to be half-integral, we have to consider the monodromy 
as living in $PSL(2)$, so that $\pm 1$ is a trivial monodromy. If $l$ is restricted to be integral, then we can take the monodromy to be valued in 
$SL(2)$, and will still be trivial. This binary choice is a manifestation of Geometric Langlands duality: $PSL(2)$ opers are dual to the $SL(2)$ Gaudin model, and viceversa.

\subsection{Miura $\lambda$-opers and singularities of trivial monodromy} \label{sec: Miura}

A \emph{Miura $\mathfrak{sl}_2$ $\lambda$-oper}, or simply \emph{Miura $\lambda$-oper}, is a connection of the form
\begin{equation} \label{lambda-Miura}
\partial_x + \begin{pmatrix} a_+(x) & P(x)\lambda^{-1}  \cr \lambda^{-1} & - a_+(x) \end{pmatrix}.
\end{equation}
This is of the general form \eqref{lambda-oper general} and therefore a Miura $\lambda$-oper defines a $\lambda$-oper with stress tensor $t(x) = a_+(x)^2 + \partial_x a_+(x)$. We refer to this as the $\lambda$-oper underlying \eqref{lambda-Miura}. It can be described as a complex Schr\"odinger operator
\begin{equation} \label{Miura Schro}
(\partial_x + a_+(x))(\partial_x - a_+(x)) - \frac{P(x)}{\lambda^2}.
\end{equation}

Crucially, the connection \eqref{lambda-Miura} is locally gauge equivalent (in $PSL(2)$) to a connection of the following alternative form
\begin{equation} \label{dual lambda-Miura}
\partial_x + \begin{pmatrix} - a_-(x) & \lambda^{-1}  \cr P(x) \lambda^{-1} & a_-(x) \end{pmatrix}
\end{equation}
where $a_+(x) + a_-(x) = - \frac12 \frac{\partial_x P(x)}{P(x)}$. We refer to this connection as the \emph{dual} of the Miura $\lambda$-oper \eqref{lambda-Miura}. Since it is of the general form \eqref{dual lambda-oper connection} it also defines a $\lambda$-oper, which we call the \emph{dual $\lambda$-oper} underlying \eqref{lambda-Miura}, with stress tensor $\tilde t(x) = a_-(x)^2 + \partial_x a_-(x)$. The latter can also be described as a complex Schr\"odinger operator of the same form as in \eqref{Miura Schro} with $a_+(x) \to a_-(x)$.

Since the Miura $\lambda$-oper \eqref{lambda-Miura} and its dual \eqref{dual lambda-Miura} are gauge equivalent, we therefore identify a crucial property of $\lambda$-opers: the pair of $\lambda$-opers with stress tensors built from $a_\pm (x)$, i.e. the $\lambda$-oper and the dual $\lambda$-oper underlying a given Miura $\lambda$-oper, have the same monodromy (in $PSL(2)$, unless $P(x)$ is a perfect square).

If at some generic point $w$ we have 
\begin{equation}
a_+(x) = \frac{1}{x-w} + O(z-w)
\end{equation}
then it follows by the above arguments for singularities of Miura opers that the Miura $\lambda$-oper built from $a_+(x)$ has an apparent singularity at $w$ while the other Miura $\lambda$-oper built from $a_-(x)$ has a regular singularity at $w$, which must necessarily have trivial monodromy (in $PSL(2)$). The same argument applies if at a point $w'$ we have
\begin{equation}
a_-(x) = \frac{1}{x-w'} + O(z-w')
\end{equation}
with the roles of the two Miura $\lambda$-opers \eqref{lambda-Miura} and \eqref{dual lambda-Miura} interchanged.

If at a zero $z$ of order $k$ of $P(x)$ we have 
\begin{equation}
a_+(x) = -  \frac{l}{x-z} + O(1)
\end{equation}
then 
\begin{equation}
a_-(x) = -  \frac{\frac{k}{2}-l}{x-z} + O(1).
\end{equation}
As long as $0 \leq l \leq \frac{k}{2}$, the pair of Miura $\lambda$-opers both have trivial monodromy around $z$. Indeed, the Miura $\lambda$-oper \eqref{lambda-Miura} is gauge equivalent to the connection
\begin{equation*}
\partial_x + \begin{pmatrix} r(x) & (x-z)^{k-2l} q(x)\lambda^{-1}  \cr (x-z)^{2l} \lambda^{-1} & - r(x) \end{pmatrix}
\end{equation*}
where we wrote $P(x) = (x-z)^k q(x)$ with $q(z) \neq 0$ and $r(x) = a_+(x) +  \frac{l}{x -z}$, which is manifestly regular at $z$ when $0 \leq l \leq \frac{k}{2}$.

A quick discussion of the term ``trivial monodromy'' is again in order here. If $l$ is allowed to be half-integral and $k$ integral, we have to consider the monodromy as living in $PSL(2)$, so that $\pm 1$ is a trivial monodromy and gauge transformations can have a sign ambiguity. 
If $l$ is restricted to be integral and $k$ even, then we can take the monodromy to be valued in $SL(2)$. This binary choice is 
presumably a manifestation of an affine Geometric Langlands duality: $PSL(2)$ $\lambda$-opers are dual to the affine $SL(2)$ Gaudin model, and viceversa.

\subsection{Opers with singularities of trivial monodromy and Bethe equations}

For a general oper, the condition for a regular singularity to have trivial monodromy is an intricate polynomial constraint 
on the coefficients of the expansion of $t(x)$ near the regular singularity. 

Given a Miura oper on $\mathbb{C}$ with a rank $1$ irregular singularity at infinity and whose other singularities are all regular with trivial monodromy, we can write
\begin{equation}
a(x) = -\alpha -  \sum_a \frac{l_a}{x-z_a} + \sum_i \frac{1}{x-w_i}.
\end{equation}
The condition that each $w_i$ is an apparent singularity reduces to the Bethe equations
\begin{equation}
-  \sum_a \frac{l_a}{w_i-z_a} + \sum_{j \neq i} \frac{1}{w_i-w_j} = \alpha.
\end{equation}
These are the Bethe equations for a $\mathfrak{sl}_2$ quantum Gaudin model with sites of spectral parameters $z_a$, supporting $\mathfrak{sl}_2$ irreps of dimension $2l_a+1$.

We call the overall residue of $a(x)$ at infinity the \emph{weight at infinity} of the Miura oper.
Since the underlying oper has trivial monodromy at all the $z_a$ and $w_i$ we refer to it as an \emph{oper with singularities of trivial monodromy}.
The eigenvalues of the quantum Gaudin Hamiltonians can be extracted from the expression of $t(x)$.

\subsection{$\lambda$-Opers with singularities of trivial monodromy and affine Bethe equations}

We are interested in the class of Miura $\lambda$-opers on $\mathbb{C}$ for which $a_\pm(x)$ take the same form
\begin{subequations}
\begin{align}
a_+(x) &= -\alpha_+ -  \sum_a \frac{l_a}{x-z_a} + \sum_i \frac{1}{x-w_i}- \sum_i \frac{1}{x-w'_i},\\
a_-(x) &= -\alpha_- -  \sum_a \frac{\frac{k_a}{2}-l_a}{x-z_a} + \sum_i \frac{1}{x-w'_i}- \sum_i \frac{1}{x-w_i}
\end{align}
\end{subequations}
and satisfy the Bethe equations
\begin{subequations} \label{Bethe eqs}
\begin{align}
 - \sum_a \frac{l_a}{w_i-z_a} + \sum_{j \neq i} \frac{1}{w_i-w_j}- \sum_j \frac{1}{w_i-w'_j} = \alpha_+,\\
 - \sum_a \frac{\frac{k_a}{2}-l_a}{w'_i-z_a} + \sum_{j \neq i} \frac{1}{w'_i-w'_j}- \sum_j\frac{1}{w'_i-w_j} = \alpha_-.
\end{align}
\end{subequations}
These ensure that the (dual) Miura $\lambda$-oper built from $a_+(x)$ (resp. $a_-(x)$) has apparent singularities at each $w_i$ (resp. $w'_i$).
The \emph{weight at infinity} of the Miura $\lambda$-oper is the pair of residues of $a_\pm(x)$ at infinity. 

These are the Bethe equations for an affine $\mathfrak{sl}_2$ quantum Gaudin model with sites of spectral parameters $z_a$, supporting $\widetilde{\mathfrak{sl}}_2$ Weyl representations induced from $\mathfrak{sl}_2$ irreps of dimension $2l_a+1$, for WZW current algebras of level $k_a$.

The $\lambda$-oper (resp. the dual $\lambda$-oper) underlying a given Miura $\lambda$-oper has regular singularities with trivial monodromies at the zeroes $z_a$ of $P(x)$ as well as at $w'_i$ (resp. $w_i$), for all values of $\lambda$. We therefore refer to the $\lambda$-oper and its dual as a \emph{pair of $\lambda$-opers with singularities of trivial monodromy}. They have interesting Stokes data at $x= \infty$ which we call the \emph{monodromy data} of the pair of $\lambda$-opers. 

The eigenvalues of the affine $\mathfrak{sl}_2$ quantum Gaudin model transfer matrices, as well as the quantum local Hamiltonians, 
can be extracted from the Stokes data in a manner described in the remainder of the paper.

\subsection{Weyl reflections}

Given some $\mathfrak{sl}_2$ oper with trivial monodromy $t(x) = a(x)^2 + \partial_x a(x)$ and $\alpha \neq 0$, 
there must be two canonical solutions which at infinity behave like $e^{\pm \alpha x}$ times some analytic functions. 
The one behaving as $e^{- \alpha x}$ is the Miura eigenline, with logarithmic derivative $a(x)$. 
The other gives a second rational solution $\tilde a(x)$ of $t(x) = \tilde a(x)^2 + \partial_x \tilde a(x)$, namely
\begin{equation}
\tilde a(x) = \alpha -  \sum_a \frac{l_a}{x-z_a} + \sum_i \frac{1}{x-\tilde w_i}
\end{equation}
with opposite weight at infinity to $a(x)$. This gives an action of the Weyl group of $\mathfrak{sl}_2$ on the collection of Miura opers with the same stress tensor $t(x)$. 
In particular, it acts as a Weyl transformation on the weight at infinity of the Miura oper.

In terms of connections of the form \eqref{Miura connection}, the above transformation $a(x) \to \tilde a(x)$ is implemented as a gauge transformation by a unipotent upper-triangular matrix which preserves the Miura form of the connection. Explicitly, a gauge transformation of the Miura oper \eqref{Miura connection} by
\begin{equation}
\begin{pmatrix} 1 & f(x) \cr 0 & 1 \end{pmatrix}
\end{equation}
transforms it as $a(x) \to \tilde a(x) = a(x) + f(x)$ provided $f(x)$ is a (rational) solution of the Riccati equation
\begin{equation}
\partial_x f(x) + f(x)^2 + 2 a(x) f(x) = 0.
\end{equation}

If we have a Miura $\lambda$-oper with trivial monodromy, such that $\alpha_\pm$ are sufficiently generic, then we have two transformations, 
$\alpha_+ \to - \alpha_+$ or $\alpha_- \to - \alpha_-$, which map it to a different 
Miura $\lambda$-oper, with the same $P(x)$ and the same monodromy data, as either one of the $\lambda$-opers is fixed by the transformations. 

Explicitly, on a Miura $\lambda$-oper \eqref{lambda-Miura} or its dual Miura $\lambda$-oper \eqref{dual lambda-Miura} we can perform a gauge transformation by, respectively,
\begin{equation}
\begin{pmatrix} 1 & f_+(x) \lambda \cr 0 & 1 \end{pmatrix} \qquad \textup{or} \qquad \begin{pmatrix} 1 & 0 \cr f_-(x) \lambda & 1 \end{pmatrix}.
\end{equation}
This produces a new pair of Miura $\lambda$-opers with $a_+(x) \to \tilde a_+(x) = a_+(x) \pm f_\pm(x)$ and $a_-(x) \to \tilde a_-(x) = a_-(x) \mp f_\pm(x)$ provided that the functions $f_\pm(x)$ are (rational) solutions of the Riccati equation
\begin{equation}
\partial_x f_\pm(x) + f_\pm(x)^2 + 2 a_\pm(x) f_\pm(x) = 0.
\end{equation}

These two reflections can be iterated to generate an interesting group: the Weyl group of $\widetilde{\mathfrak{sl}}_2$. It acts as a Weyl transformation on the weight at infinity of the Miura affine oper.
More precisely, repeated reflections act as 
\begin{align}
\cdots \longleftrightarrow (\alpha_+ + 2 \alpha_-,- \alpha_-) &\longleftrightarrow (\alpha_+, \alpha_-) \longleftrightarrow (-\alpha_+, 2 \alpha_+ + \alpha_-)\cr
&\qquad \longleftrightarrow (3 \alpha_+ + 2\alpha_-,- 2 \alpha_+ - \alpha_-)\longleftrightarrow \cdots
\end{align}
and similarly on the weight at infinity. 

\subsection{Conjectural count of Bethe solutions}
For generic values of $\alpha$, the relation between opers with singularities of trivial monodromy and the Gaudin model suggests that the number of solutions of the Bethe equations should coincide with the graded dimension of the Gaudin Hilbert space, which is the product of $\mathfrak{sl}_2$ irreps 
of dimension $2l_a+1$, graded by total weight. A priori, this statement is rather not obvious. 

We expect a similar statement for the affine opers with singularities of trivial monodromy: for generic values of $\alpha$ the number of solutions of the Bethe equations should coincide with the graded dimension of the affine Gaudin Hilbert space, which is the product of $\widetilde{\mathfrak{sl}}_2$ Weyl representations induced from $\mathfrak{sl}_2$ irreps of dimension $2l_a+1$, for WZW current algebras of level $k_a$. 

\subsection{Special values of $\alpha_\pm$}
The Weyl reflection is not well-defined for an oper with singularities of trivial monodromy when $\alpha = 0$, essentially because there isn't a canonical choice of a second solution. Any choice of solution will do, so we really get a $\mathbb{CP}^1$ family of opers with singularities of trivial monodromy.
Only one of these solutions is special, in the sense that it decreases at infinity faster than the others, and will thus have a special weight. 

In the dual Gaudin model, we are turning off a parameter which breaks the global $\mathfrak{sl}_2$ symmetry.
A whole $\mathfrak{sl}_2$ irrep of eigenstates is represented by a single special oper with singularities of trivial monodromy.

Something similar happens if $\alpha_+ = n(\alpha_+ + \alpha_-)$ for any integer $n$. One of the Weyl reflections in the chain 
breaks down, and instead we get a continuous family of solutions. In the dual affine Gaudin model, 
we are restoring one of the $\mathfrak{sl}_2$ subgroups of the total affine $\widetilde{\mathfrak{sl}}_2$ symmetry. 
A whole $\mathfrak{sl}_2$ irrep of eigenstates is represented by a single special affine oper with singularities of trivial monodromy.

If we set both $\alpha_\pm=0$, the whole Weyl chain breaks down and we get an intricate continuous family of solutions.
In the dual affine Gaudin model, we are restoring the whole total affine $\widetilde{\mathfrak{sl}}_2$ symmetry. 
Essentially, the transfer matrices commute with the total Kac-Moody currents and thus secretly live in the coset CFT. 

\subsection{WKB expansion and quasi-canonical form} \label{sec: WKB quasi-can sl2}

The Miura $\lambda$-oper \eqref{lambda-Miura} and its dual \eqref{dual lambda-Miura} are locally gauge equivalent to a connection of the more symmetric form
\begin{equation} \label{Miura affine oper sl2}
\partial_x + \begin{pmatrix} a_0(x) & \sqrt{P(x)} \lambda^{-1}  \cr \sqrt{P(x)} \lambda^{-1}  & -a_0(x)\end{pmatrix}
\end{equation}
where $a_0(x) = a_+(x) + \frac{\partial_x P(x)}{4 P(x)}$. Following \cite{Frenkel:2003qx, Feigin:2007mr}, we will refer to this as a \emph{Miura $\widetilde{\mathfrak{sl}}_2$ oper}. We can consider the equivalence class of such a connection under gauge transformations by matrices of the form
\begin{equation} \label{affine gauge tr}
\exp \begin{pmatrix} u(x; \lambda) & v_+(x; \lambda) \cr v_-(x; \lambda) & - u(x; \lambda) \end{pmatrix}
\end{equation}
for some formal power series
\begin{equation}
u(x; \lambda) = \sum_{n=0}^\infty P(x)^{-n} u_n(x) \lambda^{2n}, \qquad
v_\pm(x; \lambda) = \sum_{n=0}^\infty P(x)^{-n- \frac{1}{2}} v^\pm_n(x) \lambda^{2n+1}
\end{equation}
where $u_n(x)$ and $v^\pm_n(x)$ are rational functions.
This defines an \emph{affine $\mathfrak{sl}_2$ oper}, or more precisely an \emph{$\widetilde{\mathfrak{sl}}_2$ oper} for the untwisted affine Kac-Moody algebra $\widetilde{\mathfrak{sl}}_2$ associated with $\mathfrak{sl}_2$. We are working in the loop realisation of $\widetilde{\mathfrak{sl}}_2$ associated with the principal $\mathbb{Z}$-gradation.

\begin{figure}
\centering
\begin{tabular}{cc}
\begin{tikzcd}
& & \begin{tabular}{c}
\textup{Miura $\mathfrak{sl}_2$}\\
\textup{$\lambda$-oper}
\end{tabular}
\arrow{r} &
\textup{$\mathfrak{sl}_2$ $\lambda$-oper}
\\[-35pt]
\begin{tabular}{c}
\textup{$\widetilde{\mathfrak{sl}}_2$ oper/WKB}\\
\textup{momentum}
\end{tabular} & \begin{tabular}{c}
\textup{Miura}\\
\textup{$\widetilde{\mathfrak{sl}}_2$ oper}
\end{tabular}
\arrow[<->, dashed]{ru} \arrow[<->, dashed]{rd} \arrow{l} & &
\\[-35pt]
& & \begin{tabular}{c}
\textup{dual Miura}\\
\textup{$\mathfrak{sl}_2$ $\lambda$-oper}
\end{tabular}
\arrow{r} &
\begin{tabular}{c}
\textup{dual $\mathfrak{sl}_2$}\\
\textup{$\lambda$-oper}
\end{tabular}
\end{tikzcd}
&
\begin{tikzpicture}[baseline =0,scale=.6]
\draw[thick] (-.14,1.5) -- (-.14,-1.5);
\draw[thick] (.14,1.5) -- (.14,-1.5);
\draw[thick] (-.4,.4) -- (0,.8);
\draw[thick] (0,.8) -- (.4,.4);
\draw[thick] (-.4,-.4) -- (0,-.8);
\draw[thick] (0,-.8) -- (.4,-.4);
\filldraw[fill=white, thick] (0,1.5) circle (4mm);
\filldraw[fill=white, thick] (0,-1.5) circle (4mm);
\end{tikzpicture}
\end{tabular}
\caption{Different types of opers: The middle dotted arrows are local gauge transformations by diagonal matrices. The right arrows correspond to working modulo gauge transformations by upper and lower triangular matrices of the form \eqref{lambda-oper gauge tr} and \eqref{dual lambda-oper gauge tr}, respectively. The left arrow corresponds to working modulo gauge transformations by matrices of the form \eqref{affine gauge tr}.}
\label{fig: different opers}
\end{figure}
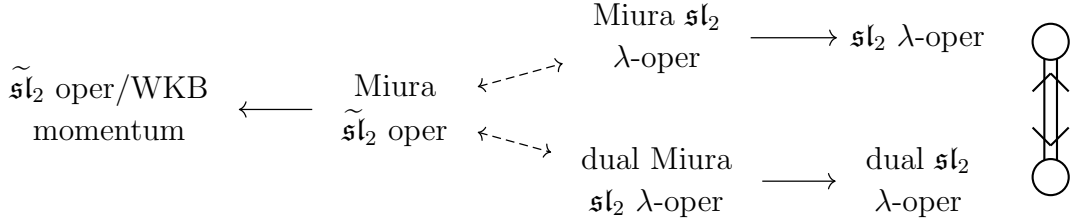

The relationship between the different opers described above is depicted in Fig. \ref{fig: different opers}. Note that the notions of $\lambda$-oper and dual $\lambda$-oper are naturally associated with the two roots of the Dynkin diagram of $\widetilde{\mathfrak{sl}}_2$.

By allowing gauge transformations as in \eqref{affine gauge tr} one can bring the connection \eqref{Miura affine oper sl2} to a \emph{quasi-canonical form} \cite{Lacroix:2018fhf}
\begin{equation} \label{oper WKB}
\partial_x + \begin{pmatrix}0& p(x;\lambda) \cr p(x;\lambda)  & 0\end{pmatrix}
\end{equation}
for some formal Laurent series
\begin{equation}
p(x; \lambda) = \frac{\sqrt{P(x)}}{\lambda} + \sum_{n = 1}^\infty P(x)^{-n+\frac{1}{2}} p_n(x) \lambda^{2n-1}.
\end{equation}
Unlike the canonical form \eqref{canonical form} of an $\mathfrak{sl}_2$ oper as in \eqref{oper sl2}, however, the quasi-canonical form \eqref{oper WKB} of an affine $\mathfrak{sl}_2$ oper is not unique. Indeed, the quasi-canonical form is preserved by residual gauge transformations of the form \eqref{affine gauge tr} with $u(x; \lambda) = 0$ and $v_-(x; \lambda) = v_+(x; \lambda)$, the effect of which is
\begin{equation} \label{quasi-can ambiguity}
p(x;\lambda) \longmapsto p(x;\lambda) + \partial_x v_+(x; \lambda).
\end{equation}

Now the quasi-canonical form \eqref{oper WKB} can be transformed to
\begin{equation}
\partial_x + \begin{pmatrix}- \frac12 \frac{\partial_x p(x;\lambda)}{p(x;\lambda)}  & \lambda p(x;\lambda)^2 \cr \lambda^{-1}  & \frac12 \frac{\partial_x p(x;\lambda)}{p(x;\lambda)}  \end{pmatrix}
\end{equation}
and by a further gauge transformation we can bring it back to the form
\begin{equation} \label{WKB vs can}
\partial_x + \begin{pmatrix}0 & \lambda \Big( p(x;\lambda)^2  + \frac34 \Big( \frac{\partial_x p(x;\lambda)}{p(x;\lambda)} \Big)^2 - \frac12 \frac{\partial^2_x p(x;\lambda)}{p(x;\lambda)} \Big) \cr \lambda^{-1} &0 \end{pmatrix}.
\end{equation}
In particular, by comparing this with the expression \eqref{lambda-oper connection} for the $\lambda$-oper underlying the Miura $\lambda$-oper we started with, we recognize the equation for the WKB momentum
\begin{equation}
\frac{P(x)}{\lambda^2} +  t(x) = p(x;\lambda)^2  + \frac34 \left( \frac{\partial_x p(x;\lambda)}{p(x;\lambda)} \right)^2 - \frac12 \frac{\partial^2_x p(x;\lambda)}{p(x;\lambda)}
\end{equation} 
which is used to study the $\lambda \to 0$ limit of the transport data of the $\lambda$-oper.

In particular, the contour integrals 
\begin{equation}
\oint p(x;\lambda)dx
\end{equation}
control the WKB asymptotics of certain transport coefficients. They are also known to match the eigenvalues of local integrals of motion for the affine Gaudin model \cite{Bazhanov:2013cua, Lacroix:2018fhf, Lacroix:2018itd}.

\section{Bethe states and transfer matrices}

\subsection{Some Kac-Moody conventions} \label{sec: WZW example}

We follow the convention from \cite{KondoGLW}. Our normalization convention for the spin basis of $\mathfrak{sl}_2$ is
\begin{equation}
t^{\pm} = \frac{1}{\sqrt{2}} (t^1\pm it^2), \quad t^0 = \frac{1}{\sqrt{2}} t^3,
\end{equation}
which satisfy the relations
\begin{equation}
[t^0,t^{\pm}] = \pm t^{\pm}, \quad [t^+, t^-] = 2t^0.
\end{equation}
The relations in the corresponding untwisted affine Kac-Moody algebra $\widetilde{\mathfrak{sl}}_2$ read
\begin{align}
\left[J_{n}^{0}, J_{m}^{0}\right]&=\frac{\kappa n}{2} \delta_{n+m, 0}\\
\left[J_{n}^{0}, J_{m}^{\pm}\right]&=\pm J_{n+m}^{\pm}\\
\left[J_{n}^{+}, J_{m}^{-}\right]&=2 J_{n+m}^{0}+\kappa n \delta_{n+m, 0},
\end{align}
for $n,m \in \mathbb{Z}$. Let $|l, \kappa \rangle$ denote the ground state in the spin $l$ module at level $\kappa$.

\subsubsection{Action of the spectral flow}

Spectral flow \cite{Feigin:1998sw,kac1990infinite} is an automorphism of $\widehat{\mathfrak{sl}}_2$ given, for $\alpha \in \mathbb{R}$, by
\begin{subequations}
\begin{align}
\mathrm{U}_{\alpha}: \quad J_{n}^{+} &\mapsto J_{n+\alpha}^{+}, \quad J_{n}^{-} \mapsto J_{n-\alpha}^{-}, \quad J_{n}^{0} \mapsto J_{n}^{0}+\frac{k}{2} \alpha \delta_{n, 0},\\
L_0 &\mapsto L_0 + \alpha J_0^0+\frac{k}{4} \alpha^2. \label{eq:spectralflow}
\end{align}
\end{subequations}

There is also an involutive automorphism induced by the Weyl group $\mathrm{W}(\mathfrak{sl}_2) = \mathbb{Z}_2$
\begin{equation}
w_1: \quad J_{n}^{+} \mapsto J_{n}^{-}, \quad J_{n}^{-} \mapsto J_{n}^{+}, \quad J_{n}^{0} \mapsto-J_{n}^{0}
\end{equation}
which satisfy
\begin{equation}
\mathrm{U}_{\alpha} \mathrm{U}_{\alpha^{\prime}}=\mathrm{U}_{\alpha+\alpha^{\prime}}, \quad \mathrm{U}_{0}=w_1^{2}=1, \quad \mathrm{U}_{\alpha} w_1=w_1 \mathrm{U}_{-\alpha}.
\end{equation}
We therefore have $\mathrm{Aut}(\widehat{\mathfrak{sl}}_2) = \mathbb{R} \rtimes \mathbb{Z}_2$. In particular, the even part is inner and corresponds to the affine Weyl group $\mathrm{W}(\widehat{\mathfrak{sl}}_2) = (2\mathbb{Z}) \rtimes \mathbb{Z}_2$. Consequently, the induced action by $\mathrm{U}_{2\mathbb{Z}}$ maps each integral highest weight representation into itself, whereas more general $\mathrm{U}_{\alpha}$ maps between the (twisted) modules. For example,
\begin{equation}
\mathrm{U}_1 : j \mapsto \frac{k}{2} -j, \quad j = 0, \frac{1}{2}, 1, \dots, \frac{k}{2}.
\end{equation}

\subsection{The Bethe equations and Bethe vectors}

Let us take the quadratic differential $P(x) = e^{2x} x^\kappa$. This should correspond to an affine Gaudin model with a single site, i.e. 
an integrable Kondo problem in the $SU(2)_\kappa$ WZW model. A generic state in the spin $l$ module of the $SU(2)_\kappa$ WZW model which is singular under the zero-mode $\mathfrak{sl}_2$ subalgebra can be described by a pair of Miura $\lambda$-opers of the form
\begin{subequations} \label{Miura WZW oper}
\begin{align}
a_+(x) &= - \frac{l}{ x} + \sum_i \frac{1}{x-w_i}- \sum_i \frac{1}{x-w'_i},\\
a_-(x) &= - 1 - \frac{\frac{\kappa}{2}-l}{ x} + \sum_i \frac{1}{x-w'_i}- \sum_i \frac{1}{x-w_i}
\end{align}
\end{subequations}
with the Bethe roots $w_i$ and $w'_i$ satisfying the Bethe equations
\begin{subequations} \label{Bethe eq WZW}
\begin{align}
 - \frac{l}{w_i} + \sum_{j \neq i} \frac{1}{w_i-w_j}- \sum_j \frac{1}{w_i-w'_j} = 0,\\
- 1 - \frac{\frac{\kappa}{2}-l}{w'_i} + \sum_{j \neq i} \frac{1}{w'_i-w'_j}- \sum_j\frac{1}{w'_i-w_j} = 0.
\end{align}
\end{subequations}

It is useful to denote the set of all Bethe roots $\{ w_i \} \cup \{ w'_i \}$ collectively as $\{ t_i \}$. To each Bethe root $w_i$ we associate the lowering operator $F_{w_i} = J^-_0$ in $\widetilde{\mathfrak{sl}}_2$ and to each Bethe root $w'_i$ the lowering operator $F_{w'_i} = J^+_{-1}$. By analogy with the finite-dimensional case \cite{Feigin:2006xs} and based on the expression for the Bethe vector in affine Gaudin models with regular singularities \cite{SV91}, we then conjecture that, associated with each solution of the Bethe equations \eqref{Bethe eq WZW} with $m$ Bethe roots, there is a corresponding Bethe vector in the spin $l$ module given by
\begin{equation} \label{Bethe vector WZW}
|\{ t_i \} \rangle = \sum_{\sigma \in S_m} \frac{F_{t_{\sigma(1)}} F_{t_{\sigma(2)}} \ldots F_{t_{\sigma(m)}} |l, \kappa \rangle}{\big(t_{\sigma(1)} - t_{\sigma(2)}\big)\big(t_{\sigma(2)} - t_{\sigma(3)}\big) \ldots \big(t_{\sigma(m-1)} - t_{\sigma(m)}\big) t_{\sigma(m)}}.
\end{equation}
This state is singular under the zero-mode $\mathfrak{sl}_2$ subalgebra. Moreover, its Virasoro level is equal to the number $\# w'$ of Bethe roots $w'_i$ and its spin is $l + \# w' - \# w$.

The leading non-trivial term in the UV expansion of the transfer matrix is proportional to the zero-mode quadratic Casimir $T^{(2)}=J^a_0 J^a_0$. Since \eqref{Bethe vector WZW} has definite spin it is an eigenvector of $T^{(2)}$ with eigenvalue $2(l + \# w' - \# w) (l +1 + \# w' - \# w)$. From the subleading term in the UV expansion we obtain the operator
\begin{equation}
T^{(3)} = \sum_{n>0} \frac{i}{2 n} f_{abc} J^a_{-n} J^b_0 J^c_{n} + \sum_{n>0} \frac{2}{n} J^a_{-n}J^a_{n}.
\end{equation}
We checked that this non-trivial operator is indeed diagonalized by the examples of Bethe vectors in Subsection \ref{sec:Examples}. It would be very nice to derive the Bethe equations directly from the diagonalization of $T^{(3)}$ with the Bethe vector ansatz \eqref{Bethe vector WZW}. Moreover, we conjecture that the expectation value of $T^{(3)}$ in the generic eigenstate \eqref{Bethe vector WZW} is
\begin{equation} \label{T3 eval conj}
\langle \{ t_i \} | T^{(3)} |\{ t_i \} \rangle = - 2 (l +1 + \# w' - \# w) \bigg( \sum_i w'_i - \sum_j w_j \bigg).
\end{equation}
A similar conjecture was made in \cite{Feigin:2017gcv} for the eigenvalue of the first non-local integral of motion of the affine $\widetilde{\mathfrak{sl}}_2$ Gaudin model describing quantum KdV theory as a coset CFT. We also expect from the UV expansion of the corresponding $\lambda$-oper in Appendix \ref{app: susy poly} that the eigenvalues of all the higher order UV expansion coefficients $T^{(n)}$ of the transfer matrix are given by supersymmetric polynomials in the Bethe roots $\{ w_i \} \cup \{ w'_j \}$. This was conjectured in \cite{Feigin:2017gcv} for the higher non-local charges of quantum KdV theory.

The expression for the next subleading term $T^{(4)}$ is much more complicated. We did check in Appendix \ref{app:UVexpansion} that the Bethe vector expectation value of the UV expansion of transfer matrices matches the Stokes data of the corresponding $\lambda$-opers.

The multichannel case can be treated similarly. For the product of WZW model we take the quadratic differential $P(x) = e^{2x} \prod_a (x-z_a)^{\kappa_a}$. We can then describe a generic state in the tensor product of spin $l_a$ modules which is singular under the total zero-mode $\mathfrak{sl}_2$ subalgebra using a pair of Miura $\lambda$-opers of the form
\begin{subequations} \label{Miura WZW oper N}
\begin{align}
a_+(x) &= - \sum_a \frac{l_a}{x - z_a} + \sum_i \frac{1}{x-w_i}- \sum_i \frac{1}{x-w'_i},\\
a_-(x) &= - 1 - \sum_a \frac{\frac{\kappa_a}{2}-l_a}{x - z_a} + \sum_i \frac{1}{x-w'_i}- \sum_i \frac{1}{x-w_i}
\end{align}
\end{subequations}
where $w_i$ and $w'_i$ are the Bethe roots satisfying the Bethe equations
\begin{subequations} \label{Bethe eq WZW N}
\begin{align}
 - \sum_a \frac{l_a}{w_i - z_a} + \sum_{j \neq i} \frac{1}{w_i-w_j}- \sum_j \frac{1}{w_i-w'_j} = 0,\\
- 1 - \sum_a \frac{\frac{\kappa_a}{2}-l_a}{w'_i - z_a} + \sum_{j \neq i} \frac{1}{w'_i-w'_j}- \sum_j\frac{1}{w'_i-w_j} = 0.
\end{align}
\end{subequations}
If we denote the set of all Bethe roots collectively as $\{ t_i \}$ then we conjecture that the corresponding Bethe vector in the tensor product of spin $l_i$ modules is given by
\begin{equation} \label{Bethe vector N WZW}
|\{ t_i \} \rangle = \sum_{\{ t_{i, j} \}}  \bigotimes_a \frac{F_{t_{a,1}} F_{t_{a,2}} \ldots F_{t_{a, m_a}} |l_a, \kappa_a \rangle}{\big(t_{a,1} - t_{a, 2} \big)\big(t_{a, 2} - t_{a, 3} \big) \ldots \big(t_{a, m_a-1} - t_{a, m_a}\big) \big( t_{a, m_a} - z_a \big)}
\end{equation}
where the sum is over all partitions of the set $\{ t_i \}$ into $N$ ordered subsets $(t_{a, 1}, \ldots t_{a, m_a})$ with $m_1 + \ldots + m_N = m$. It has Virasoro level $\# w'$ and spin $\sum_a l_a + \# w' - \# w$.

In the two-point case, the leading term in the UV expansion of the transfer matrix is proportional to
%\begin{equation} \label{T2 UV multi}
%\frac{J^a_{0,1} J^a_{0,1}}{(z-z_1)^2} + \frac{2 J^a_{0,1} J^a_{0,2}}{(z-z_1)(z - z_2)} + \frac{J^a_{0,2} J^a_{0,2}}{(z-z_2)^2}.
%\end{equation}
%Its leading term in the large $z$ expansion is equal to 
the total zero-mode quadratic Casimir $(J^a_{0,1} + J^a_{0,2})^2$ with eigenvalue $2(l_1 + l_2 + \# w' - \# w) (l_1 + l_2 +1 + \# w' - \# w)$ on the Bethe vector \eqref{Bethe vector N WZW}.
The next term in the UV expansion is proportional, in a suitable renormalization scheme, to
\begin{align*}
T^{(3)} &= \sum_{n>0} \frac{i}{2n} f_{abc} (J^a_{-n,1} + J^a_{-n,2}) (J^b_{0,1} + J^b_{0,2}) (J^c_{n,1} + J^c_{n,2})\\
&\qquad\qquad + \sum_{n>0} \frac{2}{n} (J^a_{-n,1} + J^a_{-n,2}) (J^a_{n,1} + J^a_{n,2}) - (z_1 J^a_{0,1} + z_2 J^a_{0,2}) (J^a_{0,1} + J^a_{0,2}),
\end{align*}
%where the zero-mode terms in the second line are the contributions from the subleading term in the large $z$ expansion of \eqref{T2 UV multi}.
We conjecture that its expectation value in the eigenstate \eqref{Bethe vector N WZW} is given by the supersymmetric polynomial in the Bethe roots
\begin{equation} \label{T3 N eval conj}
\langle \{ t_i \} | T^{(3)} |\{ t_i \} \rangle = - 2 (l_1 + l_2 +1 + \# w' - \# w) \bigg( z_1 l_1 + z_2 l_2 + \sum_i w'_i - \sum_j w_j \bigg).
\end{equation}

\subsection{Examples} \label{sec:Examples}

In this subsection we study the solutions to the Bethe equation \eqref{Bethe eq WZW} in the vacuum and spin $\frac{1}{2}$ WZW modules.

\subsubsection{Vacuum module}\label{sec:Ex:vacuummodule}

\paragraph{Level 1 states:}

\paragraph{\underline{$\#w' = 1$, $\#w = 0$:}} The Bethe equation \eqref{Bethe eq WZW} is just $\frac{\kappa}{w'} + 2 = 0$. This is inconsistent if $\kappa=0$. When $\kappa \neq 0$ we have $w' = - \frac{\kappa}{2}$ and the corresponding Bethe vector \eqref{Bethe vector WZW} is then proportional to
\begin{equation*}
| \{ w' \} \rangle \propto J^+_{-1} |0, \kappa \rangle.
\end{equation*}
This is singular when $\kappa = 0$, corresponding to the fact that there are no solutions to the Bethe equations in this case.

\paragraph{\underline{$\#w' = 1$, $\#w = 1$:}} No solution, as it should be since there is no singlet state at level $1$ in the vacuum module: $J^0_{-1} |0, \kappa \rangle = - \tfrac 12 J^-_0 J^+_{-1} |0, \kappa\rangle$ is a descendant.

\paragraph{Level 2 states:}

\paragraph{\underline{$\#w' = 2$, $\#w = 0$:}} The Bethe equations \eqref{Bethe eq WZW} have solutions if and only if $\kappa \neq 0, 1$. The Bethe vector corresponding to the cases $\kappa \neq 0, 1$ is proportional to
\begin{equation*}
| \{ w'_1, w'_2 \} \rangle \propto J^+_{-1} J^+_{-1} |0, \kappa \rangle.
\end{equation*}
which for $\kappa=1$ is singular and for $\kappa=0$ is a descendant of the singular vector $J^+_{-1} |0,0\rangle$.

\paragraph{\underline{$\#w' = 2$, $\#w = 1$:}} The Bethe equations have a solution if and only if $\kappa \neq 0, -1$. The corresponding Bethe vector is given by
\begin{align*}
| \{w'_1, w'_2, w_1 \} \rangle &= \bigg( \frac{1}{w'_1 - w'_2} \frac{1}{w'_2 - w_1} \frac{1}{w_1} + \frac{1}{w'_2 - w'_1} \frac{1}{w'_1 - w_1} \frac{1}{w_1} \bigg) J^+_{-1} J^+_{-1} J^-_0 |0, \kappa \rangle\\
&\quad + \bigg( \frac{1}{w'_1 - w_1} \frac{1}{w_1 - w'_2} \frac{1}{w'_2} + \frac{1}{w'_2 - w_1} \frac{1}{w_1 - w'_1} \frac{1}{w'_1} \bigg) J^+_{-1} J^-_0 J^+_{-1} |0, \kappa \rangle\\
&\qquad + \bigg( \frac{1}{w_1 - w'_1} \frac{1}{w'_1 - w'_2} \frac{1}{w'_2} + \frac{1}{w_1 - w'_2} \frac{1}{w'_2 - w'_1} \frac{1}{w'_1} \bigg) J^-_0 J^+_{-1} J^+_{-1} |0, \kappa \rangle,
\end{align*}
which is proportional to $J^+_{-2} |0, \kappa \rangle = \big( \tfrac 12 J^+_{-1} J^-_0 + J_{-1}^0 \big) J^+_{-1} |0, \kappa \rangle$. In particular, when $\kappa=0$ it is a descendant of the singular vector $J^+_{-1} |0, 0 \rangle$.

The situation when $\kappa = -1$ is more subtle since we see that the state $J^+_{-2} |0, -1 \rangle$ in the spin $0$ module of level $-1$ is not described by a solution of the Bethe ansatz. In the limit $\kappa \to -1$ of a solution of the Bethe equations for $\kappa \neq -1$, all the Bethe roots collide with the origin so that the Miura $\lambda$-oper \eqref{Miura WZW oper} becomes
\begin{equation} \label{generalised Miura}
a_+(x) = - \frac{1}{x}, \qquad
a_-(x) = - 1 + \frac{3}{2 x}.
\end{equation}
This is no longer of the form \eqref{Miura WZW oper} since the residues at the origin are not given by the pair $(-l, - \frac{\kappa}{2}+l) = (0, \tfrac 12)$ corresponding to the highest weight of the vacuum module at level $\kappa = -1$. However, the residues of \eqref{generalised Miura} do correspond to a shifted Weyl reflection of this highest weight. Indeed, the simple roots of $\widetilde{\mathfrak{sl}}_2$ act by shifted Weyl reflections on the residues at the origin as
\begin{equation*}
\cdots \longleftrightarrow (0, \tfrac 12) \longleftrightarrow (1, - \tfrac 12) \longleftrightarrow (-1, \tfrac 32) \longleftrightarrow \cdots
\end{equation*}
Therefore the pair \eqref{generalised Miura} describes a generalised Miura $\lambda$-oper in the sense of \cite{Frenkel:2003qx, Frenkel:2004qy}. We conjecture that the state $J_{-2}^+ |0, -1 \rangle$ is described by this generalised Miura $\lambda$-oper. This is checked to fourth order in the UV expansion in 
Appendix \ref{app:UVexpansion}.

\paragraph{\underline{$\#w' = 2$, $\#w = 2$:}} The Bethe equations admit a solution if and only if $\kappa\neq -2, 0$. The corresponding Bethe vector is proportional to
\begin{align*}
|\{ w'_1, w'_2, w_1, w_2 \} \rangle \propto J^a_{-1} J^a_{-1} |0, \kappa \rangle.
\end{align*}
This is singular when $\kappa=-2$, the critical level, and for $\kappa = 0$ it is a descendant of the singular vector $J^+_{-1} |0, 0\rangle$ since it can be written as
\begin{equation*}
J^a_{-1} J^a_{-1} |0, 0 \rangle = \big( \! - \tfrac 12 J^0_{-1} J^-_0 - \mbox{\small $\frac{1}{4}$} J^+_{-1} J^-_0 J^-_0 + \mbox{\small $\frac{1}{2}$} J^-_{-1} \big) J^+_{-1} |0, 0\rangle.
\end{equation*}

\subsubsection{Spin $\tfrac 12$ module}

\paragraph{Level 1 states:}

\paragraph{\underline{$\#w' = 1$, $\#w = 0$:}} The Bethe equations \eqref{Bethe eq WZW} have a solution if and only if $\kappa \neq 1$, in which case the Bethe vector is proportional to
\begin{equation*}
|\{ w'_1 \} \rangle \propto J^+_{-1} |\tfrac 12, \kappa \rangle.
\end{equation*}
This becomes singular at $\kappa=1$.

\paragraph{\underline{$\#w' = 1$, $\#w = 1$:}} The Bethe equations \eqref{Bethe eq WZW} are inconsistent when $\kappa = -2$ and for $\kappa \neq -2$ they have the unique solution $w'_1 = -(\kappa+2)$ and $w_1 = - \frac{1}{3}(\kappa+2)$.
The corresponding Bethe vector reads
\begin{align*}
|\{ w'_1, w_1 \} \rangle &= \frac{1}{w'_1 - w_1} \frac{1}{w_1} J^+_{-1} J^-_0 |\tfrac 12, \kappa \rangle + \frac{1}{w_1 - w'_1} \frac{1}{w'_1} J^-_0 J^+_{-1} |\tfrac 12, \kappa \rangle\\
&= \frac{3}{(\kappa+2)^2} (J^+_{-1} J^-_0 + J^0_{-1}) |\tfrac 12, \kappa \rangle.
\end{align*}
The vector $(J^+_{-1} J^-_0 + J^0_{-1}) |\tfrac 12, \kappa \rangle$ becomes singular when $\kappa= - 2$.

Note that when $\kappa=1$ we have the spin $\tfrac 12$ state
\begin{equation} \label{}
|\{ w'_1, w_1 \} \rangle = \mbox{\small $\frac{2}{3}$} (J^+_{-1} J^-_0 + J^0_{-1}) |\tfrac 12, 1 \rangle = 2 J^0_{-1} |\tfrac 12, 1 \rangle + \mbox{\small $\frac{2}{3}$} J^-_0 J^+_{-1} |\tfrac 12, 1 \rangle.
\end{equation}
The second term on the right hand side is a descendant of the singular vector $J^+_{-1} |\tfrac 12, 1 \rangle$ and is therefore zero in the spin $\tfrac 12$ module at level $\kappa=1$.

\paragraph{Level 2 states:}

\paragraph{\underline{$\#w' = 2$, $\#w = 0$:}} The Bethe equations \eqref{Bethe eq WZW} have no solution unless $\kappa \neq 1, 2$, in which case the corresponding Bethe vector is given by the spin $\frac{5}{2}$ state
\begin{equation*}
|\{ w'_1, w'_2 \} \rangle = \frac{1}{(\kappa-1)(\kappa-2)} J^+_{-1} J^+_{-1} |\tfrac 12, \kappa \rangle.
\end{equation*}
The state $J^+_{-1} J^+_{-1} |\tfrac 12, \kappa \rangle$ becomes singular when $\kappa=2$. When $\kappa=1$ it is a descendant of the singular vector $J^+_{-1} |\tfrac 12, 1 \rangle$.

\paragraph{\underline{$\#w' = 2$, $\#w = 1$:}} The Bethe equations \eqref{Bethe eq WZW} have two inequivalent families of solutions:
\begin{itemize}
  \item[$(i)$] The first family is valid for $\kappa \neq - \tfrac 12$ and the Bethe vector is proportional to
\begin{align*}
w_+ &= - \mbox{\small $\frac{1}{6}$} \Big( 3 + 8\kappa + \sqrt{41 + 64 \kappa + 64 \kappa^2} \Big) J^+_{-1} (J^0_{-1} + J^+_{-1} J^-_0) |\tfrac 12, \kappa \rangle\\
&\qquad\qquad\qquad\qquad\qquad\qquad\qquad\qquad\qquad\qquad + \mbox{\small $\frac{1}{3}$} (\kappa+2) J^+_{-2} |\tfrac 12, \kappa \rangle.
\end{align*}
This is singular for $\kappa = - \tfrac 12$. Moreover, it vanishes for $\kappa = -2$.
  \item[$(ii)$] The second family is valid for $\kappa \neq 1, -2$ with Bethe vector proportional to
\begin{align*}
w_- &= - \mbox{\small $\frac{1}{6}$} \Big( 3 + 8\kappa - \sqrt{41 + 64 \kappa + 64 \kappa^2} \Big) J^+_{-1} (J^0_{-1} + J^+_{-1} J^-_0) |\tfrac 12, \kappa \rangle\\
&\qquad\qquad\qquad\qquad\qquad\qquad\qquad\qquad\qquad\qquad + \mbox{\small $\frac{1}{3}$} (\kappa+2) J^+_{-2} |\tfrac 12, \kappa \rangle.
\end{align*}
When $\kappa=1$ we can rewrite this vector as $w_- = \mbox{\small $\frac{1}{3}$} \big( 5 J^0_{-1} + J^-_0 J^+_{-1} \big) J^+_{-1} |\tfrac 12, 1 \rangle$ which is thus a descendant of the singular vector $J^+_{-1} |\tfrac 12, 1 \rangle$. Likewise, when $\kappa=-2$ we obtain the state $w_- = \mbox{\small $\frac{13}{3}$} J^+_{-1} (J^0_{-1} + J^+_{-1} J^-_0) |\tfrac 12, -2 \rangle$ which is a descendant of the singular vector $(J^0_{-1} + J^+_{-1} J^-_0) |\tfrac 12, -2 \rangle$.
\end{itemize}

In conclusion, we have the following cases:
\begin{itemize}
  \item[$\bullet$] For $\kappa \neq -2, -\tfrac 12, 1$ we have two spin $\frac{3}{2}$ Bethe vectors $w_\pm$.
  \item[$\bullet$] For $\kappa = 1$ we have just one spin $\frac{3}{2}$ Bethe vector
\begin{align*}
w_+ = J^+_{-2} |\tfrac 12, 1 \rangle - 4 J^+_{-1} (J^0_{-1} + J^+_{-1} J^-_0) |\tfrac 12, 1 \rangle.
\end{align*}
  \item[$\bullet$] For $\kappa = - \tfrac 12$ we also have just one spin $\frac{3}{2}$ Bethe vector
\begin{align*}
w_- = \tfrac 12 J^+_{-2} |\tfrac 12, 1 \rangle + J^+_{-1} (J^0_{-1} + J^+_{-1} J^-_0) |\tfrac 12, 1 \rangle.
\end{align*}
  \item[$\bullet$] For $\kappa = -2$ there are no spin $\frac{3}{2}$ Bethe vectors.
\end{itemize}

\paragraph{\underline{$\#w' = 2$, $\#w = 2$:}} There are again two inequivalent families of solutions:
\begin{itemize}
  \item[$(i)$] The first family is valid for $\kappa \neq - 2, - \frac{7}{3}$ and the corresponding Bethe vector is proportional to
\begin{align*}
w_+ &= \Big( (\kappa+2) \big(\kappa+2+\sqrt{\kappa^2 + 16\kappa + 32} \big) J^+_{-2} J^-_0\\
&\qquad + \big( 5\kappa+12-\sqrt{\kappa^2 + 16\kappa + 32} \big) (J^+_{-1} J^-_{-1} + J^0_{-1} J^0_{-1})\\
&\qquad + \tfrac 12 \big( \kappa + \sqrt{\kappa^2 + 16\kappa + 32} \big) J^+_{-1}J^+_{-1}J^-_0J^-_0\\
&\qquad + \big( \kappa^2 - \kappa - 8 + (\kappa+3) \sqrt{\kappa^2 + 16\kappa + 32} \big) J^-_{-2} \Big) |\tfrac 12, \kappa \rangle.
\end{align*}
This vanishes when $\kappa=-2$. On the other hand, when $\kappa = - \frac{7}{3}$ we find
\begin{equation*}
w_+ = - J^+_{-1}J^+_{-1}J^-_0J^-_0 |\tfrac 12, \mbox{\small $-\frac{7}{3}$} \rangle
\end{equation*}
which has zero norm.
  \item[$(ii)$] The second family is valid for $\kappa \neq 1$ with Bethe vector proportional to
\begin{align*}
w_- &= \Big( \tfrac 12 \big(7\kappa+16-\sqrt{\kappa^2 + 16\kappa + 32} \big) J^+_{-2} J^-_0\\
&\qquad - \tfrac 12 \big( 3\kappa+8 + 3 \sqrt{\kappa^2 + 16\kappa + 32} \big) (J^+_{-1} J^-_{-1} + J^0_{-1} J^0_{-1})\\
&\qquad + 2 J^+_{-1}J^+_{-1}J^-_0J^-_0\\
&\qquad + \big( 5\kappa + 12 + \sqrt{\kappa^2 + 16\kappa + 32} \big) J^-_{-2} \Big) |\tfrac 12, \kappa \rangle.
\end{align*}
For $\kappa=1$ this vector can be rewritten as
\begin{equation*}
w_- = 2 \big( J^-_0 J^-_0 J^+_{-1} - 4 J^-_{-1} J^+_{-1} + 8 J^0_{-1} J^-_0 \big) J^+_{-1} |\tfrac 12, 1 \rangle
\end{equation*}
which is thus a descendant of the singular vector $J^+_{-1} |\tfrac 12, 1 \rangle$.
\end{itemize}

In conclusion, we have the following cases:
\begin{itemize}
  \item[$\bullet$] For $\kappa \neq - \frac{7}{3}, -2, 1$ we have two spin $\frac{1}{2}$ Bethe vectors $w_\pm$.
  \item[$\bullet$] For $\kappa = 1$ we have just one spin $\frac{1}{2}$ Bethe vector
\begin{equation*}
w_+ = 2 \big( 10 J^0_{-2} + 2 J^+_{-1} J^+_{-1} J^-_0 J^-_0 + 5 J^0_{-1} J^0_{-1} + 5 J^+_{-1} J^-_{-1} + 15 J^+_{-2} J^-_0 \big) |\tfrac 12, 1 \rangle.
\end{equation*}
  \item[$\bullet$] For $\kappa = - 2$ or $\kappa = - \frac 73$ we also have just one spin $\frac{1}{2}$ Bethe vector
\begin{align*}
w_- = 2 \big( 2 J^0_{-2} + J^+_{-1} J^+_{-1} J^-_0 J^-_0 - 2 J^0_{-1} J^0_{-1} - 2 J^+_{-1} J^-_{-1} \big) |\tfrac 12, \kappa \rangle.
\end{align*}
\end{itemize}

\section{The UV expansion}\label{sec:UVexpansion}
Kondo line defects in $\prod_i SU(2)_{k_i}$ WZW models are defined as the trace of the path ordered exponential
\begin{equation}
\hat{T}_{\mathcal{R}}\left(\left\{g_{i}\right\}\right):=\operatorname{Tr}_{\mathcal{R}} \mathcal{P} \exp \left(i \int_{0}^{2 \pi} \sum_i g_{i} t^{a} J_{i}^{a}(\sigma)d \sigma \right)\label{eq:KondoLineT}
\end{equation}
where $t^a$ are the generators of the Lie algebra $\mathfrak{su}(2)$ and the trace is taken in an $\mathfrak{su}(2)$ representation $\mathcal{R}$, labeled by its dimension $n$ from now on. The factor $i$ in front of the integral is $\sqrt{-1}$. The WZW currents are denoted as $J^a_i(\sigma)$ for each $SU(2)$ factor. The integration is along the compact direction.

Following \cite{KondoGLW}, we adopt the convention that the physical RG flows start from asymptotically free defects and the couplings grow in the positive real direction approaching the infrared. Therefore the UV expansion is concerned with small positive $g_i$. Perturbatively in $g_i$, we can expand the exponential
\begin{equation}
\hat{T}_{n}\left(\left\{g_{i}\right\}\right)=n + \sum_{N=1}^{\infty} i^{N} \hat{T}_{n}^{(N)},
\end{equation}
where each $\hat{T}_{n}^{(N)}$ depend on the set $\{ g_i \}$, and perform the loop integral. In doing so, one carries out a careful and lengthy renormalization procedure since the currents don't commute with each other. This was done \cite{KondoGLW} following the prescription given in \cite{Bachas:2004sy}. We are interested in the expectation value of the Kondo line operator in a generic state $|\ell\rangle$, with $\ell$ denoting the list of quantum numbers of the state. Details can be found in Appendix \ref{app:UVexpansion}.

If the twist $\alpha_+$ is nonzero, the Kondo line defect comes with a twist,
\begin{equation}
\hat{T}_{\mathcal{R}}\left(\left\{g_{i}\right\}, \alpha_+ \right):=\operatorname{Tr}_{\mathcal{R}} e^{i 2\pi\alpha_+ t^0}\mathcal{P} \exp \bigg(i \int_{0}^{2 \pi} \sum_i g_{i} t^{a} J_{i}^{a}(\sigma) d \sigma \bigg).\label{eq:KondoLineTwithtwist}
\end{equation}
Therefore the leading order is simply given by the character in the representation $\mathcal{R}$ of dimension $n$, namely
\begin{equation}
\Tr_\mathcal{R}e^{i 2 \pi\alpha_+ t^0} = \frac{\sin n \pi \alpha_+}{\sin \pi \alpha_+}. \label{eq:character}
\end{equation}
Note that in order for \eqref{eq:KondoLineTwithtwist} to make sense, the integrand inside the path ordered exponential has to be single-valued. Therefore, the inclusion of the nonzero twist in the trace forces us to work with the twisted affine Lie algebra. It is easy to see that this twisted affine algebra is precisely the one we get by acting with $\mathrm{U}_{\alpha_+}$ defined in \eqref{eq:spectralflow} on the untwisted affine algebra $\widehat{\mathfrak{sl}}_2$.

It was proposed\footnote{The vacuum module at $k=1$ and other related ODEs have been proposed and studied in \cite{Bazhanov_2003,Lukyanov_2004,Lukyanov_2004_1,Lukyanov_2006,Lukyanov_2007,Lukyanov_2007_1,Lukyanov_2013}.} in \cite{KondoGLW} that the expectation values in a state $|\ell\rangle$ of such a Kondo line defect will coincide with the transport coefficients of the Miura $\lambda$-oper, where the quadratic differential is $P(x) =e^{2x}\prod_i(x -z_i)^{k_i} $ for  $\prod_i SU(2)_{k_i}$ WZW and where $t(x)$ is constructed from solutions of the Bethe equations for the state $|\ell \rangle$. The corresponding Schr\"odinger equation reads
\begin{equation}
\partial_x^2 \psi(x) = \left(\frac{1}{\lambda^2}e^{2x}\prod_i(x -z_i)^{k_i}+ t(x) \right)\psi(x).
\end{equation}

The UV perturbative expansion is available whenever the Stokes data for the Schr\"odinger equation becomes close to the 
Stokes data for the simpler equation 
\begin{equation}
\partial_x^2 \psi(x) = e^{2x}\psi(x).
\end{equation}

In order to study the UV asymptotics ($0< g_i \ll 1$), we rewrite this as
\begin{equation}
\partial_x^2 \psi(x) = \left(e^{2\theta}e^{2x}\prod_i(1 + g_i x)^{k_i}+ t(x) \right)\psi(x) \label{eq:UVexpansionODE}
\end{equation}
where $e^{2\theta} = \frac{1}{\lambda^2} \prod_i g_i^{-k_i}$ is identified with an RG scale. When $t(x)= 0$, the Miura $\lambda$-oper describes the vacuum state whereas non-trivial $t(x) = a_+(x)^2 + \partial_x a_+(x)$ that we build from the solution to the Bethe equations \eqref{Bethe eqs} describes a more general state $|\ell\rangle$.

Let $\psi(x;\theta)$ be the unique \textit{small solution}, whose precise meaning is defined in the next section, that decays exponentially fast along the ray of large real positive $x+\theta$. The normalization is chosen to match asymptotically the WKB series in \eqref{eq:psi0WKB}. By the ODE/IM correspondence, we identify 
\begin{equation}
T_{n;\ell}(\theta) \equiv \langle \ell |\hat{T}_n|\ell \rangle = i \left(\psi\Big(x;\theta- \frac{i\pi n}{2}\Big), \psi\Big(x,\theta+\frac{i \pi n}{2}\Big) \right). \label{eq:TasWonskian}
\end{equation}
As the transfer function $T_{n;\ell}(\theta)$ is independent of $x$, we can evaluate the Wronskian in a convenient region where the explicit form of $\psi$ is accessible. For the case of $SU(2)_k$ chiral WZW model, this region is $1/g \gg -x \gg 0$. For $\prod_{i=1}^m SU(2)_{k_i}$, one can define an overall scaling parameter $g$ defined as $ \frac{1}{g} = \frac{1}{m} \sum_{i=1}^m \frac{1}{g_i}$ for which the relevant region is $1/g \gg -x \gg 0$ (see Appendix F.3 of \cite{KondoGLW} for details).
This allows for a systematic expansion in powers of the $g_i$. 

In the same vein as in \cite{Bazhanov:1998wj,Dorey:1999uk}, we can also define $Q$-functions, essentially as the coefficients of 
 $\psi(x)$ in an expansion at large negative $x$. If the twist $\alpha_+$ in $a_+(x)$ is zero, then the construction of $Q$-functions is given in \cite{KondoGLW} and reviewed in Appendix \ref{app:UVexpansion}. As a result, \eqref{eq:TasWonskian} becomes expressible as a quantum Wronskian
\begin{equation}
T_{n;\ell}(\theta)=\frac{i}{2\ell+1}\left[Q_{\ell}\left(\theta+\frac{i \pi n}{2}\right) \widetilde{Q}_{\ell}\left(\theta-\frac{i \pi n}{2}\right)-Q_{\ell}\left(\theta-\frac{i \pi n}{2}\right) \widetilde{Q}_{\ell}\left(\theta+\frac{i \pi n}{2}\right)\right], \label{eq:TasQQ}
\end{equation}
a form that is familiar from the integrability literature. 

This turns out to be a useful tool in the perturbative calculations as well. We can find the expression for $Q$ and $\tilde{Q}$ to sufficient order in $g$ by comparing to a direct perturbative evaluation of \eqref{eq:UVexpansionODE}. For the ground state in a generic spin $l$ module, the above claim has been verified in \cite{KondoGLW}. In this paper we will present a few examples of the claim for excited states in Appendix \ref{app:UVexpansion}.

\subsection{The coset scaling limit}
We can scale the variable $x$ in the $\lambda$-oper to get a slightly different parameterization
\begin{equation}
\partial_x^2 \psi(x) = \left(\frac{ \alpha^{2 + \sum_i k_i}}{\lambda^2}e^{2 \alpha x}\prod_i(x - \alpha^{-1} z_i)^{k_i}+\alpha^2 t(\alpha x) \right)\psi(x).
\end{equation}
A scaling limit $\alpha \to 0$ while keeping $\alpha^{-1} z_i$ and $\frac{ \alpha^{2 + \sum_i k_i}}{\lambda^2}$ fixed 
will bring this to a $\lambda$-oper which is naturally associated to integrable lines in a coset model.  

Physically, we are sending the $g_i$ to infinity while keeping their ratios fixed and adjusting the RG flow scale. 
We expect the Kondo lines RG flow in that limit to admit an intermediate regime where the line defects become 
effectively transparent to the overall WZW currents, so that they can be identified with defect lines in a coset model. 
It would be nice to explore this limit more carefully. 

\section{The IR expansion}\label{sec:IRexpansion}

The Kondo line defects $L_j[\theta]$, where $j$ is the spin that labels the $SU(2)$ representation and $\theta$ is the spectral parameter, 
are asymptotically free line defects defined in a product of WZW models. They have a non-trivial, possibly non-perturbative RG flow
which can be explored by looking at their action on the circle Hilbert space with the help of the ODE/IM correspondence. 
In the UV, the action is given perturbatively by the corresponding operator $\hat{T}_{j}$, defined in \eqref{eq:KondoLineT}. 

In the IR, the defects will flow to conformally-invariant defects. Because of their chiral nature, in the IR they will commute with both holomorphic and anti-holomorphic stress tensor and define topological line defects. A rich CFT such as the product of WZW models can have a very large variety of topological line defects, which commute with the stress tensor but with little else. 

The ODE/IM correspondence, though, gives immediate evidence that the IR limit of Kondo defects should be more special than that, and commute with all the Kac-Moody currents. Indeed, we will see that the far IR limit of the ODE/IM solution is controlled by a WKB leading answer which depends very little on the details of $t(z)$, 
up to the choice of $l_i$. In particular, they are blind to the details of the Bethe roots, which control which current descendants one is taking expectation values on. 

Line defects which commute with the whole current algebra of the product of WZW models are referred to as Verlinde line operators. They are labeled by the Kac labels, i.e. same as current algebra primary operator. They are products of individual Verlinde lines in each WZW factor. 

Denoted by $\mathcal{L}_j$, with $j = 0,\frac{1}{2},1, \dots, \frac{k}{2}$, their expectation value in the vacuum state, often called the quantum dimension, is given by
\begin{equation}
\langle \mathcal{L}_j\rangle  = d_{2j+1}^{(k)} \equiv \frac{\sin \frac{\pi}{k+2}(2j+1)}{\sin \frac{\pi}{k+2}}.
\end{equation}
In the ground state of spin $l$, or in any descendant of that, their expectation value is
\begin{equation}
\langle l, k| \mathcal{L}_j|l, k \rangle = d_{2j+1;l}^{(k)} \equiv \frac{\sin{\frac{\pi}{k+2}(2l+1)(2j+1)}}{\sin{\frac{\pi }{k+2}(2l+1)}}.
\end{equation}

If we go to the IR, but not to the infinitely far IR, the Kondo line defects will be described as IR free deformations of some sums of products of Verlinde lines. 
The deformation can involve any $SU(2)$-invariant local operators supported on the Verlinde lines. For generic Verlinde lines, 
there are many such operators, looking like descendants of chiral primary fields of various spins. We will see that the subleading WKB corrections 
do generically involve fractional powers of the scale $e^\theta$ which can be explained by the conformal dimension of these operators. 

As we mentioned in Section \ref{sec:affine}, something special happens when the far IR defect is the identity line, or some other Verlinde line which does {\it not}
support non-trivial chiral primaries. In such a situation, the IR free deformation must involve the integral along the line of $SU(2)$-invariant {\it bulk} chiral operators, 
starting with the stress tensor. The expectation values of these deformed identity lines are simply the exponential of the zero-modes of these 
bulk operators, which behave as local Hamiltonians for the affine Gaudin model. 

We will see below that the ODE/IM correspondence predicts such IR destiny for line defects associated with pairs of Stokes sectors which are joined by a generic 
WKB line\footnote{In the case of a single $\mathrm{SU}(2)$, there is one generic WKB line; see the end of Section \ref{sec:t(x)=0}.}. The number of such pairs is precisely the same as the number of zeroes for $\varphi(z)$, as expected from the classical affine Gaudin model. The WKB expansion of these reproduce the expectation values of the quantum local Hamiltonians.

\subsection{Vacuum state $t(x) = 0$}\label{sec:t(x)=0}
Let us first focus on the case of single $SU(2)$ and $t(x) = 0$. The Schr\"odinger equation takes the form
\begin{equation}
\partial_x^2 \psi(x)  = e^{2\theta} e^{2x} (1+gx)^k \psi(x).
\end{equation}
In this section, we are interested in the IR limit $\lambda^{-1} = e^{\theta} \rightarrow \infty$, where Voros/GMN-style WKB analysis is applicable. The analysis is essentially the same as in \cite{KondoGLW}, except that we are also interested in the sub-leading terms in the $\lambda$ expansion. 

We will briefly review the analysis and leave details in Appendix \ref{app:sec:wkb}. One starts by reading off the quadratic differential $P(x)dx^2 = e^{2x} (1+gx)^k dx^2$, which has a zero of order $k$ at $x_0 = -1/g$ and an exponential singularity at infinity. For any angle $\vartheta \in \mathbb{R}/2\pi \mathbb{Z}$, $\vartheta$-\emph{WKB lines} are curves in the complex plane where 
\begin{equation}
\mathrm{Im}\  \Big[e^{i \vartheta} \sqrt{P(x)}dx \cdot \partial_t\Big] = 0,
\end{equation}
where $\partial_t$ is the tangent vector of the curve. One such line passes through any point in the $x$ plane. Generic WKB lines go to positive infinity in both directions, joining two Stokes sectors there. Special WKB lines hit a zero of $P(x)$ such as $x = -\frac{1}{g}$ or flow to negative infinity.
\footnote{Special WKB lines are also sometimes called Stokes lines. In the situation at hand, there are two possible meaning for ``Stokes'': it may refer to the asymptotic expansion of a solution at large positive $x$, as in defining the Stokes data of the oper, 
or it may refer to the WKB asymptotic expansion at small $\lambda$. In order to avoid confusion, we use the terms ``WKB'' exclusively for the latter and ``Stokes'' for the former. \label{foot}}

The union of special WKB lines is called WKB diagram/spectral network. $\vartheta$ is chosen such that $e^\theta$ lies in the half plane centered on $e^{i \vartheta}$, where the WKB approximation gives the correct $e^{-\theta} \rightarrow 0$ asymptotics. The structure of the spectral network 
governs which solutions of the Schr\"odinger equation have a specific WKB asymptotic expansion.

\begin{figure}
	\centering
	\includegraphics[width=0.7\linewidth]{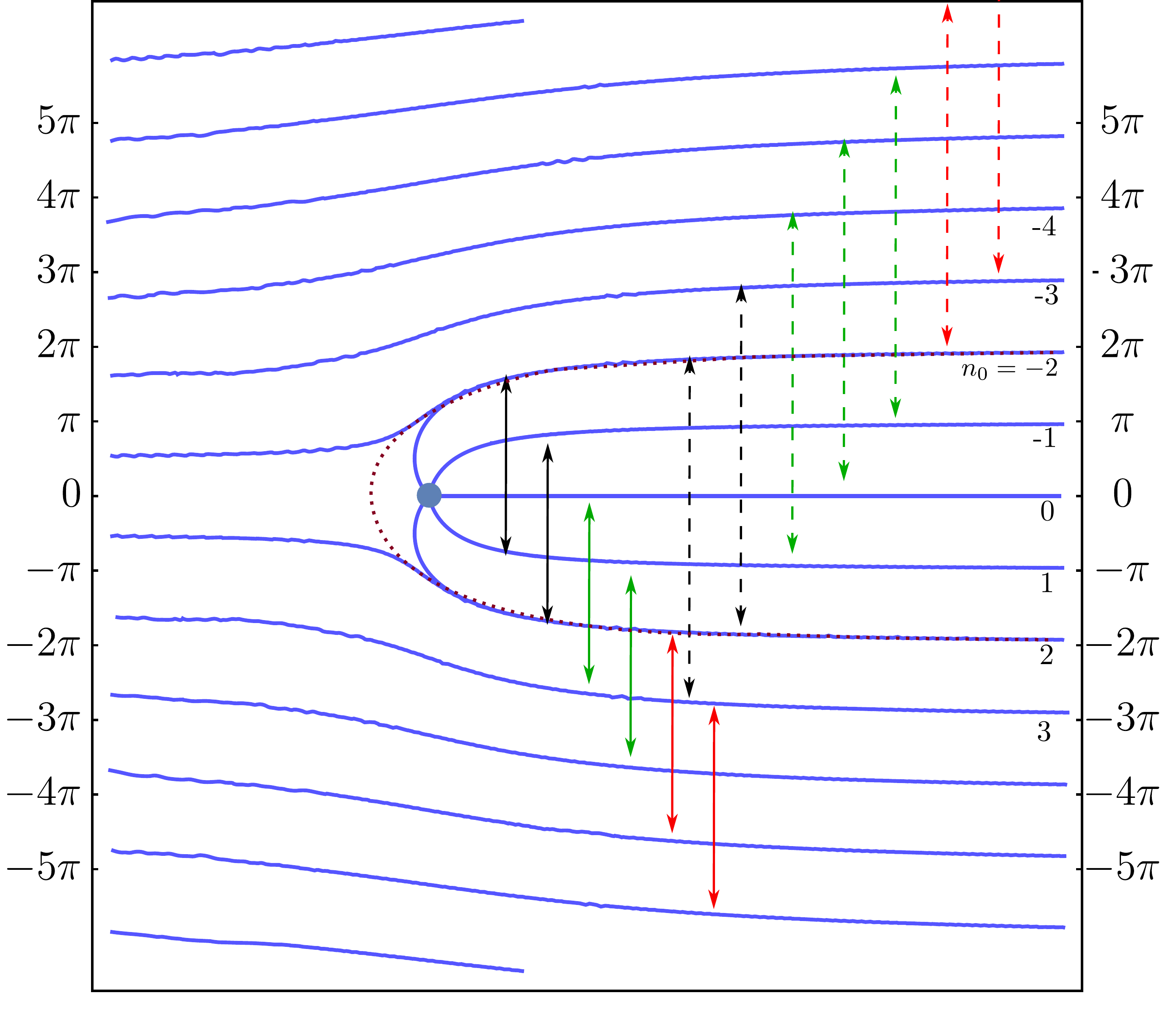}
	\caption{Blue curves are the Stokes diagram for $k=3$ and $\vartheta = 0$. There are five special WKB lines connected to the zero of order $k=3$. The rest of the special WKB lines connect to the negative infinity.  Double headed arrows indicate which two solutions are used in the Wronskian $i(\psi_{n_1},\psi_{n_2})$ with $n_2-n_1 = n=2j+1$ in different scenarios. The lower (upper) end points to the special WKB lines associated to the small solution $\psi_{n_2}$ ($\psi_{n_1}$). Solid lines ($j = 1$) depict scenarios $0 \leq j \leq \frac{k}{2}$. Colors (black, green and red) of the lines indicate three scenarios as we shift $\Im\theta$: physical strip ($L_j$ flows to $\mathcal{L}_{j}$ ), sequence of transitions ($L_j$ flows to $L^{\mathrm{IR}}_{j-\frac{s}{2}} \otimes \mathcal{L}_{\frac{s}{2}}$, $s = 2j-1, \dots, 1$) and $L_j^{\mathrm{IR}}$. Dashed lines ($j=2$) depict scenarios $j> \frac{k}{2}$ respectively. Colors (black, green and red) of the lines indicate three scenarios as we shift $\Im\theta$: physical strip ($L_j$ flows to $\mathcal{L}_{k/2}\otimes L^{IR}_{j-k/2}$ ), sequence of transitions ($L_j$ flows to $L^{\mathrm{IR}}_{j-\frac{s}{2}} \otimes \mathcal{L}_{\frac{s}{2}}$, $s = k-1, \dots, 1$) and $L_j^{\mathrm{IR}}$. The dotted burgundy line denotes the unique generic WKB line that gives rise to the local integrals of motion. }
	\label{fig:wkbexpk3}
\end{figure}

In our current example, the structure of the WKB diagram is shown in Fig. \ref{fig:wkbexpk3}. Special WKB lines go  towards positive infinity along the positive real $x+\theta+i \pi n$, $n \in \mathbb{Z}$ direction. There are $k+2$ of them that are connected to the order $k$ zero $x_0 = -\frac{1}{g}$. The remaining special WKB lines go towards negative infinity with imaginary part shifted by $\pm \frac{\pi}{2}k$. 

Next, one needs to find a set of solutions, referred to as \emph{small solutions}, which decrease exponentially fast along the Stokes lines towards positive infinity (and thus along WKB lines asymptoting to them). In particular, we define $\psi_0(x)$ to be the small solution that decreases fast along the line of large real positive $x+\theta$ and agrees with the WKB asymptotics along this line
\begin{equation}
\psi_{0}(x ; \theta) \sim \frac{1}{\sqrt{2 \partial S^{\mathrm{asym}}(x, e^{\theta})}} e^{-S^{\mathrm{asym}}(x,e^\theta)} \label{eq:psi0WKB}
\end{equation}
where $S^{\mathrm{asym}}(x,e^\theta)$ is the primitive of the WKB momentum, given by an asymptotic series in large $x$ and small $e^{-\theta}$. Although the WKB momentum is uniquely defined, its primitive needs a choice of integration constant. We choose the leading term to be
\begin{equation}
e^{\theta}\int_{-\frac{1}{g}}^{x} e^{y}(1+g y)^{\frac{k}{2}} d y= e^{\theta} e^{-\frac{1}{g}} g^{\frac{k}{2}} \int_{0}^{x+\frac{1}{g}} e^{y} y^{\frac{k}{2}} d y.
\end{equation}
Different choices clearly lead to different normalization of the Wronskians, namely $T$ functions \eqref{eq:TasWonskian}. This choice has the nice property later on that the exponent of \eqref{eq:Wronskiansjoinedatzero} is zero at the leading order. In a practical calculation involving subleading terms, one also needs to make a choice for every order in $e^{-\theta}$.

Then for all $n\in \mathbb{Z}$, we have a small solution $\psi_n(x)\equiv \psi_0(x;\theta+i\pi n)$ along the large positive real $x+\theta +i \pi n$ direction.

Next, we want to use the WKB network to evaluate the WKB asymptotics of the Wronskians. In the standard Voros/GMN-style WKB analysis,
one studies Wronskians between two of the small solutions joined by a generic WKB line. These Wronskians are controlled by the contour integral along the WKB line of the WKB one form whose leading term is $\sqrt{P(x)}dx$. 
This collection of Wronskians is incomplete, though, unless all zeroes of $P(x)$ are simple. 

As has been developed in \cite{KondoGLW} and Appendix \ref{app:sec:wkb}, WKB analysis can be generalized to study non-simple zeros. Roughly speaking, one also needs the information around the matching regions, which, in the current example, are the order $k$ zero $x_0 = -\frac{1}{g}$, and the large negative $x$. Correspondingly, one can derive WKB asymptotics for Wronskians between two of the small solutions joined by a special WKB line
to the same zero, or to negative infinity.

Following from the WKB diagram shown in Fig. \ref{fig:wkbexpk3} let us suppose, for convenience, that the numbering of the special WKB lines that are connected to the zero at $x = -\frac{1}{g}$ is $n_0, n_0+1, n_0+2, \dots, n_0+k+1$. The precise value of $n_0$ depends on the parity of $k$ and $\Im \theta$, which are given in \cite{KondoGLW} and are not important to us. There are three different scenarios:
\begin{itemize}
	\item Wronskians between two of the $k+2$ small solutions whose special WKB lines are connected to the zero, namely $i(\psi_n,\psi_{n'})$ for $n_0 \leq n < n' \leq n_0+k+1$. 
	\item Wronskians between small solutions whose special WKB lines are connected to large negative matching region (to be made precise below), namely $i(\psi_n,\psi_{n'})$ for $n < n' \leq n_0$ or $n_0 +k+1 \leq n < n'$ or $ n \leq n_0 <n_0+k+1 \leq  n'$.
	\item The remaining ones can be related to the first two scenarios using Pl\"ucker formula.
\end{itemize}

In particular, to deal with the first case, it is important to study the local behavior around the zero $x_0 = -\frac{1}{g}$. Locally around $x_0$, a zero of order $k$, the stress tensor should take the form $y^k +\dots$, with $y$ being the coordinate in the local coordinate system. Indeed, one can always find the coordinate transformation $x\rightarrow y(x)$ such that stress tensor takes the form of
\begin{equation}
y^k +a_{k-2} \gamma^{k} y^{k-2}+ \dots +  a_{j} \gamma^{2+j} y^{j} +\dots+ a_0 \gamma^{2} \label{eq:localcoordT}
\end{equation}
where $\gamma = e^{-\theta \frac{2}{k+2}}$. Importantly there are $k-1$ coefficients $a_j = a_j^{(0)} + \gamma^{k+2} a_j^{(2)} + \gamma^{2(k+2)} a_j^{(4)} +\dots$ that are uniquely fixed in $\gamma^{k+2}$ asymptotics. One can find a set of nice solutions $A_{k;i}(y)$ to this local problem and evaluate the Wronskian perturbatively. The general procedure to do this is described in Appendix \ref{app:sec:wkb}. On the other hand, Wronskians between small solutions $\psi_n(x)$ are equal to the Wronskians between the corresponding local solutions $A_{k;i}(y)$ with a careful treatment on the normalization of the solutions.
We will only quote the result here, leaving the details in Appendix \ref{app:sec:wkb},
\begin{equation}
i\left(\psi_{n}, \psi_{n^{\prime}}\right) \sim  e^{0 + O(e^{-\theta})}\Big(d_{n^{\prime}-n}^{(k)} +  O(\gamma^2)\Big) \label{eq:Wronskiansjoinedatzero}
\end{equation}
whose leading term is given by the quantum dimension defined as
\begin{equation}
d^{(k)}_n = \frac{e^{\frac{\pi i}{k+2}n}-e^{-\frac{\pi i}{k+2}n}}{e^{\frac{\pi i}{k+2}}-e^{-\frac{\pi i}{k+2}}}.
\end{equation}
Subleading terms are computable order by order in $\gamma$. See Appendix \ref{app:sec:wkb} for the general prescription. There are two exceptions $k=1,2$ where we can calculate Wronskians exactly. The important part is that the corrections to the Wronskians come in as integer power of $\gamma$ but start from $\gamma^2$ order.

In the second scenario, the special WKB lines `meet' at the large negative $x$. It turns out, for some suitably chosen $x_{-\infty}$, a shift of the coordinate $x \rightarrow \delta = x-x_{-\infty}$ will transform the quadratic differential into
\begin{align}
e^{2\delta} \left(1+ g_{\mathrm{eff}}(\theta){\delta}\right)^k.
\end{align}
The details are given in Appendix \ref{app:sec:wkb}. Here, what matters to us is that $x_{-\infty}$ has a large negative real part and in the IR limit $\theta\rightarrow \infty$ we have
\begin{equation}
x_{-\infty} \sim -\theta -\frac{1}{2} k \log (-g\theta) -\frac{k^2}{4} \frac{\log(-g\theta)}{\theta}+ O\Big( \frac{1}{\theta} \Big).
\end{equation}
The coupling $g_{\mathrm{eff}}(\theta)$ is defined by the relation 
\begin{equation}
x_{-\infty}(\theta) = \frac{1}{g_{\mathrm{eff}}(\theta)} - \frac{1}{g}
\end{equation}
and goes to $0$ in the IR limit $\theta\rightarrow \infty$. This is precisely the effective coupling for the infrared free line defect, whose physical meaning will be given below in Section \ref{sec:physicalmeaning}. For now, we only need the fact that $g_{\mathrm{eff}}(\theta)\rightarrow 0$ as $\theta\rightarrow \infty$. Therefore, we can study the Wronskians of solutions in $g_{\mathrm{eff}}$ expansion.

In the leading order, the local solutions are given by Bessel functions. Therefore by means of Bessel function identities and with normalization factors carefully taken into account, the results are
\begin{equation}
i (\psi_n, \psi_{n'}) \sim \exp {\left(\frac{(-1)^n+(-1)^{n'}}{2}  e^{i \frac{\pi k}{2}} m_k(g) e^\theta + O(e^{-\theta})\right)}\Big[(n'-n) + O(g_{\mathrm{eff}}^2) \Big]
\end{equation}
whenever $n \leq n_0$ and $n' \leq n_0$,
\begin{equation}
i (\psi_n, \psi_{n'}) \sim  \exp {\left(\frac{(-1)^n+(-1)^{n'}}{2}  e^{-i \frac{\pi k}{2}} m_k(g) e^\theta + O(e^{-\theta})\right)}\Big[(n'-n) + O(g_{\mathrm{eff}}^2) \Big]
\end{equation}
whenever $n \geq n_0 + k + 1$ and $n' \geq n_0 + k + 1$, and
\begin{equation}
i (\psi_n, \psi_{n'}) \sim   \exp \left({\frac{(-1)^n e^{i \frac{\pi k}{2}} +(-1)^{n'} e^{-i \frac{\pi k}{2}}}{2} e^\theta  m_k(g)}  + O(e^{-\theta})\right) \Big[(n'-n-k) + O(g_{\mathrm{eff}}^2)\Big]
\end{equation}
whenever $n \leq n_0$ and $n' \geq n_0 + k + 1$. 

In the third scenario, we can just use Pl\"ucker formula to reduce to the previous two scenarios. Details can be found in \cite{KondoGLW}.

Recall that the leading order of the second term agrees with the one from the UV expansion and intuitively just counts the number of spacing between different special WKB lines at the left hand side of the special WKB diagram, see e.g. Fig. \ref{fig:wkbexpk3}. The exponential factor is the non-perturbative ground state energy shift.

As we discussed in Section \ref{sec:affine} and at the beginning of this section, the vevs of local integrals of motion for the affine Gaudin model are given by the Wronskians which correspond to generic WKB lines. In the example at hand, there is only one such line depicted by the dotted burgundy line in Fig. \ref{fig:wkbexpk3}, corresponding to $\mathcal{L}_{\frac{k}{2}}$ in the infrared. Indeed, the $\mathcal{L}_{\frac{k}{2}}$ line does not support nontrivial chiral WZW primaries. Since the corresponding Wronskian $i(\psi_{-\frac{k+1}{2}},\psi_{\frac{k+1}{2}})$ is controlled by the contour integral of the WKB momentum along the generic WKB line, it doesn't involve the local analysis around the zero or negative infinity, hence it is simply organized by odd powers of $e^{-\theta}$. 

\subsection{$t(x)\neq 0$}
When $t(x) = a_+(x)^2 + \partial a_+(x) \neq 0$, the evaluation of the Wronskians via WKB analysis is basically the same as the previous section except for the following modifications.

In the first scenario of the previous section, namely, for the Wronskians of two solutions whose special WKB lines are connected at the zero $x = -\frac{1}{g}$, the local coordinate system in general has an additional piece, compared to \eqref{eq:localcoordT}
\begin{equation}
y^k +a_{k-2} \gamma^{k} y^{k-2}+ \dots +  a_{j} \gamma^{2+j} y^{j} +\dots+ a_0 \gamma^{2} + \frac{l(l+1)}{y^2} \label{eq:localcoordTnonzerot}
\end{equation}
where $-l$ is the residue of $a_+(x)$ at the zero. This will change the leading order of \eqref{eq:Wronskiansjoinedatzero} to  be
\begin{equation}
d_{2j+1;l}^{(k)} \equiv \frac{\sin{\frac{\pi}{k+2}(2l+1)(2j+1)}}{\sin{\frac{\pi }{k+2}(2l+1)}}.
\end{equation}

The nonzero regular part of $a_+(x)$ has smaller impact. It will change the coefficients $a_j$, the details of the map $x\mapsto y(x)$, and therefore the higher order corrections. But importantly, the corrections are still organized by integer powers of $\gamma$.

In the second scenario, where two special WKB lines are connected at the negative infinity, we can again go to the coordinate in $\delta = x-x_{-\infty}$ where the quadratic differential reads
\begin{align}
e^{2\delta} \left(1+ g_{\mathrm{eff}}(\theta){\delta}\right)^k + t\Big(x\mapsto \delta + \frac{1}{g_{\mathrm{eff}}}\Big).
\end{align}
We are then in a situation very similar to the UV expansion. Therefore, the higher order corrections of the Wronskians come in powers of $g_{\mathrm{eff}}$.
\subsection{Physical interpretation}\label{sec:physicalmeaning}

According to the ODE/IM correspondence \eqref{eq:TasWonskian}, which we repeat here, the expectation value of the Kondo line operator in the state $|\ell\rangle$ is given by
\begin{equation}
T_{n;\ell}(\theta) \equiv \langle \ell |\hat{T}_n|\ell \rangle = i \left(\psi_0\Big(x;\theta- \frac{i\pi n}{2}\Big), \psi_0\Big(x,\theta+\frac{i \pi n}{2}\Big) \right)
\end{equation}
where $n= 2j+1$. We showed in Section \ref{sec:UVexpansion} that the leading term in the UV expansion is given by the dimension $n$ of the representation, $T_{n;\ell}\sim n + \dots$.

We now provide the physical implication of the IR expansion we evaluated using WKB analysis previously in this section. The IR expansion of $T_{n;\ell}(\theta)$ reviews an interesting infrared structure. The leading order has been demonstrated in \cite{KondoGLW}. We will review briefly now and explain how the structure of higher order corrections we obtained in this section fits in the paradigm.

Depending on the imaginary part of $\theta$, and whether $0\leq 2j\leq k$ or $2j>k$, the RG flow takes the Kondo line operator $L_j[\theta]$ to different IR line operators. 

Firstly, if $\theta$ is real, or more precisely, valued in a strip around the real $\theta$ axis of width about\footnote{This is true up to $\pm \pi/2$, depending on the parity of $k$ and $n$.} $(n-k-1)\pi$, we have the physical RG flow\footnote{Recall that $e^\theta$ labels the RG scale, so the physical RG flow corresponds to real $\theta$.}:
\begin{itemize}
	\item For $0 \leq j\leq \frac{k}{2}$, over/exact-screening\footnote{This terminology is based on the intuitive physical picture that Kondo defect disappears in the IR because magnetic impurity spin is screened by the bulk fermions. See e.g. \cite{Affleck:1995ge} for more details.}, $T_{n;\ell}(\theta)\sim d_{2j+1;l}^{(k)}$, $L_j$ flows to $\mathcal{L}_j$,
	\item For $j>\frac{k}{2}$, under-screening, $T_{n;\ell}(\theta)\sim e^{-E(n,\ell,k)e^\theta} d_{k+1;l}^{(k)}(n-k)$, $L_j$ flows to $\mathcal{L}_{k/2}\otimes L^{IR}_{j-k/2}$.
\end{itemize}
Second, if we increase the imaginary part of $\theta$ either positively or negatively, there is an interesting sequence of transitions starting from $|\Im \theta|  \sim \frac{(n-k-1)\pi}{2}$, the edge of the strip mentioned above. Every time $|\Im \theta|$ increases by $\pi$, we trade one unit of spin for the topological defect with one unit of spin for the internal degrees of freedom. More precisely, $L_j$ flows to $L^{\mathrm{IR}}_{j-\frac{s}{2}} \otimes \mathcal{L}_{\frac{s}{2}}$, $s = k,k-1, \dots, 0$. After $s$ decreases to zero, i.e. when $|\Im \theta|$ is large enough, we will have the circular RG flows where $L_j$ flows to $L_j^{\mathrm{IR}}$.

Since $e^\theta$ labels the RG scale and $\theta \rightarrow \infty$ is the deep infrared, we have just demonstrated that the leading term of $T_{n;\ell}(\theta)$ simply tells us which infrared line defects we flow to starting from $L_j$ defined in the UV. Correspondingly, the far IR destiny of UV line defects 
can be determined from some simple combinatorics from the topology of WKB network.

Subleading terms are obviously due to the deformation that brings us away from the deep IR. More precisely, we need to look at the RG flow a bit away from the deep IR. In the case of the RG flow that flows to the Kondo line defect $L_j^{\mathrm{IR}}$, an infrared free line defect, the effective coupling $g_{\mathrm{eff}}(\theta)$ is negative and becomes smaller in the IR. Therefore, the corrections are expected to come in powers of $g_{\mathrm{eff}}$, which is small and negative. 

On the other hand, topological lines in the infrared are not free and appear in the RG flow as strongly coupled infrared fixed points\footnote{Unless the levels $k_i$ are large, in which case the RG flow to a topological line can be perturbative. See the next section.}. Nevertheless, we know a lot about the local operators that are supported on the lines. Among them, it is the least irrelevant operator that contributes the corrections the closest to the deep IR fixed point \cite{Affleck:1990iv}. On a spin $0<j< \frac{k}{2}$ Verlinde line in chiral WZW, it is the WZW descendant of the spin $1$ operator of dimension $\frac{2}{k+2}$, denoted by $J_{-1}^a \phi^a$. It is a Virasoro primary of scaling dimension $1+ \frac{2}{k+2}$ and has zero one-point function. Then a simple dimensional analysis implies that there will be corrections in powers of $\gamma = e^{-\theta\left(\frac{2}{k+2} \right)}$, starting from $\gamma^2$. Another obvious candidate $J^aJ^a$ has dimension $2$, thus contributing corrections in integer powers of $\hbar = \gamma^{\frac{k+2}{2}}$, which is more irrelevant. Note that the spin $1$ operator $\phi^a$ is not supported on the Verlinde line $\mathcal{L}_{\frac{k}{2}}$, obtained in the exact screening case. This is precisely what we found in the WKB analysis around \eqref{eq:localcoordT} and \eqref{eq:Wronskiansjoinedatzero}.

\section{Some comments about the semiclassical limit}\label{sec:semi}
The RG flow of the Kondo defects is often non-perturbative. An important exception occurs when the levels $k_i$ are large: in an appropriate RG scheme the 
couplings remain small all the way to the IR. It is useful to illustrate this in the $N=1$ case. 

We can start from the quadratic differential $\lambda^{-2} e^{2 x} x^k dx^2$. The RG prescription we used before for the couplings sets 
\begin{equation}
\lambda = e^{\frac{1}{g}} g^{- \frac{k}{2}}
\end{equation}
so that the quadratic differential becomes $e^{2 x- \frac{2}{g}} (gx)^k dx^2$ which can be mapped by a translation to $e^{2 x} (1 + gx)^k dx^2$
and treated perturbatively. In these conventions, $g$ is small as $\lambda \gg 1$ in the UV, but flows all the way to infinity in the IR  as $\lambda \ll 1$. 

When $k \gg 1$, these RG conventions are not very good. For example, even if $g \sim k^{-1}$ is small, $(1 + gx)^k \sim e^{g k x}$ is not close to $1$. 
Consider, instead, a different definition of $g$ where we employ both a translation and a scale transformation of $x$ to arrive to some $e^{2 \alpha(g) x} (1 + gx)^k dx^2$.
We can map this back to $e^{2  x- \frac{2 \alpha(g)}{g}} g^k x^k \alpha(g)^{-k-2} dx^2$ and thus to 
\begin{equation}
\lambda = e^{\frac{\alpha(g)}{g}} g^{- \frac{k}{2}}\alpha(g)^{1+\frac{k}{2}}.
\end{equation}
If we select $\alpha(g) = 1 - \frac{k}{2} g$, then $e^{2 \alpha(g) x} (1 + gx)^k dx^2 \sim e^{2 x} dx^2$ up to corrections of order $g^2 k$, which are small even if $g \sim k^{-1}$. 

This seems a more reliable way to deal with the perturbative RG flow. Now we have 
 \begin{equation}
\lambda = e^{\frac{1}{g} - \frac{k}{2}} g \Big( g^{-1} - \frac{k}{2} \Big)^{1+\frac{k}{2}},
\end{equation}
which flows from $g=0$ to $g = \frac{2}{k}$ from the UV to the IR, so that the coupling is perturbatively small all the way. 

This RG scheme also seems appropriate to make contact with the classical affine Gaudin model in a $k \to \infty$ semiclassical limit. 
In general, define 
\begin{equation}
g_i = \frac{1}{\varphi(z)} \frac{1}{z-z_i}.
\end{equation}
This definition is inspired by the classical affine Gaudin Lax matrix \eqref{Lax matrix}.

Then 
\begin{equation}
e^{\frac{2x}{\varphi(z)}} \prod_i (1+g_i x)^{k_i} dx^2
\end{equation}
can be mapped to 
\begin{equation}
e^{2x} \prod_i \Big(1+\frac{1}{z-z_i} x \Big)^{k_i} \varphi(z)^2 dx^2
\end{equation}
and then to
\begin{equation}
\frac{e^{- 2 z} \varphi(z)^2}{\prod_i (z-z_i)^{k_i}} e^{2x} \prod_i (x-z_i)^{k_i} dx^2
\end{equation}
so that we have a $z$ RG flow controlled by 
 \begin{equation}
\lambda =\varphi(z)^{-1} e^z \prod_i (z-z_i)^{\frac{k_i}{2}}
\end{equation}
and fixed points at $z \sim z_i$ where $g_i$ is finite and small, while all other $g_j$ vanish. 

\section{Generalizations: $\mathfrak{sl}_3$} \label{sec: sl3}

The discussion in Section \ref{sec: opers} of the affine $\mathfrak{sl}_2$ opers, the Bethe equations and WKB solutions can be easily generalized to higher rank Lie algebras. We demonstrate the case of $\mathfrak{sl}_3$ in this section. We leave a proper discussion of the corresponding Kondo defects and ODE/IM solutions to future work.

\subsection{Basic Definitions}
An $\mathfrak{sl}_3$ oper is a complexified third order differential operator 
\begin{equation} \label{Schro oper 3}
\partial_x^3 - t_2(x) \partial_x + t_3(x)
\end{equation}
with a natural transformation law under a change of coordinate
\begin{equation}
\partial_x^3 - t_2(x) \partial_x + t_3(x) = ( \partial_x \tilde x)^{2} \Big(\partial_{\tilde x}^3 - \tilde t_2(\tilde x) \partial_{\tilde x} + \tilde t_3(\tilde x) \Big) ( \partial_x \tilde x).
\end{equation}
We will work with $\mathfrak{sl}_3$ opers for which both $t_2(x)$ and $t_3(x)$ are rational functions.

The data of an $\mathfrak{sl}_3$ oper is equivalent to that of a flat connection 
\begin{equation} \label{canonical form3}
\partial_x + \begin{pmatrix}0 & t_2(x) &t_3(x) \cr 1 & 0 & 0 \cr 0 & 1 & 0 \end{pmatrix}.
\end{equation}
More generally, an $\mathfrak{sl}_3$ oper can be described as a flat connection of the form
\begin{equation} \label{oper sl3}
\partial_x + \begin{pmatrix}a(x) &b(x) &c(x) \cr 1 & \tilde a(x) & \tilde b(x) \cr 0 & 1 & -a(x) - \tilde a(x) \end{pmatrix}.
\end{equation}
modulo gauge transformations by unipotent upper-triangular matrices. Any $\mathfrak{sl}_3$ oper has a unique \emph{canonical form} \eqref{canonical form3} where
\begin{subequations}
\begin{align}
t_2(x) &= a(x)^2 + a(x) \tilde a(x) + \tilde a(x)^2 + 2 \partial_x a(x) + \partial_x \tilde a(x) + b(x) + \tilde b(x),\\
t_3(x) &= - (a(x) + \tilde a(x)) \big( a(x) \tilde a(x) + \partial_x a(x) + 2 \partial_x \tilde a(x) - b(x) \big) \cr
&\qquad\qquad\qquad - \partial_x^2 a(x) - \partial_x^2 \tilde a(x) - a(x) \tilde b(x) - \partial_x \tilde b(x) + c(x).
\end{align}
\end{subequations}

An \emph{$\mathfrak{sl}_3$ $\lambda$-oper} is a complexified third order differential operator with a particular dependence on the auxiliary complex parameter $\lambda$ of the form
\begin{equation} \label{sl3 lambda-oper Schro}
\partial_x^3 - t_2(x) \partial_x + t_3(x) + \frac{P(x)}{\lambda^3},
\end{equation}
where $t_2(x)$ and $t_3(x)$ are rational functions.
Equivalently, we can describe this as a flat connection
\begin{equation} \label{lambda-oper sl3}
\partial_x + \begin{pmatrix}0 & t_2(x) \lambda & P(x) \lambda^{-1} + t_3(x) \lambda^2 \cr \lambda^{-1} & 0 & 0 \cr 0 & \lambda^{-1} & 0 \end{pmatrix}.
\end{equation}
More generally, an $\mathfrak{sl}_3$ $\lambda$-oper is defined as a connection of the form
\begin{equation} \label{lambda-oper sl3 general}
\partial_x + \begin{pmatrix} a(x) & b(x) \lambda & P(x) \lambda^{-1} + c(x) \lambda^2 \cr \lambda^{-1} & \tilde a(x) & \tilde b(x) \lambda \cr 0 & \lambda^{-1} & - a(x) - \tilde a(x) \end{pmatrix}
\end{equation}
where $a(x)$, $\tilde a(x)$, $b(x)$, $\tilde b(x)$ and $c(x)$ are rational functions, modulo gauge transformations by upper-triangular matrices of the form
\begin{equation} \label{lambda-oper sl3 gauge}
\begin{pmatrix} 1 & v(x) \lambda & w(x) \lambda^2 \cr 0 & 1 & \tilde v(x) \lambda \cr 0 & 0 & 1 \end{pmatrix}
\end{equation}
where $v(x)$, $\tilde v(x)$ and $w(x)$ are rational functions.
Every $\mathfrak{sl}_3$ $\lambda$-oper has a unique \emph{canonical form} as in \eqref{lambda-oper sl3}.

If we conjugate the connection \eqref{lambda-oper sl3 general} by cyclic permutation matrices then we obtain two alternative formulations of the differential operator \eqref{sl3 lambda-oper Schro}, leading to two alternative (but equivalent) formulations of $\mathfrak{sl}_3$ $\lambda$-opers. Specifically, an $\mathfrak{sl}_3$ $\lambda$-oper can equally be described as a flat connection of the form
\begin{equation} \label{lambda-oper sl3 2}
\partial_x + \begin{pmatrix} -a(x) - \tilde a(x) & 0 & \lambda^{-1} \cr P(x) \lambda^{-1} + c(x) \lambda^2 & a(x) & b(x) \lambda \cr \tilde b(x) \lambda & \lambda^{-1} & \tilde a(x) \end{pmatrix}
\end{equation}
modulo gauge transformations by matrices
\begin{equation} \label{lambda-oper sl3 gauge 2}
\begin{pmatrix} 1 & 0 & 0 \cr w(x) \lambda^2 & 1 & v(x) \lambda \cr \tilde v(x) \lambda & 0 & 1 \end{pmatrix}.
\end{equation}
Equivalently, an $\mathfrak{sl}_3$ $\lambda$-oper can also be described as a connection of the form
\begin{equation} \label{lambda-oper sl3 3}
\partial_x + \begin{pmatrix} \tilde a(x) & \tilde b(x) \lambda & \lambda^{-1} \cr \lambda^{-1} & -a(x) - \tilde a(x) & 0 \cr b(x) \lambda & P(x) \lambda^{-1} + c(x) \lambda^2 & a(x) \end{pmatrix}
\end{equation}
modulo gauge transformations by matrices
\begin{equation} \label{lambda-oper sl3 gauge 3}
\begin{pmatrix} 1 & \tilde v(x) \lambda & 0 \cr 0 & 1 & 0 \cr v(x) \lambda & w(x) \lambda^2 & 1 \end{pmatrix}.
\end{equation}

The unique canonical form of an $\mathfrak{sl}_3$ $\lambda$-oper in the second description \eqref{lambda-oper sl3 2} is given by
\begin{equation} \label{can form sl3 2}
\partial_x + \begin{pmatrix}0 & 0 & \lambda^{-1} \cr P(x) \lambda^{-1} + t_3(x) \lambda^2 & 0 & t_2(x) \lambda \cr 0 & \lambda^{-1} & 0 \end{pmatrix},
\end{equation}
and that of an $\mathfrak{sl}_3$ $\lambda$-oper in the third description \eqref{lambda-oper sl3 3} reads
\begin{equation} \label{can form sl3 3}
\partial_x + \begin{pmatrix}0 & 0 & \lambda^{-1} \cr \lambda^{-1} & 0 & 0 \cr t_2(x) \lambda & P(x) \lambda^{-1} + t_3(x) \lambda^2 & 0 \end{pmatrix}.
\end{equation}

\subsection{Miura $\lambda$-opers and singularities of trivial monodromy}

A \emph{Miura $\mathfrak{sl}_3$ oper} is a connection of the form 
\begin{equation} \label{Miura oper sl3}
\partial_x + \begin{pmatrix} a(x) & 0& 0 \cr 1 & \tilde a(x) & 0 \cr 0 & 1 & - a(x) - \tilde a(x) \end{pmatrix}
\end{equation}
with $a(x)$ and $\tilde a(x)$ rational. Since it is of the general form in \eqref{oper sl3} it defines an $\mathfrak{sl}_3$ oper which corresponds to the differential operator $(\partial_x + a(x))(\partial_x + \tilde a(x))(\partial_x - a(x)- \tilde a(x))$.

There are two types of apparent singularities, corresponding to the two nodes of the Dynkin diagram of $\mathfrak{sl}_3$. These can be points $w$ where
\begin{equation} \label{apparent sing sl3 a}
a(x) = \frac{1}{x-w}+ d +O(x-w), \qquad \tilde a(x) = -\frac{1}{x-w}+ d+O(x-w)
\end{equation}
so that, in particular, the constant term of $a(x) - \tilde a(x)$ is zero, or points $w'$ where
\begin{equation} \label{apparent sing sl3 b}
a(x) = \frac{0}{x-w'}- 2 d +O(x-w'), \qquad \tilde a(x) = \frac{1}{x-w'}+d  +O(x-w')
\end{equation}
so that, in particular, the constant term of $a(x) + 2 \tilde a(x)$ vanishes.

If at a singularity $z$ the Miura $\mathfrak{sl}_3$ oper is of the form
\begin{equation} \label{n1 n2 reg singularity}
a(x) = -\frac13 \frac{2 n_1 + n_2}{x-z}+ O(1), \qquad \qquad \tilde a(x) = \frac13 \frac{n_1- n_2}{x-z}+ O(1)
\end{equation}
for some non-negative integers $n_1$ and $n_2$, then $z$ is a regular singularity of the $\mathfrak{sl}_3$ oper of trivial monodromy. Indeed, one can bring the Miura $\mathfrak{sl}_3$ oper to the form
\begin{equation} \label{Miura oper sl3}
\partial_x + \begin{pmatrix} r(x) & 0& 0 \cr (x-z)^{n_1} & \tilde r(x) & 0 \cr 0 & (x-z)^{n_2} & - r(x) - \tilde r(x) \end{pmatrix}
\end{equation}
where $r(x) = a(x) + \frac13 \frac{2 n_1 + n_2}{x-z}$ and $\tilde r(x) = \tilde a(x) - \frac13 \frac{n_1- n_2}{x-z}$, which are regular at $z$.

A \emph{Miura $\mathfrak{sl}_3$ $\lambda$-oper} is a connection of the form
\begin{equation} \label{lambda-Miura sl3}
\partial_x + \begin{pmatrix} a_1(x) & 0 & P(x)\lambda^{-1}  \cr \lambda^{-1} & \tilde a_1(x) & 0 \cr
0 & \lambda^{-1} & -a_1(x) - \tilde a_1(x) \end{pmatrix}
\end{equation}
where $a_1(x)$ and $\tilde a_1(x)$ are rational functions.
This is of the general form \eqref{lambda-oper sl3 general} and so a Miura $\mathfrak{sl}_3$ $\lambda$-oper  defines an $\mathfrak{sl}_3$ $\lambda$-oper with
\begin{subequations} \label{t2 t3}
\begin{align}
t_2(x) &= a_1(x)^2 + a_1(x) \tilde a_1(x) + \tilde a_1(x)^2 + 2 \partial_x a_1(x) + \partial_x \tilde a_1(x), \\
t_3(x) &= - (a_1(x) + \tilde a_1(x))\big( a_1(x) \tilde a_1(x) + \partial_x a_1(x) + 2 \partial_x \tilde a_1(x) \big) - \partial_x^2 a_1(x) - \partial_x^2 \tilde a_1(x).
\end{align}
\end{subequations}
We refer to this as the $\mathfrak{sl}_3$ $\lambda$-oper underlying \eqref{lambda-Miura sl3}. It can be described as a third order differential operator of the form
\begin{equation}
\big(\partial_x + a_1(x) \big)\big(\partial_x + \tilde a_1(x)\big)\big(\partial_x - a_1(x)- \tilde a_1(x) \big) - \frac{P(x)}{\lambda^3}.
\end{equation}

There are two other gauge equivalent ways of presenting the same Miura $\mathfrak{sl}_3$ $\lambda$-oper as in \eqref{lambda-Miura sl3}, namely
\begin{equation} \label{lambda-Miura sl3 2}
\partial_x + \begin{pmatrix} - a_2(x) - \tilde a_2(x) & 0 & \lambda^{-1}  \cr P(x) \lambda^{-1} & a_2(x) & 0 \cr
0 & \lambda^{-1} & \tilde a_2(x) \end{pmatrix}
\end{equation}
with $a_2(x) = \tilde a_1(x) - \frac{\partial_x P(x)}{3 P(x)}$ and $\tilde a_2(x) = - a_1(x) - \tilde a_1(x) - \frac{\partial_x P(x)}{3 P(x)}$, or
\begin{equation} \label{lambda-Miura sl3 3}
\partial_x + \begin{pmatrix} \tilde a_3(x) & 0 & \lambda^{-1}  \cr \lambda^{-1} & -a_3(x) - \tilde a_3(x) & 0 \cr
0 & P(x) \lambda^{-1} & a_3(x) \end{pmatrix}
\end{equation}
with $a_3(x) = - a_1(x) - \tilde a_1(x) - \frac{2 \partial_x P(x)}{3 P(x)}$ and $\tilde a_3(x) = a_1(x) + \frac{\partial_x P(x)}{3 P(x)}$.

The Miura $\mathfrak{sl}_3$ $\lambda$-oper \eqref{lambda-Miura sl3 2} is of the particular form \eqref{lambda-oper sl3 2} so it defines a second $\mathfrak{sl}_3$ $\lambda$-oper. Likewise, the Miura $\mathfrak{sl}_3$ $\lambda$-oper \eqref{lambda-Miura sl3 3} is of the form \eqref{lambda-oper sl3 3} and thus it also defines a third $\mathfrak{sl}_3$ $\lambda$-oper. Crucially, all three $\mathfrak{sl}_3$ $\lambda$-opers share the same monodromy data since they are gauge equivalent. This \emph{triality} generalises the duality of $\mathfrak{sl}_2$ $\lambda$-opers associated with a given Miura $\mathfrak{sl}_2$ $\lambda$-oper discussed in Section \ref{sec: Miura}.

There are three types of apparent singularities, corresponding to the three nodes of the Dynkin diagram of $\widetilde{\mathfrak{sl}}_3$. In particular, we can have the same types of singularities as for a Miura $\mathfrak{sl}_3$ oper in \eqref{apparent sing sl3 a}, namely points $w$ where, cf. \eqref{apparent sing sl3 a},
\begin{equation} \label{apparent sing sl3 affine 1}
a_1(x) = \frac{1}{x-w}+ d +O(x-w), \qquad  \tilde a_1(x) = -\frac{1}{x-w}+ d+O(x-w)
\end{equation}
so that $a_1(x) - \tilde a_1(x)$ has vanishing constant term, or points $w'$ where, cf. \eqref{apparent sing sl3 b},
\begin{equation} \label{apparent sing sl3 affine 2}
a_1(x) = \frac{0}{x-w'}- 2 d +O(x-w'), \qquad \tilde a_1(x) = \frac{1}{x-w'}+d  +O(x-w')
\end{equation}
so that $a_1(x) + 2 \tilde a_1(x)$ has no constant term. Both of these singularities are absent from the $\mathfrak{sl}_3$ $\lambda$-oper underlying \eqref{lambda-Miura sl3}.
The third type of apparent singularity is at points $w''$ where
\begin{equation} \label{apparent sing sl3 affine 3}
a_1(x) = -\frac{1}{x-w''}+ d +O(x-w''), \qquad \tilde a_1(x) = \frac{0}{x-w''}-2d  +O(x-w'')
\end{equation}
so that $2 a_1(x) + \tilde a_1(x)$ has no constant term. The singularity \eqref{apparent sing sl3 affine 3} is not erased in the canonical form of the Miura $\mathfrak{sl}_3$ $\lambda$-oper \eqref{lambda-Miura sl3}. However, since it is of the form \eqref{n1 n2 reg singularity} with $n_1 = n_2 = 1$, by the above arguments for Miura $\mathfrak{sl}_3$ opers it follows that the $\mathfrak{sl}_3$ $\lambda$-oper underlying \eqref{lambda-Miura sl3} has trivial monodromy at $w''$.

In fact, singularities of both types \eqref{apparent sing sl3 affine 1} and \eqref{apparent sing sl3 affine 3} are absent in the canonical form of the second Miura $\mathfrak{sl}_3$ $\lambda$-oper in \eqref{lambda-Miura sl3 2}. Likewise, both singularities \eqref{apparent sing sl3 affine 2} and \eqref{apparent sing sl3 affine 3} are absent in the canonical form of the third Miura $\mathfrak{sl}_3$ $\lambda$-oper \eqref{lambda-Miura sl3 2}.

If $P(x)$ has a zero of order $k$ at a singularity $z$ of the Miura $\mathfrak{sl}_3$ $\lambda$-oper with
\begin{equation}
a_1(x) = -\frac13 \frac{2 n_1 + n_2}{x-z}+ O(1), \qquad \tilde a_1(x) = \frac13 \frac{n_1- n_2}{x-z}+ O(1)
\end{equation}
then provided $n_1, n_2 \geq 0$ and $n_1 + n_2 \leq k$, the underlying $\mathfrak{sl}_3$ $\lambda$-oper has trivial monodromy. Indeed, one can bring \eqref{lambda-Miura sl3} to the form
\begin{equation}
\partial_x + \begin{pmatrix} r_1(x) & 0 & (x-z)^{k-n_1-n_2} q(x) \lambda^{-1} \cr
(x-z)^{n_1} \lambda^{-1} & \tilde r_1(x) & 0 \cr
0 & (x-z)^{n_2} \lambda^{-1} & - r_1(x) - \tilde r_1(x) \end{pmatrix}
\end{equation}
where we have written $P(x) = (x-z)^k q(x)$ with $q(z) \neq 0$, $r_1(x) = a_1(x) + \frac{1}{3} \frac{2n_1 + n_2}{x - z}$ and $\tilde r_1(x) = \tilde a_1(x) - \frac{1}{3} \frac{n_1 - n_2}{x - z}$, which are clearly regular at $z$.

The behaviour of the second Miura $\mathfrak{sl}_3$ $\lambda$-oper \eqref{lambda-Miura sl3 2} at $z$ is given by
\begin{equation}
a_2(x) = \frac13 \frac{- k + n_1 - n_2}{x-z}+ O(1), \qquad \tilde a_2(x) = \frac13 \frac{- k + n_1 + 2n_2}{x-z}+ O(1)
\end{equation}
while the third Miura $\mathfrak{sl}_3$ $\lambda$-oper \eqref{lambda-Miura sl3 3} behaves as
\begin{equation}
a_3(x) = \frac13 \frac{- 2k + n_1 + 2n_2}{x-z}+ O(1), \qquad \tilde a_2(x) = \frac13 \frac{k - 2 n_1 - n_2}{x-z}+ O(1).
\end{equation}

\subsection{$\lambda$-Opers with singularities of trivial monodromy and affine Bethe equations}

A Miura $\mathfrak{sl}_3$ oper on $\mathbb{C}$ with a rank $1$ irregular singularity at infinity and whose other singularities are all regular with trivial monodromy is of the form
\begin{align}
a(x) &= - \frac{2 \alpha_1 + \alpha_2}{3} -\frac13 \sum_a \frac{2 n_{1,a} + n_{2,a}}{x-z_a} + \sum_i \frac{1}{x-w_i}, \\
\tilde a(x) &= \frac{\alpha_1 - \alpha_2}{3} + \frac13 \sum_a \frac{n_{1,a}- n_{2,a}}{x-z_a} - \sum_i \frac{1}{x-w_i} + \sum_i \frac{1}{x-w'_i}
\end{align}
where the apparent singularities $w_i$ and $w'_i$ satisfy the Bethe equations
\begin{align}
- \sum_a \frac{n_{1,a}}{w_i-z_a} + \sum_{j \neq i} \frac{2}{w_i-w_j} - \sum_j \frac{1}{w_i-w'_j} &= \alpha_1\\
- \sum_a \frac{n_{2,a}}{w'_i-z_a} - \sum_j \frac{1}{w'_i-w_j} + \sum_{j \neq i} \frac{2}{w'_i-w'_j} &= \alpha_2.
\end{align}

We are interested in the case of a Miura $\mathfrak{sl}_3$ $\lambda$-oper with a rank 1 irregular singularity at infinity and whose other singularities are all regular with trivial monodromy. This can be written as
\begin{subequations} \label{1st Miura lambda}
\begin{align}
a_1(x) &= - \frac{2 \alpha_1 + \alpha_2}{3} -\frac13 \sum_a \frac{2 n_{1,a} + n_{2,a}}{x-z_a} + \sum_i \frac{1}{x-w_i} - \sum_i \frac{1}{x-w''_i}, \\
\tilde a_1(x) &= \frac{\alpha_1 - \alpha_2}{3} + \frac13 \sum_a \frac{n_{1,a}- n_{2,a}}{x-z_a} - \sum_i \frac{1}{x-w_i} + \sum_i \frac{1}{x-w'_i}.
\end{align}
\end{subequations}
With $P(x) = e^{(\alpha_1 + \alpha_2 + \alpha_3) x} \prod_a(x - z_a)^{k_a}$, the second Miura $\mathfrak{sl}_3$ $
\lambda$-oper then reads
\begin{subequations} \label{2nd Miura lambda}
\begin{align}
a_2(x) &= - \frac{2 \alpha_2 + \alpha_3}{3} + \frac13 \sum_a \frac{-k_a + n_{1,a} - n_{2,a}}{x-z_a} - \sum_i \frac{1}{x-w_i} + \sum_i \frac{1}{x-w'_i}, \\
\tilde a_2(x) &= \frac{\alpha_2 - \alpha_3}{3} + \frac13 \sum_a \frac{- k_a + n_{1,a} + 2 n_{2,a}}{x-z_a} - \sum_i \frac{1}{x-w'_i} + \sum_i \frac{1}{x-w''_i}.
\end{align}
\end{subequations}
The condition that $w_i$ and $w'_i$ are apparent singularities for the first Miura $\mathfrak{sl}_3$ $\lambda$-oper \eqref{1st Miura lambda} and that $w''_i$ are apparent singularities for the second Miura $\mathfrak{sl}_3$ $\lambda$-oper \eqref{2nd Miura lambda} leads to the Bethe equations
\begin{align}
- \sum_a \frac{n_{1,a}}{w_i-z_a} + \sum_{j \neq i} \frac{2}{w_i-w_j} - \sum_j \frac{1}{w_i-w'_j} - \sum_j \frac{1}{w_i-w''_j} &= \alpha_1,\\
- \sum_a \frac{n_{2,a}}{w'_i-z_a} - \sum_j \frac{1}{w'_i-w_j} + \sum_{j \neq i} \frac{2}{w'_i-w'_j} - \sum_j \frac{1}{w'_i-w''_j} &= \alpha_2,\\
- \sum_a \frac{k_a - n_{1,a} - n_{2,a}}{w''_i-z_a} - \sum_j \frac{1}{w''_i-w_j} - \sum_j \frac{1}{w''_i-w'_j} + \sum_{j \neq i} \frac{2}{w''_i-w''_j} &= \alpha_3.
\end{align}

\subsection{WKB expansion and quasi-canonical form}

\begin{figure}
\centering
\begin{tabular}{cc}
\begin{tikzcd}
& & \begin{tabular}{c}
\textup{Miura $\mathfrak{sl}_3$}\\
\textup{$\lambda$-oper {\rm I}}
\end{tabular}
\arrow{r} &
\textup{$\mathfrak{sl}_3$ $\lambda$-oper {\rm I}}
\\[-18pt]
\begin{tabular}{c}
\textup{$\widetilde{\mathfrak{sl}}_3$ oper/WKB}\\
\textup{momenta}
\end{tabular} &
\begin{tabular}{c}
\textup{Miura}\\
\textup{$\widetilde{\mathfrak{sl}}_3$ oper}
\end{tabular}
\arrow[<->, dashed]{ru} \arrow[<->, dashed]{rd} \arrow[<->, dashed]{r} \arrow{l} &
\begin{tabular}{c}
\textup{Miura $\mathfrak{sl}_3$}\\
\textup{$\lambda$-oper {\rm II}}
\end{tabular} \arrow{r} & \textup{$\mathfrak{sl}_3$ $\lambda$-oper {\rm II}}
\\[-18pt]
& & \begin{tabular}{c}
\textup{Miura $\mathfrak{sl}_3$}\\
\textup{$\lambda$-oper {\rm III}}
\end{tabular}
\arrow{r} &
\textup{$\mathfrak{sl}_3$ $\lambda$-oper {\rm III}}
\end{tikzcd}
&
\begin{tikzpicture}[baseline =0,scale=.6]
\draw[thick] (0,2.8) -- (0,-2.8);
\draw[thick] (0,2.8) to[out=-40,in=40] (0,-2.8);
\filldraw[fill=white, thick] (0,2.8) circle (4mm);
\filldraw[fill=white, thick] (0,0) circle (4mm);
\filldraw[fill=white, thick] (0,-2.8) circle (4mm);
\end{tikzpicture}
\end{tabular}
\caption{The three different types of (Miura) $\mathfrak{sl}_3$ $\lambda$-opers, labelled I, II and III, associated with the three nodes of the Dynkin diagram of $\widetilde{\mathfrak{sl}}_3$. They all share a common $\widetilde{\mathfrak{sl}}_3$ oper which describes the WKB momenta of the third order differential operator \eqref{sl3 lambda-oper Schro}.}
\label{fig: different opers sl3}
\end{figure}
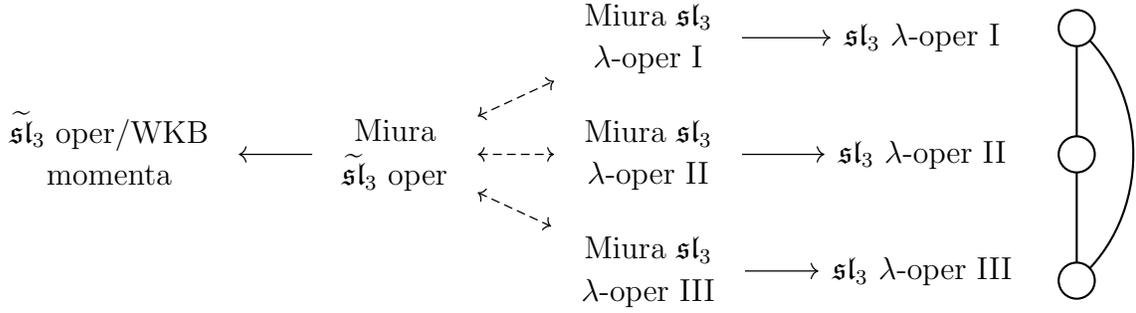

The three Miura $\mathfrak{sl}_3$ $\lambda$-opers \eqref{lambda-Miura sl3}, \eqref{lambda-Miura sl3 2} and \eqref{lambda-Miura sl3 3} are locally gauge equivalent to a connection of the more symmetric form
\begin{equation} \label{Miura affine sl3}
\partial_x + \begin{pmatrix} a(x) & 0 & P(x)^{\frac{1}{3}}\lambda^{-1}  \cr P(x)^{\frac{1}{3}} \lambda^{-1} & \tilde a(x) & 0 \cr
0 & P(x)^{\frac{1}{3}} \lambda^{-1} & -a(x) - \tilde a(x) \end{pmatrix}
\end{equation}
where $a(x) = a_1(x) + \frac{\partial_x P(x)}{3 P(x)}$ and $\tilde a(x) = \tilde a_1(x)$. We refer to \eqref{Miura affine sl3} as a \emph{Miura $\widetilde{\mathfrak{sl}}_3$ oper}. The underlying \emph{$\widetilde{\mathfrak{sl}}_3$ oper}, or \emph{affine $\mathfrak{sl}_3$ oper}, is then defined as its equivalence class under gauge transformations by matrices of the form
\begin{equation} \label{affine gauge tr sl3}
\exp \begin{pmatrix} u(x; \lambda) & v_+(x; \lambda) & w_+(x; \lambda) \cr v_-(x; \lambda) & \tilde u(x; \lambda) & \tilde v_+(x; \lambda) \cr
w_-(x; \lambda) & \tilde v_-(x; \lambda) & -u(x; \lambda) - \tilde u(x; \lambda) \end{pmatrix}
\end{equation}
where the various functions have the following formal power series expansions
\begin{alignat*}{2}
u(x; \lambda) &= \sum_{n=0}^\infty P(x)^{-n} u_n(x) \lambda^{3n}, &\qquad
\tilde u(x; \lambda) &= \sum_{n=0}^\infty P(x)^{-n} \tilde u_n(x) \lambda^{3n},\\
v_+(x; \lambda) &= \sum_{n=0}^\infty P(x)^{-n - \frac{1}{3}} v^+_n(x) \lambda^{3n+1}, &\qquad
\tilde v_+(x; \lambda) &= \sum_{n=0}^\infty P(x)^{-n - \frac{1}{3}} \tilde v^+_n(x) \lambda^{3n+1},\\
v_-(x; \lambda) &= \sum_{n=0}^\infty P(x)^{-n - \frac{2}{3}} v^-_n(x) \lambda^{3n+2}, &\qquad
\tilde v_-(x; \lambda) &= \sum_{n=0}^\infty P(x)^{-n - \frac{2}{3}} \tilde v^-_n(x) \lambda^{3n+2}, \\
w_+(x; \lambda) &= \sum_{n=0}^\infty P(x)^{-n - \frac{2}{3}} w^+_n(x) \lambda^{3n+2}, &\qquad
w_-(x; \lambda) &= \sum_{n=0}^\infty P(x)^{-n - \frac{1}{3}} w^-_n(x) \lambda^{3n+1}
\end{alignat*}
with $u_n(x)$, $\tilde u_n(x)$, $v^\pm_n(x)$, $\tilde v^\pm_n(x)$ and $w^\pm_n(x)$ rational functions.
As we will show below, the $\widetilde{\mathfrak{sl}}_3$ oper controls the WKB asymptotics of the $\lambda$-oper \eqref{sl3 lambda-oper Schro} underlying the Miura $\mathfrak{sl}_3$ $\lambda$-oper \eqref{lambda-Miura sl3}; see Fig. \ref{fig: different opers sl3}.

Recall first that an $\widetilde{\mathfrak{sl}}_3$ oper can be brought to the \emph{quasi-canonical form} \cite{Lacroix:2018fhf}
\begin{equation} \label{quasi-can sl3}
\partial_x + \begin{pmatrix} 0 & q_1(x; \lambda) & q_2(x; \lambda) \cr q_2(x; \lambda) & 0 & q_1(x; \lambda) \cr
q_1(x; \lambda) & q_2(x; \lambda) & 0
\end{pmatrix}
\end{equation}
where the coefficients are given by the formal Laurent series
\begin{subequations}
\begin{align}
q_1(x; \lambda) &= \sum_{n=0}^\infty P(x)^{-n-\frac{1}{3}} q_{1,n}(x) \lambda^{3n+1}, \\
q_2(x; \lambda) &= \frac{P(x)^{\frac{1}{3}}}{\lambda} + \sum_{n=0}^\infty P(x)^{-n-\frac{2}{3}} q_{2,n}(x) \lambda^{3n+2}.
\end{align}
\end{subequations}
In the case of the $\widetilde{\mathfrak{sl}}_3$ oper underlying \eqref{Miura affine sl3}, the first few orders explicitly read
\begin{subequations}
\begin{align}
q_{1,0}(x) = \frac{t_2(x)}{3} + \frac{\big( \partial_x P(x) \big)^2}{27 P(x)}, \qquad
q_{2,0}(x) = \frac{t_3(x)}{3} + \frac{t_2(x) \partial_x P(x)}{9 P(x)}
\end{align}
\end{subequations}
where $t_2(x)$ and $t_3(x)$ are given by \eqref{t2 t3}.

Just as in the $\mathfrak{sl}_2$ case considered in Section \ref{sec: WKB quasi-can sl2}, the quasi-canonical form \eqref{quasi-can sl3} of an affine $\mathfrak{sl}_3$ oper is not unique. Indeed, it is preserved by residual gauge transformations of the form \eqref{affine gauge tr sl3} with $u(x; \lambda) = \tilde u(x; \lambda) = 0$ and $v_\pm(x; \lambda) = \tilde v_\pm(x; \lambda) = w_\mp(x; \lambda)$, the effect of which is to transform the quasi-canonical form as
\begin{subequations}
\begin{align}
q_1(x;\lambda) \longmapsto q_1(x;\lambda) + \partial_x v_+(x; \lambda),\\
q_2(x;\lambda) \longmapsto q_2(x;\lambda) + \partial_x v_-(x; \lambda).
\end{align}
\end{subequations}

We look for flat sections of the $\mathfrak{sl}_3$ $\lambda$-oper \eqref{sl3 lambda-oper Schro}, i.e. solutions of the third order differential equation
\begin{equation*}
\bigg( \partial_x^3 - t_2(x) \partial_x + t_3(x) + \frac{P(x)}{\lambda^3} \bigg) \psi(x; \lambda) = 0
\end{equation*}
in the form of the WKB ansatz
\begin{gather*}
\psi_1(x; \lambda) = \frac{1}{\sqrt[3]{A(x; \lambda)}} e^{\int p_1(x; \lambda) dx}, \qquad
\psi_2(x; \lambda) = \frac{1}{\sqrt[3]{A(x; \lambda)}} e^{\int p_2(x; \lambda) dx},\\
\psi_3(x; \lambda) = \frac{1}{\sqrt[3]{A(x; \lambda)}} e^{- \int p_1(x; \lambda) dx - \int p_2(x; \lambda) dx}
\end{gather*}
where the normalization factor, fixed by requiring the Wronskian of the three solutions $\psi_1$, $\psi_2$ and $\psi_3$ to be 1, is given by
\begin{align}
A(x; \lambda) &= 2 p_1(x; \lambda)^3 - 2 p_2(x; \lambda)^3 + 3 (p_1(x; \lambda) - p_2(x; \lambda)) p_1(x; \lambda) p_2(x; \lambda) \cr
&\qquad\qquad + 3 p_2(x; \lambda) \partial_x p_1(x; \lambda) - 3 p_1(x; \lambda) \partial_x p_2(x; \lambda).
\end{align}
Working perturbatively in $\lambda$ we find WKB momenta of the form
\begin{subequations}
\begin{align}
\label{p1 expansion} p_1(x; \lambda) &= - \frac{P(x)^{\frac{1}{3}}}{\lambda} - \sum_{n=1}^\infty p_{1,n}(x) \lambda^n\\
\label{p2 expansion} p_2(x; \lambda) &= - \frac{P(x)^{\frac{1}{3}}}{e^{- \frac{2 \pi i}{3}} \lambda} - \sum_{n=1}^\infty p_{2,n}(x) \big( e^{- \frac{2\pi i}{3}} \lambda \big)^n.
\end{align}
\end{subequations}
The first few coefficients of the expansion \eqref{p1 expansion} are given explicitly by
\begin{subequations}
\begin{align}
p_{1,1}(x) &= P(x)^{- \frac{1}{3}} q_{1,0}(x) + \partial_x \bigg( \frac{2 \partial_x P(x)}{9 P(x)^{\frac{4}{3}}} \bigg), \\
p_{1,2}(x) &= P(x)^{- \frac{2}{3}} q_{2,0}(x) - \partial_x \bigg( \frac{\partial_x^2 P(x)}{9 P(x)^{\frac{5}{3}}} - \frac{7 \big( \partial_x P(x) \big)^2}{54 P(x)^{\frac{8}{3}}} \bigg), \\
p_{1,3}(x) &= \partial_x \bigg( \frac{2 \partial_x t_2(x)}{9 P(x)} - \frac{4 t_2(x) \partial_x P(x)}{27 P(x)^2} + \frac{4 \partial_x^3 P(x)}{27 P(x)^2} \cr
&\qquad\qquad\qquad\qquad - \frac{16 \partial_x P(x) \partial_x^2 P(x)}{27 P(x)^3} + \frac{112 \big( \partial_x P(x) \big)^3}{243 P(x)^4} \bigg)
\end{align}
\end{subequations}
and those of the expansion \eqref{p2 expansion} read
\begin{subequations}
\begin{align}
p_{2,1}(x) &= P(x)^{- \frac{1}{3}} q_{1,0}(x) + \partial_x \bigg( \frac{2 \partial_x P(x)}{9 P(x)^{\frac{4}{3}}} \bigg), \\
p_{2,2}(x) &= P(x)^{- \frac{2}{3}} q_{2,0}(x) + \partial_x \bigg( \frac{t_2(x)}{3 P(x)^{\frac{2}{3}}} + \frac{\partial_x^2 P(x)}{9 P(x)^{\frac{5}{3}}} - \frac{7 \big( \partial_x P(x) \big)^2}{54 P(x)^{\frac{8}{3}}} \bigg), \\
p_{2,3}(x) &= e^{- \frac{\pi i}{6}} \partial_x \bigg( \frac{\partial_x t_2(x)}{3 \sqrt{3} P(x)} - \frac{2 t_2(x) \partial_x P(x)}{9 \sqrt{3} P(x)^2} + \frac{2 \partial_x^3 P(x)}{9 \sqrt{3} P(x)^2} \cr
&\qquad\qquad\qquad\qquad\qquad - \frac{8 \partial_x P(x) \partial_x^2 P(x)}{9 \sqrt{3} P(x)^3} + \frac{56 \big( \partial_x P(x) \big)^3}{81 \sqrt{3} P(x)^4} \bigg).
\end{align}
\end{subequations}

From this we find that the pair of WKB momenta $p_1(x; \lambda)$ and $p_2(x; \lambda)$ are related to the coefficients of the quasi-canonical form \eqref{quasi-can sl3} by
\begin{subequations}
\begin{align}
p_1(x; \lambda) &= - q_1(x; \lambda) - q_2(x; \lambda) + \partial_x f_1(x; \lambda),\\
p_2(x; \lambda) &= - q_1\big( x; e^{-\frac{2\pi i}{3}}\lambda \big) - q_2\big( x; e^{-\frac{2\pi i}{3}} \lambda \big) + \partial_x f_2(x; \lambda)
\end{align}
\end{subequations}
for some functions $f_1(x; \lambda)$ and $f_2(x; \lambda)$.

\section{Future directions and open problems} \label{sec:open}
\begin{enumerate}
	\item {\bf Higher rank}
	
	It would be natural to extend our work to other Lie algebras \cite{Suzuki_2000,Dorey_2000,dorey2007abcd}. The definition of the Kondo defects is valid for any Lie algebra, with an important caveat: 
	the matrices $t^a$ do not have to be generators of the Lie algebra, they only need to transform in the adjoint representation of the global symmetry
	 group. The RG flow will thus involve extra couplings, controlling the specific choice of $t^a$. 4d Chern-Simons gauge theory predicts integrability for
	 choices of couplings related to representations of the Yangian. It would be nice to understand how these structures manifest themselves in the affine
	  Gaudin description. The definition of affine opers with singularities of trivial monodromy should be possible for general gauge groups and straighforward in type $A$.
	   The precise correspondence between the Stokes data and the UV labels of Kondo defects is less obvious. The IR WKB analysis is still possible, but 
	   will likely require some more refined technology such as spectral networks \cite{Gaiotto:2012rg}.	
	   
	   \item {\bf Non-integral levels}
	
	In sections \ref{sec:UVexpansion} and \ref{sec:IRexpansion}, we made the assumption that all the level $k_i$ in 
	\begin{equation}
	P(x) = e^{2x} \prod_i(x -x_i)^{k_i}
	\end{equation}
	are non-negative integers, which corresponds to the WZW model for the product group $\prod_i SU(2)_{k_i}$. We can generalize to Kac-Moody algebras via replacing $k_i$ by $\kappa_i\in \mathbb{C}$. This forces us to study opers on a logarithmic covering space of the complex plane. (It reduces to a finite covering when $\kappa_i$ is rational.) Consequently, we have more Wronskians of small solutions to study, both between the small solutions on the same sheet and across different sheets. It would be interesting to work out some examples and understand their relationship with the expectation values of the local integrals of motions in the affine Gaudin model, which are conjecturally given by the integrals of the WKB momentum \cite{Lacroix:2018fhf}.
	
	\item {\bf Bulk deformation}
	
	Throughout this article, we considered Kondo line defects in CFTs in the bulk. It is well-known, however, that the bulk theory can be deformed in such a way that the integrable structure remains. Examples of such deformations are given by $J^a\bar{J}^a$ in the WZW model and $\Phi_{1,3}$ in minimal models. The transfer matrices $\hat{T}_{j}[\theta]$ can be naturally extended as well in such a way that commutativity and the fusion rules are still satisfied. Furthermore, there is evidence \cite{Lukyanov:2010rn,Dorey:2012bx,massiveODElecture} that the ODE/IM correspondence can be generalized as well to the so-called `massive ODE/IM correspondence'. It is natural to expect that the $P$-sinh-Gordon equation with an appropriate generalization of the potential $P(x) = e^{2x}\prod_i(x-x_i)^{k_i}$ should correspond to the integrable Gross-Neveu model.

	\item {\bf Coset limits and other models}
	
	 We only sketched the limiting procedure $e^{2x}\prod_i(x-x_i)^{k_i} \to \prod_i(x-x_i)^{k_i}$
	mapping the Kondo problems to integrable defects in coset models. It would be nice to develop the relation further. 
	Other choices of $P(x)$ should be relevant for integrable line defects in other 2d CFTs. A dictionary may be developed along the lines of \cite{Costello:2019tri} from a 4d Chern-Simons gauge theory perspective, or equivalently \cite{Vicedo:2019dej}, along the lines of \cite{Vicedo:2017cge} from the point of view of affine Gaudin models. 
	\item {\bf Non-isotropic generalization}
	
	 Here we only considered the case of Kondo problems which preserve the global $SU(2)$ symmetry of the 
	WZW models. Anisotropic Kondo problems are also very interesting and often still integrable. In \cite{Lukyanov_2007}, an ODE/IM solution was 
	proposed for the ground states of the anisotropic defects in a single WZW model. It would be very interesting to find a full solution to the problem. 
\end{enumerate}

\acknowledgments
We thank K.Costello for participation in early stages of the project and countless in depth discussions. The research of D.G., J.H.L., and J.W. is supported in part by a grant from the Krembil foundation by the Perimeter Institute for Theoretical Physics. Research at Perimeter Institute is supported in part by the Government of Canada through the Department of Innovation, Science and Economic Development Canada and by the Province of Ontario through the Ministry of Economic Development, Job Creation and Trade.

\appendix
\section{Details of UV expansion}\label{app:UVexpansion}
In this appendix, we verify the claim \eqref{eq:TasWonskian} for a few examples explicitly by working perturbatively in $g$ to $O(g^4)$. Next, based on these examples, we summarize a general recipe in the case of zero twist $\alpha_+ = 0$. We then end this section with some remarks and checks with nonzero twist.

\subsection{Examples of UV perturbative matching}

Let us perform a perturbative UV analysis of the proposed excited state Schr\"odinger equations and compare them to $SU(2)_k$ WZW line defects evaluated between certain excited states. We use the Chevalley basis rather than the orthonormal basis used in \cite{KondoGLW}. To establish notation, we write the change of basis explicitly:
\begin{align}
J_{n}^a{J}_{m}^a &= J_{n}^+{J}_{m}^- + J_{n}^-{J}_{m}^+ +2 J_{n}^0{J}_{m}^0\\
f^{abc}J_{n}^{a}J_{m}^{b}J_{l}^{c} & = 12i(J_{n}^{[+}J_{m}^{-}J_{l}^{0]}).
\end{align}

We will consider the following states as examples:
\begin{equation}
	J_{-1}^+ |0, k\rangle, \quad J_{-2}^+|0, k\rangle.
\end{equation}
From the discussion in Subsection \ref{sec:Ex:vacuummodule}, for generic values of $k$, they are described by Miura $\lambda$-opers with
\begin{align}
\label{a1 gen oper}	a_+^{(1)}(x) &=  - \frac{1}{x-w'},\\
				a_+^{(2)}(x) &=  \frac{1}{x-w}-  \frac{1}{x-w'_1} -\frac{1}{x-w'_2},
\end{align}
respectively.
Note, however, that the state $J_{-2}^+|0, -1\rangle$ is a special case of \eqref{a1 gen oper} since, as we conjectured in \ref{sec:Ex:vacuummodule}, it corresponds to the generalized Miura $\lambda$-oper
\begin{equation}
a_+(x) = -\frac{1}{x}.
\end{equation}

The left hand side of the claim \eqref{eq:TasWonskian} is an obvious, alebit tedious, task, namely to compute the $g$ expansion of the expectation value of the operator $\hat{T}_n(\theta)$ in a certain state. This operator was calculated in the Appendix E of \cite{KondoGLW}. Note that if we normalize the expectation values of the line defect as\footnote{In this appendix only, we write the excited expectation values with angled brackets to emphasize that $\langle T_n (\theta) \rangle_r$ are quantities which are directly evaluated from the defect operator computed in \cite{KondoGLW}. The quantities we write as $T_n(\theta)$ in the perturbative analysis below correspond to the Wronskian result we obtain from the Schr\"odinger equation.}
\begin{equation} \label{T expectation}
\langle T_n (\theta) \rangle_r := \frac{1}{r \kappa} \langle 0, k | J_{r}^- \hat{T}_n(\theta)  J_{-r}^+ | 0, k \rangle
\end{equation}
then the ordinary form of the Hirota bilinear relations
\begin{equation}
\Big\langle T_n\Big(\theta+ \frac{i \pi}{2}\Big) \Big\rangle_r \Big\langle T_n\Big(\theta - \frac{i \pi}{2}\Big) \Big\rangle_r = 1 + \langle T_{n+1}(\theta) \rangle_r \langle T_{n-1}(\theta) \rangle_r
\end{equation}
holds.\footnote{One could choose not to use the normalization $\frac{1}{r k}$ in \eqref{T expectation}, but the Hirota relations would then have $(r k)^2$ in place of $1$ and the following perturbative analyses acquire extra normalization factors.} 

Finding the $g$ expansion of the Wronskian on the right hand side of \eqref{eq:TasWonskian} is not obvious. In \cite{KondoGLW}, we proposed an approach based on a combination of various limits. The basic idea consists of two steps: (1) analyze the solution of the Schr\"odinger equation in the regime $1/g \gg -x$ and $-x \gg 0$ separately and (2) match the solutions in the intermediate region $1/g \gg -x \gg 0$. 

We can recycle many of the result from \cite{KondoGLW} as modifications of the Schr\"odinger equation by the potential terms do not contribute to the perturbative expansion of the differential equation at $O(1)$ and $O(g)$. Hence, we may use the direct evaluation result of the wavefunction in \cite{KondoGLW} given by
\begin{equation}
\psi = \psi^{(0)} + g \psi^{(1)} + \cdots
\end{equation}
with
\begin{align*}
\psi^{(0)}(x;\theta) &\sim  -\frac{1}{\sqrt{\pi}} (x + \theta + \gamma - \log 2) \\
\psi^{(1)}(x;\theta) &\sim -\frac{k}{4 \sqrt{\pi}} (x - \theta - 2 - \gamma + \log 2)
\end{align*}
at large negative $x$. Also, we introduce the effective coupling $g_{\text{eff}}(\theta)$ defined by \cite{KondoGLW}
\begin{equation}
g_{\text{eff}}(\theta)^{k/2} e^{-1/g_{\text{eff}}(\theta)} = g^{k/2} e^{-1/g} e^{\theta}
\end{equation}
with the $g$ expansion
\begin{equation}
g_{\text{eff}}(\theta) = g + \theta g^2 + \theta \Big(\theta - \frac{k}{2}\Big) g^3 + \theta \Big(\theta^2 - \frac{5}{4} k \theta + \frac{k^2}{4} \Big) g^4 + \cdots.
\end{equation}

\subsubsection{$J_{-1}^+ |0, k\rangle$, one $w'$, no $w$}

For the case with one $w'$ and no $w$, the excited state equation is
\begin{equation}
\partial_x^2 \psi(x;\theta) = \Bigg[ e^{2\theta} g e^{-2/g}  x^k e^{2 x} + \frac{8}{(k + 2 x)^2} \Bigg] \psi(x;\theta).
\end{equation}
We propose that this equation encodes the excited line defect expectation value $\langle T_n (\theta) \rangle_1$. The perturbative analysis as an expansion in $g$ is done after shifting $x \to x + 1/g$.

The asymptotics of $\psi$ can be determined by carefully analyzing the solutions of the Schr\"odinger equation in the regime $1/g \gg -x \gg 0$. Doing so, the wavefunction can be parametrized as
\begin{equation}
\psi(x;\theta) \sim - \frac{1}{3 g^2} Q(g_{\text{eff}}) [1 + (k+2x)g]  - \frac{g}{3} \tilde{Q}(g_{\text{eff}}) \Big[1 - \frac{1}{2} (k + 2x)g\Big],
\end{equation}
where
\begin{subequations} \label{Q tilde Q expr}
\begin{align}
    Q(g_{\text{eff}}) &= \frac{g_{\text{eff}}(\theta)}{\sqrt{\pi}} (1 + q_{1} g_{\text{eff}}(\theta) + \cdots),\\
    \tilde{Q}(g_{\text{eff}}) &= \frac{g_{\text{eff}}(\theta)^{-1}}{\sqrt{\pi}} \Big(-\frac{1}{g_{\text{eff}}(\theta)} + \tilde{q}_{0} + \tilde{q}_{1} g_{\text{eff}}(\theta) + \cdots \Big).
\end{align}
\end{subequations}
Before the shift $x \to x + 1/g$, the dependence of $\psi$ on $g$ and $\theta$ combine into $g_{\text{eff}}(\theta)$ such that $\psi$ becomes a function of $x$ and $g_{\text{eff}}$ only. The explicit $g$ dependence in the parametrization of $\psi$ above comes purely from the shift $x \to x + 1/g$.

As the general analysis in the next subsection will suggest, the difference $\# w' - \# w$ is responsible for the various factors present above. The overall coefficient $1/3$ in the asymptotics of $\psi$ arises via the combination $(2(\# w' - \# w)+1)^{-1}$, as does the overall power of $g_{\text{eff}}^{\pm (\# w' - \# w)}$ in the expression of $Q$, $\tilde{Q}$.

As usual, the T-function can be expressed as the quantum Wronskian
\begin{equation} \label{Tn from Q tilde Q}
T_n(\theta) = \frac{i}{3} \bigg[ Q\Big(\theta + \frac{i \pi n}{2}\Big) \tilde{Q}\Big(\theta - \frac{i \pi n}{2}\Big) - Q\Big(\theta - \frac{i \pi n}{2}\Big) \tilde{Q}\Big(\theta + \frac{i \pi n}{2}\Big) \bigg].
\end{equation}
The coefficients $q_i$ can be expressed in terms of $\tilde{q}_i$ by imposing the condition $T_1 = 1$. The explicit expression of $T_n(\theta)$ up to $O(g^4)$ requires the knowledge of $\tilde{q}_0$ and $\tilde{q}_1$. Comparing the parameterization of $\psi$ in terms of $ Q, \tilde{Q} $ with the direct perturbative evaluation up to $O(g)$, we obtain
\begin{align*}
    \tilde{q}_0 &= \frac{1}{4} (-5 k+8 \gamma -8 \log 2) \\
    \tilde{q}_1 &= -\frac{15 k^2}{32}-\frac{1}{4} k (5+\log 2)+\gamma \Big(\frac{k}{4}+\log4\Big)-\frac{\pi ^2}{6}-\gamma ^2-(\log2)^2.
\end{align*}

The full expression for $T_n(\theta)$ to $O(g^4)$ is then
\begin{align*}
    T_n(\theta) &= n - g^2 \Big[\frac{1}{3} \pi ^2 n (n^2-1) \Big] + g^3 \Big[ \frac{1}{12} \pi ^2 n (n^2-1) (7 k-8 \theta-8 \gamma +\log 256) \Big] \\
    &- g^4 \Big[\frac{1}{288} \pi ^2 n (n^2-1) [195 k^2-120 k (5 \theta+1-5 \log 2)-24 \gamma  (25 k \\
    &-24 \theta+\log 16777216)+8 \pi ^2 (10-3 n^2)+288 (\theta-\log 2)^2+288 \gamma ^2] \Big] + O(g^5).
\end{align*}
This expression satisfies the Hirota bilinear relations and has been verified to match the perturbative line defect computation for $\langle T_n(\theta) \rangle_1$ in a suitable renormalization scheme, where $g$ is shifted as
\begin{align*}
    g &\to \lambda g + \lambda^2 g^2 \Big[ k-2 \log \epsilon -2 \gamma -2 \log \pi \Big] \\
    &+ \lambda^3 g^3 \Big[ 2 \log \epsilon  (-3 k+2 \log \epsilon +4 \gamma +4 \log \pi )+(k-4) k \\
    &-6 k \log \pi +\gamma  (8\log \pi -6 k)+\frac{1}{6} \pi ^2 (n^2+1) +4 \gamma ^2+4 (\log \pi )^2 \Big] + O(g^4)
\end{align*}
with $\lambda = -1/2$.

\subsubsection{$J_{-2}^+|0, k\rangle$, two $w'$, one $w$}

For the case with two $w'$ and one $w$, the excited state equation is
\begin{equation}
\partial_x^2 \psi(x;\theta) = \Bigg[ e^{2\theta} g e^{-2/g}  x^k e^{2 x} + \frac{8 (k^2+k (4 x+5)+4 (x^2+x+1))}{(k^2+4 k x+k+4 x
   (x+1))^2} \Bigg] \psi(x;\theta).
\end{equation}
We propose that this equation encodes the excited line defect expectation value $\langle T_n (\theta) \rangle_2$.

As in the previous case, the wavefunction can be parametrized
\begin{equation}
\psi(x;\theta) \sim - \frac{1}{3 g^2} Q(g_{\text{eff}}) [1 + (1+k+2x)g]  - \frac{g}{3} \tilde{Q}(g_{\text{eff}}) [1 - \frac{1}{2} (1+k + 2x)g],
\end{equation}
where $Q(g_{\text{eff}})$ and $\tilde{Q}(g_{\text{eff}})$ are as in \eqref{Q tilde Q expr}.
The expression for $T_n(\theta)$ in terms of $Q$, $\tilde{Q}$ remains the same as in \eqref{Tn from Q tilde Q}. Comparing the parameterization of $\psi$ in terms of $Q$, $\tilde{Q}$ with the direct perturbative evaluation up to $O(g)$, we obtain
\begin{align*}
    \tilde{q}_0 &= \frac{1}{4} (-5 k+8 \gamma -4-8 \log 2),\\
    \tilde{q}_1 &= \gamma \Big(\frac{k}{4}+1+\log 4\Big)+\frac{1}{96} \Big( -3 k (15 k+64+\log 256) -8\big(2 \pi ^2+3 (1+\log 4)^2\big) \Big)-\gamma ^2.
\end{align*}

The full expression for $T_n(\theta)$ to $O(g^4)$ is
\begin{align*}
	T_n(\theta) &= n- \frac{1}{3} \pi ^2 g^2 n \left(n^2-1\right) + \frac{1}{12} \pi ^2 g^3 n \left(n^2-1\right) (-8 \gamma -8 \theta +7 k+4+\log 256)\\
	& + \frac{1}{288} \pi ^2 g^4 n \left(n^2-1\right) \Big(-72 (-2 \gamma -2 \theta +1+\log 4)^2-195 k^2 \Big.\\
	&\Big.\phantom{asfaf} +120 k (5 \gamma +5 \theta -1-5 \log 2)+8 \pi ^2 \left(3 n^2-10\right)\Big) 
\end{align*}
This expression satisfies the Hirota bilinear relations and has been verified to match the perturbative line defect computation for $\langle T_n(\theta) \rangle_2$ in a suitable renormalization scheme, where $g$ is shifted as
\begin{align*}
    g &\to \lambda g + \lambda^2 g^2 \Big[ k-2 \log (\pi  \epsilon )-2 \gamma -\Big(1+\frac{i}{2}\Big) \Big] \\
    &+ \lambda^3 g^3 \Big[ k^2+\frac{1}{24} \Big((-198+3 i) k+4 \pi ^2 (n^2+1)+3 \big(32 (\log \pi)^2 +(59+52 i)\big)\Big) \\
    &+\big( \! -6 k+8 \gamma +(4+2 i) \big) \log (\pi  \epsilon )-6 \gamma  k+4 \log (\epsilon) \log(\pi ^2 \epsilon )+4 \gamma ^2+(4+2 i) \gamma \Big] + O(g^4)
\end{align*}
with $\lambda = -1/2$.

\subsubsection{$J_{-2}^+|0, -1\rangle$}
With \begin{equation}
	a_+(x) = -\frac{1}{x}
\end{equation}
we have the ODE
\begin{equation}
	\partial_x^2 \psi(x;\theta) = \Bigg[ e^{2\theta} g e^{-2/g}  x^k e^{2 x} + \frac{2}{x^2} \Bigg] \psi(x;\theta).
\end{equation}
\begin{equation}
	\psi(x;\theta) \sim - \frac{1}{3 g^2} Q(g_{\text{eff}}) [1 + 2gx]  - \frac{g}{3} \tilde{Q}(g_{\text{eff}}) [1 - gx],
\end{equation}
where again $Q(g_{\text{eff}})$ and $\tilde{Q}(g_{\text{eff}})$ are as in \eqref{Q tilde Q expr}.
We can similarly find the Wronskian 
\begin{align*}
	T_n(\theta) &= n -\frac{1}{3} \pi ^2 g^2 n \left(n^2-1\right) -\frac{1}{12} \pi ^2 g^3 n \left(n^2-1\right) (8 \gamma +8 \theta +3-8 \log2)\\
	& \frac{1}{288} \pi ^2 g^4 n \left(n^2-1\right) \Big[+8 \pi ^2 \left(3 n^2-10\right)-288 \gamma ^2-24 \gamma  (24 \theta +13-24 \log 2)\Big.\\
	&\Big.+3 \left(-104 \theta -96 (\theta -\log 2)^2-49+104 \log2\right)	\Big]
\end{align*}
with 
\begin{align*}
	\tilde{q}_0 &= \frac{1}{4} +2(\gamma-\log2),\\
	\tilde{q}_1 &= \frac{1}{96} \left(-16 \left(6 (\log2-\gamma )^2+\pi ^2\right)+24 (3 \gamma +5-3\log 2)+3\right).
\end{align*}

\subsection{A general perturbative prescription} \label{app: susy poly}
One can easily see the pattern of the steps in the previous example and might wonder if it is possible to perform the perturbative analysis in a uniform way without specifying particular values for $w'_i$ and $w_i$, i.e. without solving the Bethe equations explicitly. Here in this section, we provide such a recipe for zero spin and zero twist.

In considering the Schr\"odinger equation in the regime $1/g \gg -x \gg 0$, one first solves the equation with just the potential terms and then expands in $g$ until one has $O(x)$ contributions. That is, for $(\# w', \# w) = (p,q)$ we take (after shifting $x \mapsto x + 1/g$) the $g$ expansion of the solution to
\begin{equation}
\partial_x^2 \psi(x;\theta) = \left[ a(x)^2 + a'(x) \right] \psi(x;\theta)
\end{equation}
where
\begin{equation}
a(x) = -\sum_{a=1}^{p} \frac{1}{x-w'_a} + \sum_{i=1}^{q} \frac{1}{x-w_i}.
\end{equation}
Notice that one exact solution of such an equation is
\begin{equation}
\frac{\prod_{i=1}^q (x + 1/g - w_i)}{\prod_{a=1}^p (x + 1/g - w'_a)} = g^{p-q} \Big[ 1 - g \Big( (p-q)x - \sum_{a=1}^{p} w'_a + \sum_{i=1}^{q} w_i \Big) + \cdots \Big].
\end{equation}
Expanding up to the order shown is sufficient for $p>q$, but one needs to consider an $O(g^2)$ term for $p=q$. Here, we concern ourselves with just the $p>q$ cases as the $p=q$ case is treated in a similar but more involved way.

The solution above is that proportional to $\tilde{Q}$ of the previous subsection. An exact second solution turns out to be more elusive, but this can be overcome by the fact that we only need the perturbative form of second solution. The second solution can then be determined by imposing that
\begin{equation}
T_n(\theta) = i \left( \psi\Big(x;\theta - \frac{i \pi n}{2}\Big) , \psi\Big(x;\theta + \frac{i \pi n}{2}\Big) \right),
\end{equation}
i.e. the shifted Wronskian of the wavefunction, is normalized as
\begin{equation}
T_n(\theta) = \frac{i}{2(p-q)+1} \bigg[ Q\Big(\theta + \frac{i \pi n}{2}\Big) \tilde{Q}\Big(\theta - \frac{i \pi n}{2}\Big) - Q\Big(\theta - \frac{i \pi n}{2}\Big) \tilde{Q}\Big(\theta + \frac{i \pi n}{2}\Big) \bigg].
\end{equation}

Doing so, one finds that the parametrization of the wavefunction for the cases $p>q$ becomes
\begin{align*}
\psi(x;\theta) \sim -\frac{1}{2(p-q)+1} \Bigg\{& \frac{Q(g_{\text{eff}})}{g^{p-q+1}} \Big[ 1 + \frac{p-q+1}{p-q} g \Big( (p-q) x - \sum_{a=1}^{p} w'_a + \sum_{i=1}^{q} w_i \Big) \Big] \\
&+ g^{p-q} \tilde{Q}(g_{\text{eff}}) \Big[ 1 - g \Big( (p-q)x - \sum_{a=1}^{p} w'_a + \sum_{i=1}^{q} w_i \Big) \Big] \Bigg\}
\end{align*}
with
\begin{align*}
    Q(g_{\text{eff}}) &= \frac{g_{\text{eff}}^{p-q}}{\sqrt{\pi}} (1 + q_{1} g_{\text{eff}}(\theta) + \cdots) \\
    \tilde{Q}(g_{\text{eff}}) &= \frac{g_{\text{eff}}^{-(p-q)}}{\sqrt{\pi}} \Big(-\frac{1}{g_{\text{eff}}(\theta)} + \tilde{q}_{0} + \tilde{q}_{1} g_{\text{eff}}(\theta) + \cdots \Big).
\end{align*}
We see that $\tilde{q}_0$ and $\tilde{q}_1$, determined by comparison to the direct perturbative $g$ expansion of the Schr\"odinger equation, will only depend on the difference of sums of $w'$'s and $w$'s.

The analysis in this subsection can be generalized to other spin modules and those with twists.

\subsection{nonzero twist $\alpha_+ \neq 0$} \label{app:sec:nonzerotwist}
Mimicking the case of $\alpha_+ = 0$, we proceed with the UV expansion as follows. We are interested in the ODE
\begin{equation}
\partial_{x}^{2} \psi (x)=\left[ e^{2 x+2\theta}(1+g x)^{k}+ t(x)\right] \psi (x) \label{app:eq:ODE_nonzeroalpha}
\end{equation}
Consider the matching region $\frac{1}{g} \gg -x \gg 0$. The first inequality means we are reduced to 
\begin{equation}
\partial_{x}^{2} \psi_{\mathrm{I}} (x)=\left[ e^{2 x+2\theta} + \alpha_+^2\right] \psi_{\mathrm{I}} (x)
\end{equation}
The solution is given by
\begin{equation}
\psi_{\mathrm{I}} (x) = K_{\alpha_+}(e^{x+\theta}) \sim \Gamma(-\alpha_+)2^{-1-\alpha_+} e^{\alpha_+ (x+\theta)} +\Gamma(\alpha_+) 2^{\alpha_+-1} e^{-\alpha_+ (x + \theta)},\quad x\rightarrow -\infty
\end{equation}
Recall Miura part $t(x) = a_+(x)^2 +\partial_x a_+(x)$, takes the form
\begin{equation}
a_+(x) = -\alpha_+ -\frac{l}{x+1/g} + \dots
\end{equation}
where $\alpha_+$ is generic enough. 

The second inequality means 
\begin{equation}
\partial_{x}^{2} \psi_{\mathrm{II}} (x)=\left[a_+(x)^2 +\partial_x a_+(x) \right] \psi_{\mathrm{II}} (x)
\end{equation}
whose solutions are given by 
\begin{equation}
\psi_{\mathrm{II}}(x) = c_1 \Gamma(\alpha_+)2^{\alpha_+-1}e^{\int^x dx'\ a_+(x')} + c_2 \Gamma(-\alpha_+)2^{-1-\alpha_+} e^{\int^x dx'\ \tilde{a}_+(x')}
\end{equation}
where $\tilde{a}_+(x)$ is defined to be the Weyl reflection of $a_+(x)$, i.e. $\tilde{a}_+(x) = a_+(x) + f[a_+(x)]$, where for any given function $a(x)$, we define 
\begin{equation}
f[a(x)] \equiv \frac{e^{-2\int^{x} dx'a(x')}}{\int^x dx' e^{-2\int^{x'} dx''a(x'')}}.
\end{equation}
Since both $\psi_\mathrm{I}(x)$ and $\psi_{\mathrm{II}}(x)$ are approximate solutions to \eqref{app:eq:ODE_nonzeroalpha} in the matching region $\frac{1}{g} \gg -x \gg 0$, we can take the coefficient
\begin{align}
c_1  &\sim e^{\alpha (\theta-\frac{1}{g})} g^l,\\
c_2  &\sim  e^{-\alpha (\theta-\frac{1}{g})} g^{-l}.
\end{align}
Recall that $\frac{1}{g_{\mathrm{eff}}} = \frac{1}{g}-\theta +\dots$. Thus it is very natural to define 
\begin{equation}
\begin{split}
Q[g_{\mathrm{eff}}] &=e^{-\frac{\alpha_+}{g_{\mathrm{eff}}}} g_{\mathrm{eff}}^l \big[1+ q_1 g_{\mathrm{eff}}+q_2 g_{\mathrm{eff}}^2 +\dots \big],\\
\tilde{Q}[g_{\mathrm{eff}}] &=  e^{\frac{\alpha_+}{g_{\mathrm{eff}}}} g_{\mathrm{eff}}^{-l} \big[-1+ \tilde{q}_1 g_{\mathrm{eff}}+\tilde{q}_2 g_{\mathrm{eff}}^2 +\dots\big],\\
T_n &=  \frac{i}{2\sin \pi \alpha_+} \Big(Q^{(+n)}[g_{\mathrm{eff}}]\tilde{Q}^{(-n)}[g_{\mathrm{eff}}] -Q^{(-n)}[g_{\mathrm{eff}}]\tilde{Q}^{(+n)}[g_{\mathrm{eff}}] \Big). \label{eq:definitionofQwithalpha}
\end{split}
\end{equation}
One can calculate $T_{n;l} = \langle l| \hat{T}_{n} | l\rangle$ to be
\begin{equation}
T_{n;l} = \frac{ \sin(n\pi \alpha_+)}{\sin\pi\alpha_+} + g\Big(n\pi (2l-k\alpha_+)\cos(n\pi\alpha_+) + 2(q_1 -\tilde{q}_1) \sin(n \pi \alpha_+) \Big) +\dots
\end{equation}
Note that the leading term is precisely the character 
\begin{equation}
\Tr_\mathcal{R}e^{i 2 \pi\alpha_+ t^0} = \frac{\sin n \pi \alpha_+}{\sin \pi \alpha_+}.
\end{equation}

\section{WKB analysis}\label{app:sec:wkb}
\subsection{General remarks and organizations} \label{WKB gener}
Given a $\mathfrak{sl}_2$ $\lambda$-oper, the goal of this appendix is to develop techniques to evaluate the Stokes data, namely the Wronskians between certain solutions (referred to as \textit{small solutions} defined below) to the corresponding Schr\"odinger equation. Our formalism is based on the WKB analysis, which culminates in the Voros analysis \cite{aoki2005virtual,AIF_1993__43_1_163_0, VOROS1983, silverstone1985jwkb, kawai2005algebraic, takei2017wkb,iwaki2014exact,10.1007/3-540-09532-2_85, voros1982spectre,GMN:2009hg,Gaiotto:2012rg}. We refer readers to the appendix of \cite{KondoGLW} and references therein for a review of WKB analysis and other related recent progress. Contrary to the common wisdom, we will find that in the general situation including the examples in this article, the standard WKB analysis will not provide a complete collection of Stokes data, which we briefly review below. 

The WKB analysis is concerned with the holomorphic differential equation
\begin{equation}
\partial_x^2 \psi =\left[\frac{P(x)}{\hbar^2}+ t(x) \right] \psi \equiv T(x) \psi(x) \label{eq:opergeneral}
\end{equation}
defined on a Riemann surface by using the oper coordinate transformation between patches
\begin{equation}
\psi(x) = \frac{1}{\sqrt{\partial_x \tilde x(x)}}\tilde \psi(\tilde x(x)).
\end{equation}
Here we start with brief definitions of two main players, Stokes diagrams and small solutions. Other relevant notions will be explained wherever needed.

$T(x)$ and $P(x)dx^2$ are referred to as the stress tensor and the quadratic differential respectively. We will briefly review the analysis. For any angle $\vartheta \in \mathbb{R}/2\pi \mathbb{Z}$, $\vartheta$-\emph{WKB lines} are curves in the complex plane where 
\begin{equation}
	\mathrm{Im}\  \Big[e^{i \vartheta} \sqrt{P(x)}dx \cdot \partial_t\Big] = 0 
\end{equation}
where $\partial_t$ is the tangent vector of the curve. One such line passes through any point in the $x$ plane. a WKB curve is {\it special} if it ends on a zero of $P$ or on an asymptotic region of exponentially fast decrease for $P$.
The union of all special WKB lines is called a WKB diagram/spectral network. See Footnote \ref{foot}.

For each singularity of $P(x)$, we can define a set of \emph{small solutions}. In particular,  for each special WKB line coming into the singularity, we define a small solution to be the unique solution (up to normalization) that decays exponentially fast approaching the singularity along the corresponding special WKB line. We will choose the normalization such that asymptotically near the singularity, it matches with the WKB solutions $\psi^{\mathrm{WKB}}_\pm(x)$. The definition of the WKB solutions and the associated WKB coordinate system will be given in Section \ref{sec:WKBcoord}.

We would like to find the Wronskians between small solutions. The standard GMN/Voros-style WKB analysis is interested in contour integrals along the generic WKB lines connecting every pairs of Stokes sectors, while, importantly, staying away from the zeros of the quadratic differential $P(x)$. They are indeed all we need to provide a complete collection of Wronskians when all zeroes of the quadratic differential $P(x)$ are simple. However, more generally, the collection is not complete and we also need information from the local analysis of the \textit{matching region}. 

An example where we need a generalized WKB analysis would be a polynomial oper with non-simple zeros. And the zeros are the \textit{matching region} we need to understand. More precisely, for each zero of $P(x)$, we can define a \emph{local} coordinate system and nice local solutions therein. Let's consider first $t(x) = 0$. For each zero $x_0$ of order $n$, we want to find a coordinate system $\tilde{y}(x)$ such that the oper of interest
\begin{equation}
\frac{P(x)}{\hbar^2}
\end{equation}
becomes
\begin{equation}
\frac{\tilde{y}^n}{\hbar^2} + O(\hbar^0)
\end{equation}
where $\tilde{y} (x_0) =  0 + O(\hbar^2)$. We also require the (possibly nonzero) subleading $\hbar$ terms in $\tilde{y}(x)$ are regular at $x_0$.

Generically one can't find $\tilde{y}(x)$ such that the stress tensor takes exactly the form $\frac{\tilde{y}^n}{\hbar^2}$. Instead, it turns out the best one can do is 
\begin{equation}
\frac{\tilde{y}^n}{\hbar^2} +  a_{n-2} \tilde{y}^{n-2} +\dots + a_1 \tilde{y} + a_0. \label{eq:localTtildey}
\end{equation}
The constant coefficients can be determined order by order in $\hbar$,
\begin{equation}
a_{m} = a_{m}^{(0)} + \hbar^2 a_{m}^{(2)} + \hbar^4 a_{m}^{(4)} + \dots
\end{equation}
In particular when $n=1$, namely, around a simple zero, there always exists a coordinate system such that the stress tensor takes the form $\frac{\tilde{y}}{\hbar^2}$. We can also write $y = \hbar^{-\frac{2}{n+2}} \tilde{y}$, in which the stress tensor takes the form
\begin{equation}
y^n + a_{n-2} \hbar^{\frac{2n}{n+2}} y^{n-2}+ \dots +  a_{j} \hbar^{\frac{2j+4}{n+2}} y^{j} +\dots+ a_0 \hbar^{\frac{4}{n+2}}. \label{eq:localTy}
\end{equation}
We will discuss how to find such a local coordinate system and nice local solutions defined in there, as well as the cases with $t(x)\neq 0$, in Subsection \ref{sec:localcoord}. 

Another example of the \textit{matching region} is the negative infinity of the oper with exponential potential. The details on how to deal with such matching region is given in Subsection \ref{app:sec:matchingaround_negative_inf}.

We will first describe three different coordinate systems in detail and the transformation between them in Section \ref{app:sec:coordsystem}. Next in Section \ref{app:sec:recipeWronskian}, we will explain why the WKB analysis away from the zeros is not enough and what information from the local analysis is crucial to give a complete collection of Wronskians/Stokes data. We then give a general recipe for evaluating the Wronskians by incorporating the missing local information while deferring the local perturbative analysis of extracting the local information to Section \ref{sec:localWronskian}.

In order to verify our proposed recipe, we carry out numerical computations. The method of numerical implementation is given in Section \ref{app:sec:numerics}.

Finally, we apply the generalized WKB analysis to the nice examples at hand including polynomial oper with non-simple zeros  in Section \ref{app:sec:polynomial} and the exponential case that describes the chiral WZW model in Section \ref{app:sec:WZW}.

\subsection{Coordinate systems}\label{app:sec:coordsystem}
We would like to define interesting local coordinate systems with good $\hbar \to 0$ asymptotics. There are a few useful coordinate systems we will use frequently in this paper:
\begin{itemize}
	\item the original one, denoted as $x$, where $T = \frac{1}{\hbar^2}P(x) +t(x)$,
	\item the local coordinate around a zero, $y$ or $\tilde{y}$, where $T = y^n +\dots$ or $\frac{\tilde{y}^n}{\hbar^2} + \dots$,
	\item the WKB coordinate $s_{ab}$ where $T = \frac{1}{4}$.
\end{itemize}
\subsubsection{WKB coordinate systems}\label{sec:WKBcoord}
Near each (Stokes sector of a) singularity $a$ of the quadratic differential $P(x) dx^2$ we can find a solution 
$\psi^{\mathrm{WKB}}_a$ which decreases exponentially fast approaching the singularity. It is given as a specific asymptotic expansion near the singularity,
\begin{equation}
\psi_a^{\mathrm{WKB}} = \frac{1}{\sqrt{\pm \partial_x s_a^{\mathrm{asy}}(x)}} e^{\mp\frac12 s_a^{\mathrm{asy}}(x)}
\end{equation}
where $s_a^{\mathrm{asy}}(x)$ is a primitive of twice the \emph{WKB momentum} $p_a^{\mathrm{asy}}(x) = \frac12\partial_x s_a^{\mathrm{asy}}(x)$, which 
satisfies the differential equation 
\begin{equation}
p_a^{\mathrm{asy}}(x)^2  + \frac34 \left(\frac{\partial_x p_a^{\mathrm{asy}}(x)}{p_a^{\mathrm{asy}}(x)}\right)^2   -\frac12 \frac{\partial_x^2 p_a^{\mathrm{asy}}(x)}{p_a^{\mathrm{asy}}(x)}  =  T(x). \label{eq:Ricattiforpasym}
\end{equation}
The one form $p_a^{\mathrm{asy}}(x) dx$ is also referred to as the WKB one form. If we write $p_a^{\mathrm{asy}}(x)$ in $\hbar$ asymptotics as
\begin{equation}
p_a^{\mathrm{asy}}(x;\hbar) = \frac{p^{\mathrm{asy}}_{-1}(x)}{\hbar} + \hbar p^{\mathrm{asy}}_1(x) + \hbar^3 p^{\mathrm{asy}}_3(x) + \cdots \label{eq:momentumexpand}
\end{equation}
the equation \eqref{eq:Ricattiforpasym} can then be written as a recursive equation for $p_n(x)$. The leading term is given by $p^{\mathrm{asy}}_{-1}(x)\sim \sqrt{P(x)}$.

For a polynomial singularity at $x = \infty$, the asymptotic expansion $s_a^{\mathrm{asy}}(x)$
involves increasingly negative powers of $x$, with coefficients which are Laurent polynomials in $\hbar$. {In order to fix the normalization at $x = \infty$ we only need to worry about powers of $x$ greater than $-1$, so the last two terms are unimportant.} For a singularity of odd degree, the expansion involves fractional powers of $x$ and thus $s_a^{\mathrm{asy}}(x)$ can be chosen unambiguously to have no constant term. For a singularity of even degree, there will be a $\log x$ term and we will have to make some choice. Of course, sometimes there are some natural nonzero choices of the constant as well. See, for example, Section \ref{sec:Exx22a}. For more generic cases, we will fix the constant terms of $s_a^{\mathrm{asy}}(x)$ on a case-by-case basis. 

Once we fix the constant term in $s_a^{\mathrm{asy}}(x)$, the normalization of the WKB solutions $\psi_a^{\mathrm{WKB}}$ is then fixed. This then further fixes the normalization of the small solutions, which, as we defined in section \ref{WKB gener}, are normalized to match the WKB solutions asymptotically near the singularity. 

Although we defined $s_a^{\mathrm{asy}}(x)$ only as an asymptotic series, it is easy to produce actual functions which 
have such an asymptotic expansion: given any other solution\footnote{Unless we state explicitly otherwise, by a \emph{solution}, we mean an actual function, as opposed to a WKB solution, which is an asymptotic expansion.} $\psi_*$ which is not proportional to $\psi_a$, 
we could define 
\begin{equation}
e^{s_{a,*}(x)} = \frac{1}{(\psi_a,\psi_*)} \frac{\psi_*(x)}{\psi_a(x)}
\end{equation} 
Redefinitions of $\psi_*$ by multiples of $\psi_a$ would just shift $e^{s_{a,*}(x)}$ by an $x$-independent constant.

For generic $x$, we sit along a generic WKB line associated to a pair of small solutions 
$\psi_a(x)$ and $\psi_b(x)$, with some normalization determined at the corresponding
directions at infinity or singularities. 

These solutions have good $\hbar \to 0$ asymptotics in a certain sector of width $\pi$ in the 
$\hbar$ plane. So does their ratio, which defines an useful local coordinate
\begin{equation}
z_{ab}(x) \equiv e^{s_{ab}(x)} = \frac{1}{(\psi_a,\psi_b)} \frac{\psi_b(x)}{\psi_a(x)}
\end{equation} 
Then 
\begin{equation}
\partial_x z_{ab} = \frac{1}{\psi_a^2}
\end{equation}
so that the solutions $\psi_a$ and $\psi_b$ map to $1$ and $(\psi_a,\psi_b)z_{ab}$ in the $z_{ab}$ coordinate. Or equivalently, the stress tensor in $z_{ab}$ coordinate is $T(z_{ab}) = 0$.

Also, 
\begin{equation}
\partial_x s_{ab} \equiv 2 p_{ab}(x)= \frac{(\psi_a,\psi_b)}{\psi_a(x) \psi_b(x)}
\end{equation}
and thus the solutions $\psi_a$ and $\psi_b$ map to $e^{-\frac12 s_{ab}(x)}$ and $(\psi_a,\psi_b) e^{\frac12 s_{ab}(x)}$ in the $s_{ab}$ coordinate. Equivalently, the stress tensor in $s_{ab}$ coordinate is $T(s_{ab}) = \frac14$

$s_{ab}(x)$ can be determined in the following way. Similar to \eqref{eq:Ricattiforpasym}, the WKB momentum $p_{ab}(x)\equiv  \frac12 \partial_x s_{ab}$ satisfies the differential equation
\begin{equation}
p_{ab}(x)^2 + \frac34 \left(\frac{\partial_x p_{ab}(x)}{p_{ab}(x)}\right)^2 -\frac12 \frac{\partial_x^2 p_{ab}(x)}{p_{ab}(x)}  =  T(x) 
\end{equation}
which can be solved recursively to find the WKB expansion of $p(x)$ away from zeroes of $P(x)$:
\begin{equation}
p(x) = \frac{1}{\hbar} \sqrt{P(x)}+ \hbar \frac{16 P^2(x) t(x) - 5  P'(x)^2+ 4 P(x)  P''(x)}{32 P(x)^{\frac52}}+ \cdots
\end{equation}
It is easy to see that $p_{n>0}(x)$ will involve increasingly negative powers of $p_{-1}(x) = \sqrt{P(x)}$. Therefore the zeros of $P(x)$ remain the places where the WKB approximation breaks down, regardless of $t(x)$.

In order to compute $s_{ab}$ from $p_{ab}$, we need to fix the integration constants. This is done by 
expanding $p(x)$ near the singularity and comparing term-by-term with $s_a^{\mathrm{asy}}(x)$.

It is easy to express $s_{ab}$ as a regularized contour integral. We can write 
\begin{equation}
s_{ab}(x) = s_{ab}(\bar x) + \int_{\bar x}^x p_{ab}(u) du 
\end{equation}
and send $\bar x$ towards the singularity $x_a$:
\begin{equation}
s_{ab}(x) = \mathrm{lim}_{\bar x \to x_a} \left[s_a^{\mathrm{asy}}(\bar x) + \int_{\bar x}^x p_{ab}(u) du \right].
\end{equation}

Another useful observation is that 
\begin{equation}
e^{s_{ab}(x)+ s_{ba}(x)} = \frac{1}{(\psi_a,\psi_b)(\psi_b,\psi_a)} 
\end{equation} 
and $p_{ab} +p_{ba} = 0$, so that 
\begin{equation}
\log \left[(\psi_a,\psi_b)(\psi_b,\psi_a) \right] = - \mathrm{lim}_{\bar x \to x_a} \mathrm{lim}_{\bar x' \to x_b} \left[s_a^{\mathrm{asy}}(\bar x) +s_b^{\mathrm{asy}}(\bar x') + \int_{\bar x}^{\bar x'} p_{ab}(u) du \right]
\end{equation}
with the integral taken along a path equivalent to the WKB line between $x_a$ and $x_b$. The overall sign can be determined by computing the Wronskian explicitly: 
\begin{equation}
(\psi_a,\psi_b) = e^{-\frac12 s_{ab}(x)-\frac12 s_{ba}(x)} \frac{\sqrt{p_{ab}(x)}}{\sqrt{p_{ba}(x)}}.
\end{equation} 
\subsubsection{Local coordinate system around zeros}\label{sec:localcoord}
Here in this section, we explain how one can find explicitly the coordinate system in which $\frac{1}{\hbar^2}P(x)$ around a zero of order $n$ takes the form of 
\begin{equation}
\frac{\tilde{y}^n}{\hbar^2} +  a_{n-2} \tilde{y}^{n-2} +\dots + a_1 \tilde{y} + a_0 
\end{equation}
where $a_i = a_i^{(0)} + a_i^{(2)}\hbar^2 +a_i^{(4)}\hbar^4+\dots$. Or with $y = \hbar^{-\frac{2}{n+2}} \tilde{y}$, 
\begin{equation}
y^n + a_{n-2} \hbar^{\frac{2n}{n+2}} y^{n-2}+ \dots +  a_{j} \hbar^{\frac{2j+4}{n+2}} y^{j} +\dots+ a_0 \hbar^{\frac{4}{n+2}}. \label{eq:Tlocalcoordy}
\end{equation}
In this section, we will mostly use $\tilde{y}$ so that only integer power of $\hbar$ will appear.

When we have nontrivial $t(x) = a(x)^2 + \partial_x a(x)$, and $a(x)$ has residue $-l$ at $x_0$, 
\begin{equation}
\frac{\tilde{y}^n}{\hbar^2} +  a_{n-2} \tilde{y}^{n-2} +\dots + a_1 \tilde{y} + a_0 + \frac{l(l+1)}{\tilde{y}^2}
\end{equation}
or 
\begin{equation}
y^n + a_{n-2} \hbar^{\frac{2n}{n+2}} y^{n-2}+ \dots +  a_{j} \hbar^{\frac{2j+4}{n+2}} y^{j} +\dots+ a_0 \hbar^{\frac{4}{n+2}} +\frac{l(l+1)}{y^2}
\end{equation}
In particular if $t(x)$ is nontrivial but regular at the zero of interest $x_0$, we should find the stress tensor takes the local form \eqref{eq:localTtildey} and \eqref{eq:localTy}.

Below we will see that $\tilde{y}(x)$ can be uniquely determined with no ambiguity and that it is unavoidable to have the coefficients $a_i$ by providing two different ways of finding such $\tilde{y}(x)$. 
By doing so we will explain why the coefficients $a_m$ are unavoidable. (1) In comparing the local coordinate and other coordinates, we need to have some parameters to adjust so that the cross ratios defined from the local solutions coincide with the same cross-ratio built from $\psi_n(x)$. (2) $a_m$ need to be specific values to make the coordinate transformation $\tilde{y}(x)$ non-divergent at the interested zero.

\noindent{\bf \emph{local} coordinate transformation}

We call this a local map since it is only valid in the neighborhood of the zero
\begin{equation}
\hbar\ll \tilde{y}^2 \ll \tilde{y} \sim (x-x_0) \ll 1.
\end{equation}
Suppose we are interested in the zero of order $n$, then apparently one can always do a shift in the coordinate such that
\begin{equation}
P(x) = t_1 x + t_2 x^2 + t_3 x^3 + \dots + t_N x^N
\end{equation}
where $t_1 = t_2 = \dots = t_{n-1} = 0$ and $x=0$ is the order $n$ zero of interest. By solving the following equation
\begin{equation}
\left(\frac{\partial \tilde{y}}{\partial x}\right)^2\Big(\frac{\tilde{y}^n}{\hbar^2} +  a_{n-2} \tilde{y}^{n-2} +\dots + a_1 \tilde{y} + a_0 \Big) -\frac12 \{\tilde{y},x\} =  \frac{P(x)}{\hbar^2} 
\end{equation}
order by order in $\hbar$ and $x$, we can determine the coordinate transformation $\tilde{y}(x)$ and more importantly $\{a_m\}$ explicitly as functions of $\{t_n, t_{n+1}, \dots, t_N \}$. 

\noindent{\bf \emph{non-local} coordinate transformation}

To state again, we want to find a coordinate transformation $\tilde{y}(x)$ with respect to a zero $x_0$ of order $n$ satisfying
\begin{equation}
\left(\frac{\partial \tilde{y}}{\partial x}\right)^2\Big(\frac{\tilde{y}^n}{\hbar^2} +  a_{n-2} \tilde{y}^{n-2} +\dots + a_1 \tilde{y} + a_0 \Big) -\frac12 \{\tilde{y},x\} =  \frac{P(x)}{\hbar^2}
\end{equation}
where $\tilde{y}(x) = \tilde{y}_0(x) + \hbar \tilde{y}_1(x) + \hbar^2 \tilde{y}_2(x) + \dots $. This equation can be solved recursively order by order in $\hbar$, and at each order we have a first order differential equation. The leading order equation $ \tilde{y}_0'^2\tilde{y}_0^n = P(x)$ is solved by
\begin{equation}
\tilde{y}_0(x) = \left(\frac{n+2}{2} \int_{x_0}^{x} \sqrt{P(x')} dx'  \right)^{\frac{2}{n+2}}.
\end{equation}
The integration constant is fixed by choosing the integration starting point from the interested zero $x_0$ of order $n$, such that locally around $x_0$, $\tilde y_0(x)$ start from linear order in $(x-x_0)$ and vanishes at $x_0$.

At the order of $\hbar^{-1}$, we have a homogeneous differential equation $n \tilde{y}_1y_0' + 2\tilde{y}_0 \tilde{y}_1' = 0$, which is solved by 
\begin{equation}
\tilde{y}_1(x) = \frac{c_1}{\tilde{y}_0(x)^{n/2}}.
\end{equation}
Since $\tilde{y}_0(x)$ vanishes at the zero $x_0$ and we require $\tilde{y}(x)$ to be regular there, the only choice is to choose $c_1 = 0$, hence $\tilde{y}_1(x) = 0$. The same is true for ever odd order in $\hbar$.

At the order of $\hbar^0$, we end up with the an inhomogeneous first order differential equation for $\tilde{y}_2(x)$. We fix the homogeneous part $\frac{1}{\tilde{y}_0(x)^{n/2}}$ of the solution by choosing the lower limit of the integral at $x_0$. This renders $\tilde y_2(x)$ regular, and in general nonzero, at $x_0$. For example
\begin{align}
\textrm{simple zero: }\tilde{y}_2(x) & = \frac{1}{\sqrt{\tilde{y}_0(x)}} \int_{x_0}^x \sqrt{\tilde{y}_0(x')} \frac{-3 (\tilde{y}_0'')^2 + 2\tilde{y}_0'\tilde{y}_0'''}{8 \tilde{y}_0(\tilde{y}_0')^3} dx', \\
\textrm{double zero: }\tilde{y}_2(x) & = \frac{1}{\tilde{y}_0(x)} \int_{x_0}^x \tilde{y}_0(x') \frac{-4 a^{(0)}_0 (\tilde{y}_0')^4 -3 (\tilde{y}_0'')^2 + 2\tilde{y}_0'\tilde{y}_0'''}{8 \tilde{y}_0^2(\tilde{y}_0')^3} dx', \\
\textrm{cubic zero: }\tilde{y}_2(x) &= \frac{1}{\tilde{y}_0(x)^{3/2}} \int_{x_0}^{x} \tilde{y}_0^{3/2} \frac{-4 a_0^{(0)} \tilde{y}_0'^4-4 a_1^{(0)} \tilde{y}_0 \tilde{y}_0'^4-3 \tilde{y}_0''^2+2 \tilde{y}_0' \tilde{y}_0''' }{8 \tilde{y}_0^3 \tilde{y}_0'^3}.
\end{align}
Furthermore, one can also understand the necessity of the coefficients $a_i$ in \eqref{eq:Tlocalcoordy} from the fact that we need them in order to have $\tilde{y}_2(x)$ non-divergent at $x_0$.
\subsubsection{Relating two coordinate systems}
\noindent {\bf Near a simple zero}

There are three special WKB lines emanating from a simple zero. Therefore, near a simple zero, we can access three small solutions $\psi_a$, $\psi_b$, $\psi_c$, decaying exponentially along the three special WKB lines respectively. Apparently there must be a linear relation among them, sometimes referred to as Pl\"ucker relations. 
\begin{equation}
(\psi_a, \psi_b) \psi_c + (\psi_c, \psi_a) \psi_b + (\psi_b, \psi_c) \psi_a =0.
\end{equation}

On the other hand, there exists a \emph{local} coordinate system $y$ associated to this simple zero, in which the stress tensor reads $T(y) = y$. The coordinate $y$ satisfies the differential equation
\begin{equation}
\left(\frac{\partial y(x)}{\partial x} \right)^2 y(x) +\frac34 \left(\frac{\partial^2_x y(x)}{\partial_x y(x)}\right)^2 -\frac12 \frac{\partial_x^3 y(x)}{\partial_x y(x)}   = T(x) 
\end{equation}
and can be expanded as $\hbar \to 0$ to be $y(x) = \hbar^{-\frac23}(\tilde{y}_0+\hbar^2 \tilde{y}_2+ \hbar^4 \tilde{y}_4+\dots)$. Then the nice local solutions in this local coordinate system are given by  \eqref{eq:localsolA1a},
\begin{equation}
\mathrm{Ai}_{a}(y) = \sqrt{2\pi} e^{-\frac{\pi i a}{3}}\mathrm{Ai}(e^{\frac{2 \pi i a}{3}}y).
\end{equation}
In particular, we have
\begin{equation}
\mathrm{Ai}_{-1}(y) -\mathrm{Ai}_{0}(y) +\mathrm{Ai}_{1}(y) = 0.
\end{equation}
Define $y(x)$ by 
\begin{equation}
\frac{\mathrm{Ai}_1(y(x))}{\mathrm{Ai}_{-1}(y(x))} = \frac{(\psi_c, \psi_a) \psi_b}{(\psi_a, \psi_b) \psi_c}.
\end{equation}
This obviously satisfies also 
\begin{equation}
-\frac{\mathrm{Ai}_0(y(x))}{\mathrm{Ai}_{-1}(y(x))} = \frac{(\psi_b, \psi_c) \psi_a}{(\psi_a, \psi_b) \psi_c}, \quad -\frac{\mathrm{Ai}_1(y(x))}{\mathrm{Ai}_{0}(y(x))} = \frac{(\psi_c, \psi_a) \psi_b}{(\psi_b, \psi_c) \psi_a}
\end{equation}
which gives us the relation between small solutions $\psi_\bullet(x)$ and the local solutions $\mathrm{Ai}_\bullet(y(x))$. For example,
\begin{equation}
\partial_x y(x) \psi^2_a(x) = \frac{(\psi_c, \psi_a) (\psi_a,\psi_b)}{i (\psi_b, \psi_c)} \mathrm{Ai}_{0}(y(x))^2.
\end{equation}
We can now relate $y$ to $s_{ab}$ and to the other WKB coordinates nearby: 
\begin{equation}
\frac{\mathrm{Ai}_1(y(x))}{\mathrm{Ai}_{0}(y(x))} = - \frac{(\psi_a,\psi_b)(\psi_c, \psi_a)}{(\psi_b, \psi_c)}e^{s_{ab}(x)}.
\end{equation}
The coordinate $s_{ab}(x)$ depend on the normalization of the local solutions $\mathrm{Ai}_a(y)$ and small solutions $\psi_a(x)$. One can define an alternative local WKB coordinate, i.e. an alternative primitive of $p_{ab}(x)$, that is independent of the normalization of the solutions, given as follows
\begin{equation}
e^{s_{abc}(x)} \equiv  \frac{(\psi_c, \psi_a) \psi_b}{(\psi_b, \psi_c) \psi_a} =\frac{(\psi_a,\psi_b)(\psi_c, \psi_a)}{(\psi_b, \psi_c)}e^{s_{ab}(x)} = -\frac{\mathrm{Ai}_1(y(x))}{\mathrm{Ai}_{0}(y(x))}.
\end{equation} 
If we expand the right hand side $-\frac{\mathrm{Ai}_1(y(x))}{\mathrm{Ai}_{0}(y(x))}$ in an asymptotic expansion at large $y(x)$ using 
\begin{equation}
\mathrm{Ai}_a(y) \sim \frac{1}{\sqrt{\pm 2 \partial_y S(y)}}e^{\mp S(y)}
\end{equation}
where the function $S(y) = \frac{2}{3} y^{3/2} + \frac{5}{48} \frac{1}{y^{3/2}}+ \dots$. we get
\begin{equation}
S(y) \sim  \frac{2\tilde{y}_0^{3/2}}{3\hbar} + \hbar \left(\frac{5}{48 \tilde{y}_0^{3/2}} +\tilde{y}_0^{1/2} \tilde{y}_2 \right) + \hbar^3 \dots
= \frac{1}{\hbar} \int_{x_0}^x \sqrt{P(x')} dx' + \hbar (\dots) \label{eq:localS(y)1st}
\end{equation}
then we can obtain the $\hbar$ expansion of $s_{abc}(x)$,
\begin{align}
{s_{abc}(x)} &= \frac{\pi}{2} +   2 \left(\frac23 y(x,\hbar)^{3/2} + \frac{5}{48} \frac{1}{y(x,\hbar)^{3/2}} + \frac{1105}{9216} \frac{1}{y(x,\hbar)^{9/2}} + \dots\right)\\
& = \frac{\pi}{2} + 2\left(\frac{2\tilde{y}_0(x)^{3/2}}{3\hbar} + \left(\frac{5}{48\tilde{y}_0(x)^{3/2}} + \tilde{y}_0(x)^{1/2}\tilde{y}_2(x) \right)\hbar +\dots\right).
\end{align}
We can think about this as a regularized version of the integral of $2p_{ab}$ from the zero to $x$. More precisely, as all the ingredients of the definition above have good WKB asymptotics, we will show below that the $\hbar$ expansion exactly coincide with the contour integral of $\frac12 p_{ab}$ from $x$ to $x$ along a path $\gamma_{abc}$ which winds around the zero while keeping away from it: namely, up to some multiple of $i\frac{\pi}{2}$ we have
\begin{equation}
s_{abc}(x) = \int_{\gamma_{abc}(x)} p_{ab}(u) du. \label{eq:sabcasintofp}
\end{equation}
We should really keep track of factors of $\pm i$ in front of the exponents: 
\begin{equation}
e^{s_{abc}(x)} = \frac{\frac{\sqrt{p_{ab}(x)}}{\sqrt{p_{ba}(x)}} \frac{\sqrt{p_{ca}(x')}}{\sqrt{p_{ac}(x')}} }{ \frac{\sqrt{p_{bc}(x'')}}{\sqrt{p_{cb}(x'')}}  } e^{ \frac12 s_{ab}(x)-\frac12 s_{ac}(x')} e^{-\frac12 s_{ca}(x')+\frac12 s_{cb}(x'')}e^{\frac12 s_{bc}(x'')- \frac12 s_{ba}(x)}.
\end{equation} 
We now show explicitly why \eqref{eq:sabcasintofp} is true at the first two orders. Recall that the WKB momentum is
\begin{equation}
p(x,\hbar) = \frac{1}{\hbar} \sqrt{P(x)} + \hbar \Big[\frac{-5 P'^2+4 PP''}{32 P^{5/2}} \Big] +\dots
\end{equation} 
The leading order is easy since $\sqrt{P(x)}$ is integrable at the zero and we can just pinch the contour to the zero  
\begin{equation}
2\frac{2\tilde{y}_0(x)^{3/2}}{3\hbar} = \frac{2}{\hbar} \int_{x_0}^{x} \sqrt{P(x')}dx'.
\end{equation}
The order $\hbar$ is not as obvious 
\begin{equation}
\frac{5}{48}\tilde{y}_0(x)^{-3/2} + \tilde{y}_0(x)^{1/2}\tilde{y}_2(x)  = \frac{5}{48} \frac{1}{\int_{x_0}^x \sqrt{P(x)}} + \int_{x_0}^x \tilde{y}_{0}^{-1/2}(x') Q(x')dx'
\end{equation}
where 
\begin{equation}
{Q}(x) = \frac{-3 \tilde{y}_0''^2 + 2\tilde{y}_0'\tilde{y}_0'''}{8\tilde{y}_0'^3}
\end{equation}
is generically nonzero and regular as $x\rightarrow x_0$. Let's now try to understand how this realizes a primitive of the WKB one form at the order of $\hbar$, which is $\frac{4 PP''-5 P'^2}{32 P^{5/2}}$. Apparently, it is not integrable at the zero $x_0$ because it is divergent there. However, we can rewrite it as
\begin{align}
p_{1}(x) & = \frac{4 PP''-5 P'^2}{32 P^{5/2}} = \left[\frac{4 PP''-5 P'^2}{32 P^{5/2}} +\frac{5}{32} \frac{\sqrt{P(x)}}{\tilde{y}_0^3}\right] - \frac{5}{32} \frac{\sqrt{P(x)}}{\tilde{y}_0^3}  \\
&= \frac{{Q}(x)}{\sqrt{\tilde{y}_0}} - \frac{5}{32} \frac{\sqrt{P(x)}}{\tilde{y}_0^3}.
\label{eq:separatep}
\end{align}
Note that generically around the zero $x_0$, the behavior of $\tilde{y}_0(x)$ is
\begin{equation}
\tilde{y}_0(x) = \left(\frac{3}{2} \int_{x_0}^{x} \sqrt{P(x')} dx'  \right)^{\frac{2}{3}} \overset{x\rightarrow x_0}{\rightarrow} \#(x-x_0) + \dots
\end{equation}
So the first term in \eqref{eq:separatep} is integrable, and we can shrink the contour $\gamma_{abc}(x)$ to  the zero. The second term is badly divergent but a total derivative
\begin{equation}
-\frac{5}{32} \frac{\sqrt{P(x)}}{\tilde{y}_0^3} = \left(\frac{5}{48} \frac{1}{\tilde{y}_0^{3/2}}\right)' =\left( \frac{10}{72} \frac{1}{\int_{x_0}^x\sqrt{P(x')dx'}}\right)'.
\end{equation}
Therefore, we have found a good primitive of $p_{1}(x)$, which is the order $\hbar$ part of $p(x,\hbar)$
\begin{align}
\int_{x_0}^x \frac{{Q}(x')}{\sqrt{\tilde{y}_0}}dx' + \frac{5}{48} \frac{1}{\tilde{y}_0(x)^{3/2}}.
\end{align}
This finishes the proof of \eqref{eq:sabcasintofp} at the order of $\hbar$. This way of separating divergent total derivative from the less divergent part in \eqref{eq:separatep} is very reminiscent of a more well-known way \cite{PhysRevA.38.1679,PhysRevD.16.1740} to evaluate the contour integral, which goes as follows. Write $P(x)= V(x) -E$.
\begin{align}
p_{1}(x) & = \frac{4 PP''-5 P'^2}{32 P^{5/2}} = \frac{4 (V-E)V''-5 V'^2}{32 (V-E)^{5/2}}\\
& = \Big[\frac{1}{8} \frac{V''}{(V-E)^{3/2}} -\frac{5}{48}  \frac{V''}{(V-E)^{3/2}}\Big] +\frac{5}{48}\frac{d}{dx} \frac{V'}{(V-E)^{3/2}}\\
& = \frac{1}{24} \frac{d}{dE} \frac{V''}{(V-E)^{1/2}}  +\frac{5}{48}\frac{d}{dx} \frac{V'}{(V-E)^{3/2}}.
\end{align}
Notice that the first term is integrable at the zero and the second term is a total derivative.

\noindent {\bf Near a double zero}

\begin{figure}
	\centering
	\includegraphics[width=0.7\linewidth]{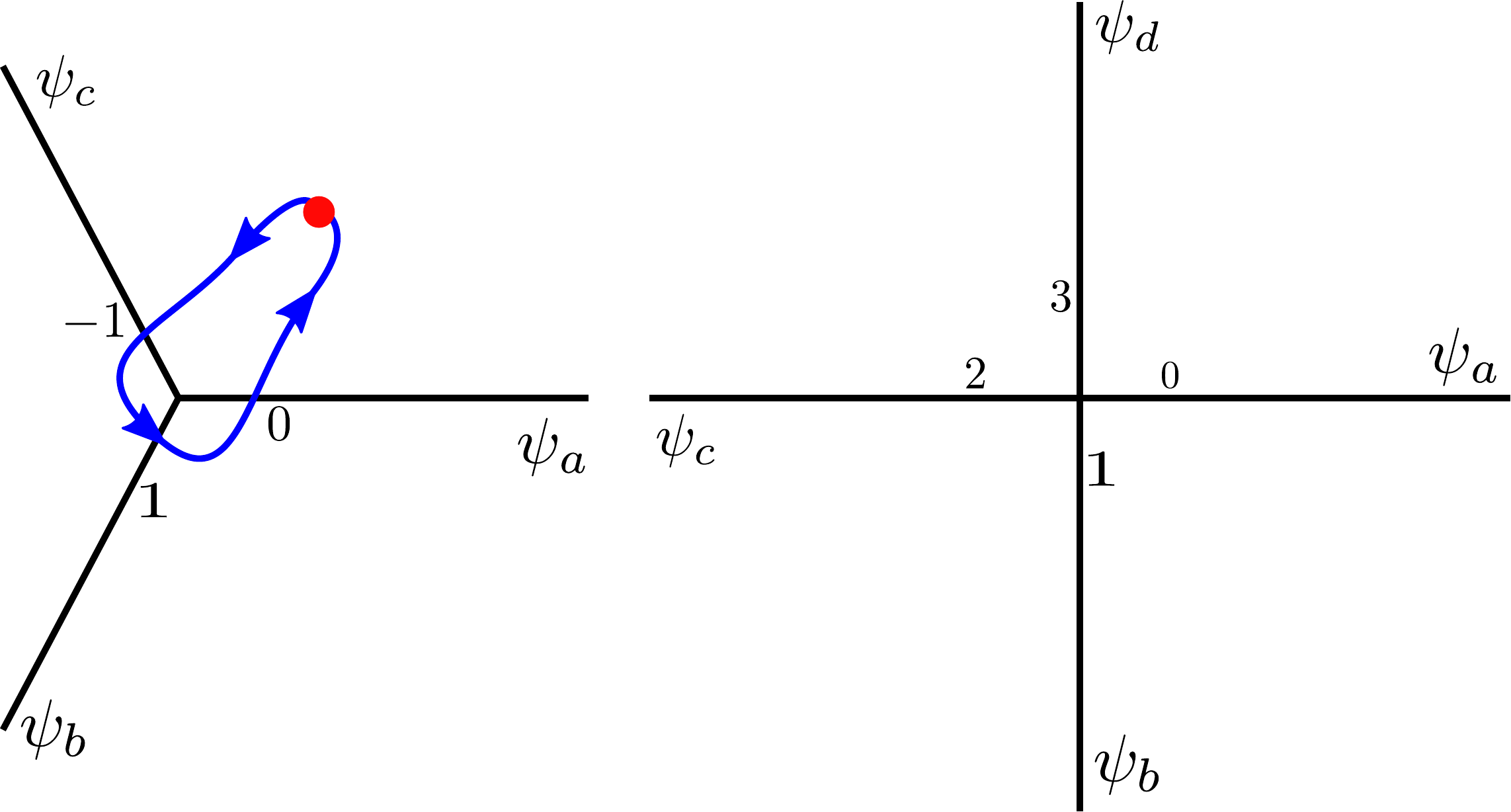}
	\caption{The Stokes diagrams in the local coordinate system around the simple zero and the double zero. The numbers close to the origin label the numbering for the nice local solutions $A_i(y)$ we use in this section, whereas $\psi_\bullet$ label the corresponding small solutions.}
	\label{fig:localcoord}
\end{figure}
If the quadratic differential $P(x)$ has a double zero at $x_0$, then there are four special WKB lines emanating from $x_0$, around which we can access four small solutions, which are denoted as $\psi_a$, $\psi_b$, $\psi_c$, $\psi_d$ in the Fig. \ref{fig:localcoord}. We now have linear relations 
\begin{equation}
\begin{split}
(\psi_a, \psi_b) \psi_c + (\psi_c, \psi_a) \psi_b + (\psi_b, \psi_c) \psi_a =0,\\
(\psi_b, \psi_c) \psi_d + (\psi_d, \psi_b) \psi_c + (\psi_c, \psi_d) \psi_b =0,\\
(\psi_c, \psi_d) \psi_a + (\psi_a, \psi_c) \psi_d + (\psi_d, \psi_a) \psi_c =0,\\
(\psi_d, \psi_a) \psi_b + (\psi_b, \psi_d) \psi_a + (\psi_a, \psi_b) \psi_d =0.
\end{split}
\end{equation}
One can define a nontrivial cross-ratio out of the four solutions $\chi = \frac{(\psi_a, \psi_b)(\psi_c, \psi_d)}{(\psi_b, \psi_c)(\psi_d, \psi_a)}$, which is close to $1$ as $\hbar \to 0$. Note that the Wronskians between non-adjacent solutions are not accessible to the standard WKB analysis. One of the goals of this section is to provide a way to evaluate such Wronskians by relating to the local coordinate system around this double zero. 

Recall that from the general expression \eqref{eq:Tlocalcoordy}, the Schr\"odinger equation in this local coordinate system takes the form
\begin{equation}
\partial_y^2 A(y) = \left[y^2 + \hbar a(\hbar)\right]A(y) \label{eq:ODEdoublezero}
\end{equation}
with $a(\hbar) = a_0 + \hbar^2 a_2+\hbar^4 a_4 + \dots$ being some function of $\hbar$ that can be determined in $\hbar\rightarrow 0$ asymptotics from the differential equation
\begin{equation}
\left(\partial_x y(x) \right)^2 \left[y(x)^2 +\hbar a(\hbar)  \right] + \frac{3}{4}\left(\frac{\partial^2_x y(x)}{\partial_x y(x)}\right)^2  -\frac12 \frac{\partial_x^3 y(x)}{\partial_x y(x)}  =  T(x).
\end{equation}

Nice local solutions to \eqref{eq:ODEdoublezero} are given in \eqref{eq:localsolAn0}, \eqref{eq:localsolAna} and \eqref{eq:localsolA2n}, which we denote as $A_i$, $i \in \mathbb{Z}$. To relate these local solutions to the small solutions $\psi_a$, $\psi_b$, $\psi_c$, $\psi_d$,  we now require
\begin{equation}
\frac{ (A_2,A_0) A_1(y(x))}{(A_1,A_2) A_0(y(x))} = \frac{(\psi_c, \psi_a) \psi_b}{(\psi_b, \psi_c) \psi_a} \equiv e^{s_{abc}(x)}
\end{equation}
which also implies easily 
\begin{equation}
\frac{ (A_2,A_0) A_1(y(x))}{(A_0,A_1) A_2(y(x))} = \frac{(\psi_c, \psi_a) \psi_b}{(\psi_a, \psi_b) \psi_c} \equiv e^{s_{cba}(x)}.
\end{equation}
On the other hand, if the cross-ratio of $A_i$ is adjusted to be $\chi$, that can be written as 
\begin{equation}
\frac{ (A_2,A_3) A_1(y(x))}{(A_3,A_1) A_2(y(x))} = \frac{(\psi_c, \psi_d)\psi_b}{(\psi_d, \psi_b) \psi_c} \equiv e^{s_{cbd}(x)}
\end{equation}
etcetera.

In short, we have a good coordinate in all four sectors. In particular, that means we could determine this way the asymptotic expansion of the cross-ration $\chi$. 
On the other hand, that asymptotic expansion is already computable from a contour integral of $ p(x)$
on a contour wrapping around the double zero while keeping away from it. 

We can now relate $y$ to $s_{ab}$ and to the other WKB coordinates nearby: 
\begin{equation}
\frac{ (A_2,A_0) A_1(y(x))}{(A_1,A_2) A_0(y(x))} = e^{s_{abc}(x)} =\frac{(\psi_c, \psi_a) \psi_b}{(\psi_b, \psi_c) \psi_a}. \label{eq:sabcDoubleZero}
\end{equation}
If we expand the left hand side in an asymptotic expansion at large $y(x)$, we can obtain the $\hbar$ expansion of $e^{s_{abc}(x)}$ as follows. Asymptotically at large $y$, $A_0(y)\sim \frac{1}{\sqrt{2\partial_y S(y)}} e^{-S(y)}$, with 
\begin{align}
S(y) &\sim \frac{y^{2}}{2}+\frac{1}{2} a \hbar \log y+\frac{3+a^{2} \hbar^{2}}{16 } y^{-2}+ \left({-\frac{19 a \hbar}{64}-\frac{a^{3} \hbar^{3}}{64}}\right) y^{-4} + O(y^{-6})\\
&\sim \frac{\tilde{y}_0^2}{2\hbar} + \hbar \Big[\frac{3}{16} \frac{1}{\tilde{y}_0^2} +\tilde{y}_0 \tilde{y}_2 +\frac{1}{2}a^{(0)}  \log \tilde{y}_0 -\frac14 a^{(0)} \log \hbar \Big] + O(\hbar^3)
\end{align}
where we have parameterized the $\hbar\rightarrow 0$ asymptotics $y(x) = \hbar^{-1/2}(\tilde{y}_0(x) + \hbar^2 \tilde{y}_2(x) + O(\hbar^4))$ and $a(\hbar) = a^{(0)} + \hbar^2 a^{(2)} + O(\hbar^4)$. We then obtain the $\hbar$ expansion of $s_{abc}(x)$
\begin{align*}
&\log\frac{(A_2,A_0)}{(A_1,A_2)} -\frac{\pi i}{2} +  2 \left(\frac{\tilde{y}_0(x)^2}{2\hbar} + \hbar \Big[\frac{3}{16} \frac{1}{\tilde{y}_0(x)^2} +\tilde{y}_0(x) \tilde{y}_2(x) \right.\\
&\qquad\qquad\qquad\qquad\qquad\qquad\qquad\qquad \left. + \frac{1}{2}a^{(0)}  \log \tilde{y}_0(x) -\frac14 a^{(0)} \log \hbar \Big] + O(\hbar^3) \right).
\end{align*}

Let's note a crucial point of \eqref{eq:sabcDoubleZero}. While $(\psi_c, \psi_a)$ cannot be computed by a naive WKB contour integral away from the zeroes, everything else in \eqref{eq:sabcDoubleZero} can be in principle evaluated: $(A_1,A_2)$ is normalized to be $-i$; $(A_2,A_0)$ will be computed in \eqref{Am1 A2}; $(\psi_b,\psi_c)$ is controlled by a WKB contour integral. Therefore \eqref{eq:sabcDoubleZero} provides an interesting prediction of $(\psi_c, \psi_a)$.  So the relation above should really be written as 
\begin{equation}
(\psi_c, \psi_a)  = (A_2,A_0)\frac{(\psi_b, \psi_c)}{(A_1,A_2)} \frac{ A_1(y(x))}{\psi_b(x)}\frac{\psi_a(x)}{A_0(y(x))} 
\end{equation}
Note that $A_1(y(x))$ and $A_0(y(x))$ have to be proportional to $\sqrt{y'(x)}\psi_b(x)$ and $\sqrt{y'(x)}\psi_a(x)$, respectively. Therefore to figure out $(\psi_c,\psi_a)$, which is $x$ independent, we just need to figure out the constants of proportionality. We can do so by comparing the $\hbar$ expansions and read out the $x$ independent terms. 

\subsection{Recipe for evaluating Wronskians}\label{app:sec:recipeWronskian}
\begin{enumerate}
	\item If we have $\psi(x)$ either exact or numerical solution, we can just evaluate $(\psi_n,\psi_m)$. Normalization of the solutions are not important if one is only interested in the cross ratios.
	\item If two solutions $\psi_n$ and $\psi_m$ are connected to the same zero of order $k$, we have to choose a branch of $\tilde{y}_0 = \left(\frac{k+2}{2}\int \sqrt{P}dx\right)^{\frac{2}{k+2}}$. This is equivalent to choosings how the local solutions $A_{k;a}(y)$ correspond to the small solutions $\psi_a(x)$. For each pair $\psi_n(x) \propto [\partial_x y(x)]^{-1/2} A_{k;a}(y)$, $\psi_n(x) \propto [\partial_x y(x)]^{-1/2} A_{k;a}(y)$, we can just read out the constant of proportionality from the large $x$ asymptotics in the corresponding directions. If we denote the constants of proportionality as $C_n(\hbar)$ and $C_m(\hbar)$, the Wronskians are just
	\begin{equation}
	(\psi_n,\psi_m) = C_n(\hbar) C_m(\hbar) (A_{k;a},A_{k,b}).
	\end{equation}
	\item If two solutions $\psi_n$ and $\psi_m$ are not connected via special WKB lines to the same zero, we can use Pl\"ucker relation to reduce to the previous case.
	\item What remains is to figure out the Wronskians between local solutions $(A_{k;a},A_{k,b})$. This will be done via perturbation theory in Section \ref{sec:localWronskian}.
\end{enumerate}
\subsection{Numerical implementation}\label{app:sec:numerics}
Here in this section, we explain how we implement the numerics. In particular, given an ODE
\begin{equation}
\partial_x^2 \psi(x) = \left(\frac{P(x)}{\hbar^2}+ t(x)\right) \psi
\end{equation}
we would like to find the corresponding small solutions and evaluate the Wronskians between them. 

Let's first consider the case where $P(x)$ is a polynomial of degree $n$ and $t(x) = 0$. In this case the ODE is regular everywhere on the complex plane with $n+2$ asymptotic direction towards the irregular singularity at infinity. Then the task would be to find the decaying solutions along each asymptotic direction. However, initial value problems are more natural in numerics, where one usually specifies the initial condition (the value of $\psi$ and $\partial_x\psi$ at a chosen initial point) and numerically integrate outward along a certain direction. An obvious way to proceed is the so-called shooting method, which reduces the boundary value problems to initial value problems and one adjusts the initial condition until the desired decaying asymptotics is reached. However, it turns out that, in practice, large $x$ asymptotics is very sensitive to the initial condition at small $x$ thus it is very hard to reach a decent accuracy.

Instead, we employ the inward integration approach where the boundary condition at a certain large value of $x$ is provided by the chosen WKB solutions. Intuitively this works better for us because the unwanted dominant solution is suppressed by the inward integration. We will see more examples of the numerical calculation below.

It's not hard to imagine that dealing with small $\hbar$ is challenging for numerics since it exponentially suppresses the solution. This can be easily resolved by a rescaling of the coordinate. For example, under a change of coordinate $y = \hbar^{-2/5} x$
\begin{equation}
\frac{x^3- a x^2}{\hbar^2} \Leftrightarrow y^3 - a \hbar^{-\frac{2}{5}} y^2
\end{equation}
Therefore, the result only depends on the combination $a\hbar^{-\frac{2}{5}}$. In the numerics we will study the latter and vary $a$. Wronskians and cross ratios will be invariant under the rescaling of the coordinate. If one wants to study the wavefunctions, we can easily restore the $\hbar$ dependence by going back to the original coordinate. 

There are other numerical implementation methods available\footnote{We thank Andy Neitzke for the helpful correspondence.}. See, e.g. \cite{Dumas:2020zoz} for a recent study. 
\subsection{Solutions near zeros and their Wronskians}\label{sec:localWronskian}
In general, given a local form around a zero of generic integer order $k$ where the stress tensor is regular
\begin{equation}
\frac{\tilde{y}^k}{\hbar^2} +  a_{k-2} \tilde{y}^{k-2} +\dots + a_1 \tilde{y} + a_0
\end{equation}
which becomes under $\tilde{y} = \hbar^{\frac{2}{k+2}}y$,
\begin{equation}
T_{k}^{\mathrm{local}}(y) = y^k +a_{k-2} \gamma^{k} y^{k-2}+ \dots +  a_{j} \gamma^{2+j} y^{j} +\dots+ a_0 \gamma^{2}
\end{equation}
where $\gamma = \hbar^{\frac{2}{k+2}}$ and
\begin{equation}
a_{m} = a_{m}^{(0)} + \gamma^{k+2} a_{m}^{(2)} + \gamma^{2(k+2)} a_{m}^{(4)} + \dots
\end{equation}
Let's attempt to solve the ODE perturbatively in $\gamma$. At the leading order, the Schr\"odinger operator is just $\partial_y^2-y^k$. A set of nice solutions has been given in \cite{KondoGLW}, which we now review. We choose the solution that decays along the positive real axis which takes the form 
\begin{equation}
A_{k;0}^{(0)}(y) = \sqrt{\frac{2 y}{\pi(k+2)}} K_{\frac{1}{k+2}}\left(\frac{2}{k+2} y^{1+\frac{k}{2}}\right) \label{eq:localsolAn0}
\end{equation} 
with large $y$ asymptotics 
\begin{equation}
A_{k;0}^{(0)}(y) \sim \frac{1}{\sqrt{2} y^{k/4}} e^{-\frac{2}{k+2} y^{1+\frac{k}{2}}}. \label{eq:localsolAk0asym}
\end{equation} 
We can produce more solutions by a rotation 
\begin{equation}\label{eq:localsolAna}
A_{k;a}^{(0)}(y) = e^{-\frac{\pi i}{k+2} a} A_{k;0}^{(0)}(e^{\frac{2 \pi i}{k+2} a} y).
\end{equation}
It deserves some remarks here. The definition for \eqref{eq:localsolAn0} is obvious because we want $\frac{1}{\sqrt{2\partial_y S(y)}} e^{-S(y)}$ type of asymptotics. Because of the rotational symmetry $y \rightarrow e^{\frac{2\pi i}{k+2}} y$, the definition \eqref{eq:localsolAna} is equivalent to
\begin{equation}
A_{k;a}^{(0)}(y = e^{-\frac{2 \pi i}{k+2} a}R) \equiv e^{- \frac{ \pi i}{k+2} a} A_{k;0}^{(0)}(R), R \in \mathbb{R}_+
\end{equation}
The inclusion of the factor $e^{- \frac{ \pi i}{k+2} a}$ is such that asymptotically along the ray of $e^{- \frac{2 \pi i}{k+2} a}$
\begin{equation}
A_{k;a}^{(0)}(y) \sim \frac{1}{\sqrt{2 e^{\pi i a} y^{k/2}}} e^{-e^{\pi i a}\frac{2}{k+2} y^{1+\frac{k}{2}}}.\label{eq:localsolAkaasym}
\end{equation}
As a result, all neighbouring Wronskians $(A_{k;a}^{(0)},A_{k;a+1}^{(0)}) = -i$. This brings a side effect that $A_{k;a}^{(0)} = -A_{k;a+k}^{(0)}$. Had we defined $A_{k;a}^{(0)}(y)$ in \eqref{eq:localsolAna} without the factor $e^{- \frac{ \pi i}{k+2} a}$, we would have $A_{k;a}^{(0)} = A_{k;a+k}^{(0)}$. Furthermore, thanks to the identity
\begin{equation}
A_{k;a-1}^{(0)}(y) + A_{k;a+1}^{(0)}(y) = \left(e^{\frac{\pi i}{k+2}} + e^{- \frac{\pi i}{k+2}}\right)A_{k;a}^{(0)}(y)\label{eq:connectionformula}
\end{equation} 
we can compute for any $a$ and $b$, $i(A_{k;a}^{(0)},A_{k;b}^{(0)}) = d^{(k)}_{b-a}$,  where
\begin{equation}
d^{(k)}_n = \frac{e^{\frac{\pi i}{k+2}n}-e^{-\frac{\pi i}{k+2}n}}{e^{\frac{\pi i}{k+2}}-e^{-\frac{\pi i}{k+2}}}.
\end{equation}

Setting $k=1$, we are reduced to the Airy functions 
\begin{equation}\label{eq:localsolA1a}
A_{1;a}^{(0)}(y) \equiv \mathrm{Ai}^{(0)}_a(y) \equiv \sqrt{2\pi} e^{-\frac{\pi i a}{3}} \mathrm{Ai}(e^{\frac{2 \pi i a}{3}} y).
\end{equation} 

To obtain higher order corrections, we parameterize the solution as 
\begin{equation}
A(y)  = \sum_{n\geq 0} \gamma^n A^{(n)}(y)
\end{equation}
where for now we suppressed the subscript. The differential equation $\partial_y^2 A(y) = T_{k}^{\mathrm{local}}(y) A(y)$ is expanded order by order in $\gamma$ as
\begin{equation}
\begin{split}
\gamma^0: & \partial^2 A^{(0)}-y^k A^{(0)} = 0, \\
\gamma^1: & \partial^2 A^{(1)}-y^k A^{(1)} = 0, \\
\gamma^2: & \partial^2 A^{(2)}-y^k A^{(2)} = a_0^{(0)} A^{(0)}, \\
\gamma^3: & \partial^2 A^{(3)}-y^k A^{(3)} = a_1^{(0)}y A^{(0)} + a_0^{(0)} A^{(1)}, \\
\gamma^4: & \partial^2 A^{(4)}-y^k A^{(4)} = a_2^{(0)}y^2 A^{(0)}+ a_1^{(0)} y A^{(1)} + a_0^{(0)} A^{(2)}.
\end{split} \label{eq:recursiveODEbygamma}
\end{equation}
We fix the normalization of the solutions by matching with the WKB asymptotics, which is uniquely defined as 
\begin{equation}
\frac{1}{\sqrt{\pm 2\partial_y S}} e^{\mp S}\label{eq:WKBsol}
\end{equation}
where $S = \frac{2}{k+2} y^{\frac{k+2}{2}} + \dots$ is a (fractional) power series of $y$, with no constant term. When the zero of even order, there is also $\log y$, which we choose to be the principal branch. For example,
\begin{equation}
\arraycolsep=1.0pt\def\arraystretch{1.2}
\begin{array}{ll}
T_{k}^{\mathrm{local}}(y)  = y, &S = \frac{2}{3} y^{3/2} +\frac{5}{48} \frac{1}{y^{3/2}} +\frac{1105}{9216} \frac{1}{y^{9/2}} +\dots\\
T_{k}^{\mathrm{local}}(y)  = y^2 +a\gamma^2, &S = \frac{1}{2} y^2 + \frac{a\gamma^2}{2}\log y + \frac{3+a^2\gamma^4}{16} \frac{1}{y^2} +\dots\\
T_{k}^{\mathrm{local}}(y)  = y^3 + b\gamma^3 y + a\gamma^2, &S = \frac{2}{5}y^{5/2} + b\gamma^3y^{1/2} -a\gamma^2 \frac{1}{y^{1/2}}+\dots\\
T_{k}^{\mathrm{local}}(y) = y^4 + c \gamma^4 y^2 + b \gamma^3 y+ a \gamma^2, \quad &S = \frac{1}{3}y^3 + \frac{c\gamma^4}{2} y + \frac{b\gamma^3}{2} \log y + \frac{c^2 \gamma^8-4a \gamma^2}{8}\frac{1}{y} +\dots
\end{array}
\end{equation}
We can now give the prescription for determining the solutions order by order in $\gamma$. The small $A_{k;a}(y)$, which decays along the ray of $\exp(- \frac{2\pi i}{k+2}a)$, is given by
\begin{align}
A_{k;a}(y;\gamma,\{a_i \}) = A_{k;a}^{(0)}(y) +\gamma A_{k;a}^{(1)}(y) + \gamma^2 A_{k;a}^{(2)}(y) +\dots
\end{align}
$A_{k;a}$ can be obtained recursively in the following way. We already defined the leading order $A_{k;a}^{(0)}$ above. Obviously WKB solutions \eqref{eq:WKBsol} expand in power of $\gamma$ starting from $\gamma^2$, so we need to choose $A_{k;a}^{(1)} = 0$. And for each $A_{k;a}^{(n)}$ with $n\geq 2$, we need to solve an inhomogeneous ODE with two integration constants to fix, one of which is fixed by requiring the solution to decay along the chosen direction and the other one is fixed to match the normalization of the WKB solution \eqref{eq:WKBsol}. In practice, as seen in the examples below, this is achieved by choosing the lower limit of the integration to be at infinity.

Given two small solutions $A_{k;a} = A_{k;a}^{(0)} + \gamma^2 A_{k;a}^{(2)} + \gamma^3 A_{k;a}^{(3)}+\dots$ and $A_{k;b} = A_{k;b}^{(0)} + \gamma^2 A_{k;b}^{(2)} + \gamma^3 A_{k;b}^{(3)}+\dots$, their Wronskian reads
\begin{align}
(A_{k;a},A_{k;b}) &= (A_{k;a}^{(0)},A_{k;b}^{(0)}) + \gamma^2 \Big[(A_{k;a}^{(2)},A_{k;b}^{(0)})+(A_{k;a}^{(0)},A_{k;b}^{(2)}) \Big] +\gamma^3 \Big[(A_{k;a}^{(3)},A_{k;b}^{(0)})+(A_{k;a}^{(0)},A_{k;b}^{(3)}) \Big] \notag\\
&\qquad\qquad +\gamma^4 \Big[(A_{k;a}^{(4)},A_{k;b}^{(0)})+(A_{k;a}^{(0)},A_{k;b}^{(4)})  + (A_{k;a}^{(2)},A_{k;b}^{(2)})\Big] +\dots
\end{align} 
In particular, we can verify that $(A_{k;a},A_{k;a+1}) = (A_{k;a}^{(0)},A_{k;a+1}^{(0)}) =-i$ with no higher order corrections. More interesting ones are the Wronskians between non-consecutive small solutions. We will give concrete examples below for the double zero and cubic zero, since it is trivial for simple zero.

If, on the other hand, we are interested in the local coordinate system around a zero that is a singularity of trivial monodromy, we would have
\begin{equation}
\frac{\tilde{y}^k}{\hbar^2} +  a_{k-2} \tilde{y}^{k-2} +\dots + a_1 \tilde{y} + a_0 + \frac{l(l+1)}{\tilde{y}^2}
\end{equation}
or 
\begin{equation}
y^k + a_{k-2} \hbar^{\frac{2k}{k+2}} y^{k-2}+ \dots +  a_{j} \hbar^{\frac{2j+4}{k+2}} y^{j} +\dots+ a_0 \hbar^{\frac{4}{k+2}} +\frac{l(l+1)}{y^2}
\end{equation}
where $t(x) = a(x)^2 + \partial_x a(x)$, and $a(x)$ has residue $-l$ at $x_0$. The perturbative solutions can be found in a similar procedure as above except that we need to start with different solutions at the leading order. Since the Schr\"odinger operator at the leading order is $\partial^2- y^k - \frac{l(l+1)}{y^2}$, the solutions that agree with the asymptotics \eqref{eq:localsolAk0asym} and \eqref{eq:localsolAkaasym} are given by
\begin{align}
A_{k,l;0}^{(0)}(y) &= \sqrt{\frac{2 y}{\pi(k+2)}} K_{\frac{1+2l}{k+2}}\left(\frac{2}{k+2} y^{1+\frac{k}{2}}\right)\\
A_{k,l;a}^{(0)}(y) &= e^{-\frac{\pi i}{k+2} a} A_{k,l;0}^{(0)}(e^{\frac{2 \pi i}{k+2} a} y) 
\end{align}
whose Wronskians are given by
\begin{equation}
i(A_{k,l;a}^{(0)},A_{k,l;b}^{(0)}) = \frac{\sin{\frac{\pi}{k+2}(2l+1)(b-a)}}{\sin{\frac{\pi }{k+2}(2l+1)}}.
\end{equation}
One sanity check is to look at $i(A_{k;a}^{(0)},A_{k;a+k+2}^{(0)})$. One would like this to be zero since there is a unique decaying solution along a certain ray and therefore two must be proportional to each other. This is indeed mostly true since $2l+1\in \mathbb{Z}$. It fails when $2l+1$ is an integer multiple of $k+2$, where we have 
\begin{equation}
i(A_{k;a}^{(0)},A_{k;a+k+2}^{(0)}) = (-1)^{(k+1)\frac{2l+1}{k+2}} (k+2).
\end{equation}
This is another manifestation of the requirement that $2l \leq k$.

\subsubsection{Example: double zero}\label{app:sec:localcoorddoublezero}

Consider the ODE
\begin{equation}
\partial_{\tilde{y}}^2\tilde{A}(\tilde{y}) = (\frac{\tilde{y}^2}{\hbar^2}+a)\tilde{A}(\tilde{y}) \label{eq:ODE_local_ytilde}
\end{equation}
where $a = a^{(0)} + \hbar^2 a^{(2)} +\hbar^4 a^{(4)} +\dots$. To find perturbative solution, we rewrite using $\gamma = \hbar^{1/2}$ and $\tilde{y} = \gamma y$, and we have
\begin{equation}
\partial_y^2 A(y) =  \left(y^2 + \gamma^2 a\right) A(y).
\end{equation}
It is not hard to see that all the equations at the odd power of $\gamma$ are homogeneous and WKB solutions only involve even powers of $\gamma$, so solutions $A = \sum \gamma^j A^{(j)}$ only involve even powers of $\gamma$, namely $A = \sum \gamma^{2j} A^{(j)}  = \sum \hbar^{j} A^{(j)}$, where the leading order is given by
\begin{align}
A_{2;n}^{(0)} & = \sqrt{\frac{y}{2\pi}} K_{\frac{1}{4}}\left(\frac{1}{2}y^2 e^{\pi i n} \right), \label{eq:localsolA2n}
\end{align}
One of the immediate consequences is that the Wronskians $(A_{n},A_m) = (A_{n}^{(0)},A_{m}^{(0)})+\dots$ are corrected by integer powers of $\hbar$.

On the other hand, the ODE\footnote{The reason we solve \eqref{eq:ODE_local_ytilde} instead of the one in $y$ coordinate is because we have $\hbar^2$ in the former, so that we can apply the trick $\hbar \rightarrow \hbar e^{-i \pi n}$ to find other solutions. Due to the same reason, we don't shift $\hbar \rightarrow \hbar e^{-i \pi n}$ in the Jacobian part $\hbar^{-1/4}$ if one wants to go back to $y$ coordinate.} \eqref{eq:ODE_local_ytilde} can be solved exactly by using 
\begin{align}
\tilde{A}_0(\tilde{y},\hbar) &=\left(\frac{\hbar}{2}\right)^{\frac{1-a\hbar}{4}} D_{-\frac{1+a\hbar}{2}}\bigg(\frac{\sqrt{2}\tilde{y}}{\sqrt{\hbar}} \bigg)\\
\tilde{A}_n(\tilde{y},\hbar) & =  \tilde{\psi}_0(\tilde{y},\hbar e^{- i \pi n}).
\end{align}
Their Wronskians can be evaluated easily. For example, $i(\tilde{A}_n,\tilde{A}_{n+1}) = 1$,
\begin{equation} \label{Am1 A2}
i(\tilde{A}_{-1},\tilde{A}_{1}) = \frac{\sqrt{2\pi}}{\Gamma(\frac12-\frac{a\hbar}{2})} \left(\frac{\hbar}{2}\right)^{\frac{a\hbar}{2}}, \quad i(\tilde{A}_{-1},\tilde{A}_2) = e^{i a \pi \hbar}.
\end{equation}
The cross ratio 
\begin{equation}
\chi \equiv \frac{(\tilde{A}_0,\tilde{A}_1)(\tilde{A}_{-1},\tilde{A}_2)}{(\tilde{A}_{-1},\tilde{A}_{0})(\tilde{A}_{1},\tilde{A}_{2})} = e^{i \pi a \hbar}
\end{equation}
which exactly coincides with the contour integral $\exp \oint p(x,\hbar)dx $ around this double zero of the WKB momentum which has a pole
\begin{equation}
p(x,\hbar) = \frac{x}{\hbar}+\frac{a \hbar}{2 x}-\frac{3 \hbar+a^{2} \hbar^{3}}{8 x^{3}}+\frac{19 a \hbar^{3}+a^{3} \hbar^{5}}{16 x^{5}} +\dots
\end{equation}

\subsubsection{Example: cubic zero}\label{app:sec:localcoordcubiczero}
\begin{figure}
	\centering
	\includegraphics[width=0.73\linewidth]{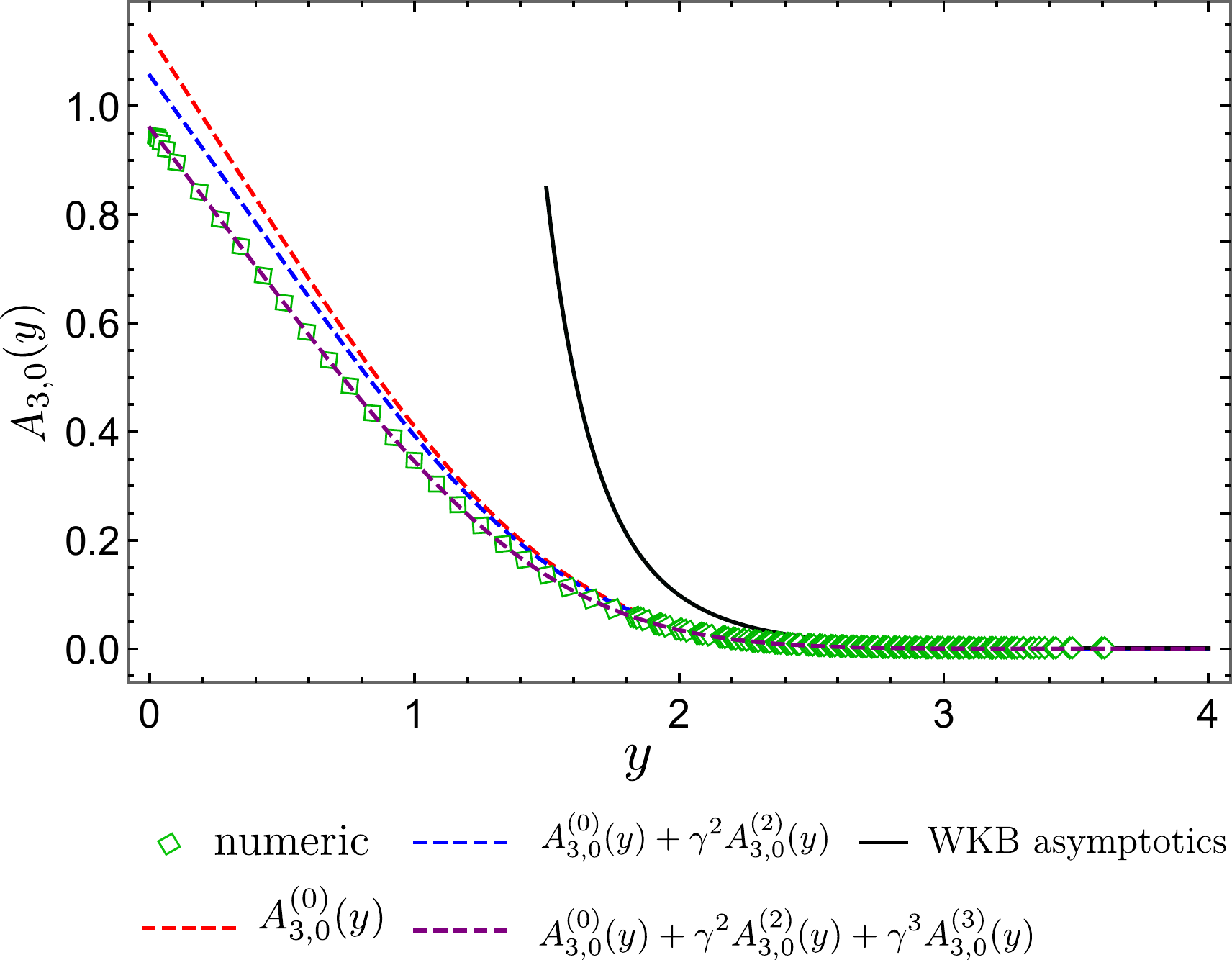}
	\includegraphics[width=0.73\linewidth]{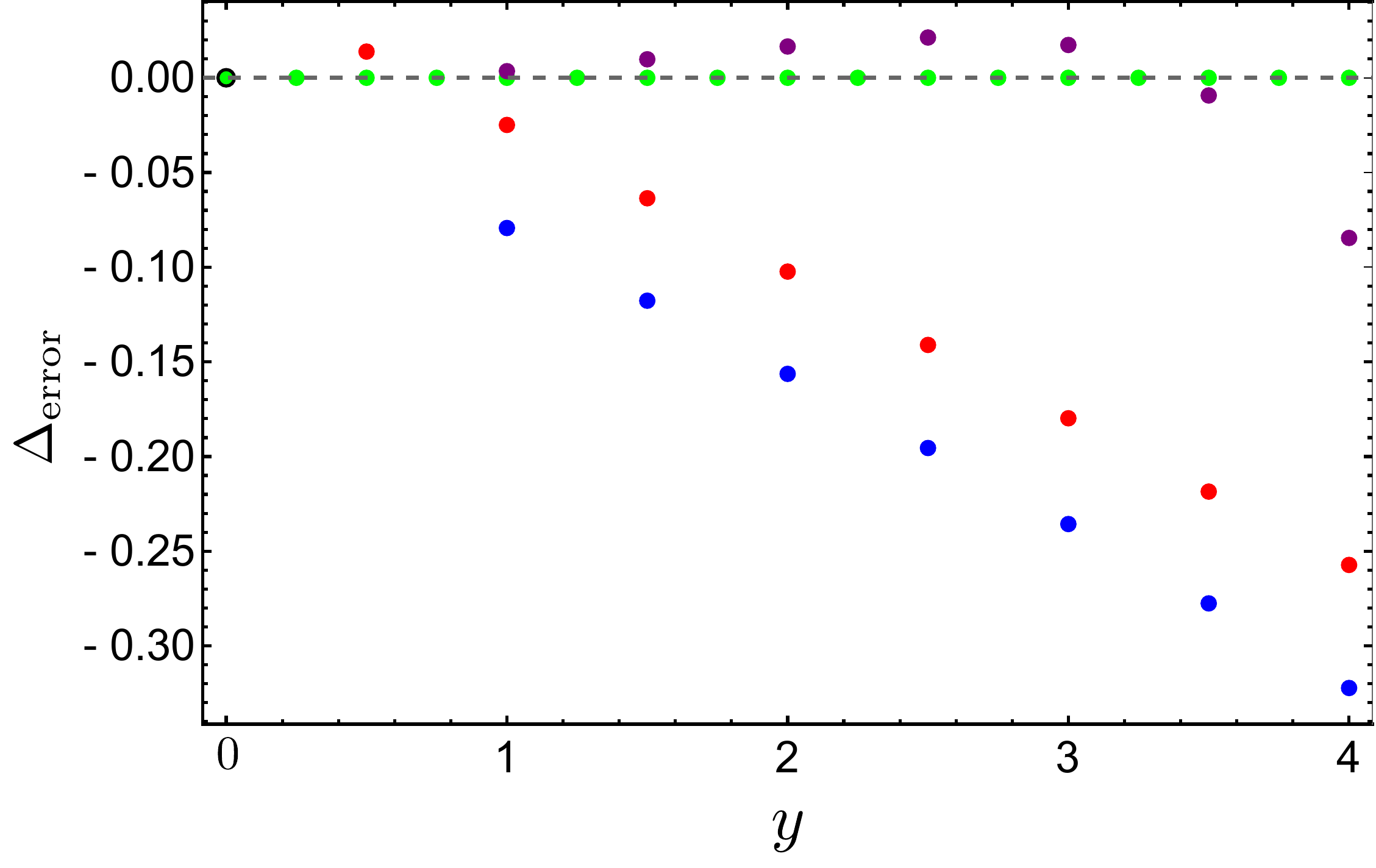}
	\caption{Numerical and analytic evaluation in the local coordinate system around a cubic zero defined at the beginning of Subsection \ref{app:sec:localcoordcubiczero}. Parameters used are chosen in a rather generic way: $\hbar = \frac{1}{5}$, $a = -\frac{4}{21}$, $b = \frac{1}{2}$. (Top) Various approximate evaluations of $A_{3,0}(y)$. Approximation gets better with higher corrections included. (Bottom) $\Delta_{\mathrm{error}} \equiv \frac{\partial_x^2\psi(x)}{\psi(x)}-\frac{1}{\hbar^2}P(x)$. We don't show the $\Delta_{\mathrm{error}}$ for the WKB asymptotic solution since the error is too big. The legend of coloring is shared in both diagrams.}
	\label{fig:cubicsolcompare}
\end{figure}
In the local coordinate system around a cubic zero, $T(\tilde{y}) = \frac{\tilde{y}^3}{\hbar^2} + b \tilde{y} +a$, where 
\begin{align}
b = b^{(0)} + \hbar^2 b^{(2)} +\hbar^4 b^{(4)} +\dots\\
a = a^{(0)} + \hbar^2 a^{(2)} +\hbar^4 a^{(4)} +\dots
\end{align}
or equivalently the stress tensor is $y^3 + b \gamma^3 y + a \gamma^2$ with $\gamma = \hbar^{2/5}$. The leading order solutions are defined in \eqref{eq:localsolAn0} and \eqref{eq:localsolAna}. Since here in this section we only flesh out details for three solutions, for convenience we write $\phi_0 = A_{3;0}^{(0)}$, $\phi_1 = A_{3;1}^{(0)}$, $\phi_{-1} = A_{3;-1}^{(0)}$.

Let's now solve the ODE perturbatively using the prescription described in \ref{sec:localWronskian}. Let's denote the small solution along the ray $e^{-i 2\pi n/5}$ as
\begin{equation}
A_{k;n}(y) = \sum_{n\geq 0}\gamma^j A_{k;n}^{(j)}(y).
\end{equation}
Wronskians between these functions are easily evaluated
\begin{align}
(A_{3,0},A_{3,1}) &= 1 \\
(A_{3,-1},A_{3,1}) &= (A_{3,-1}^{(0)},A_{3,1}^{(0)}) + \gamma^2 a^{(0)} \big[(A_{3,-1}^{(0)},A_{3,1}^{(2)})+(A_{3,-1}^{(2)},A_{3,1}^{(0)}) \big]+ O(\gamma^3)\dots
\end{align}

\subsection{Examples: polynomial oper}\label{app:sec:polynomial}
\subsubsection{Ex: $P(x) = x^2-2a$}\label{sec:Exx22a}
The stress tensor is $T(x) = \frac{x^2-2a}{\hbar^2} $. There are four small solutions, given by the parabolic functions
\begin{equation}
\psi_0(x) = \left( \frac{\hbar}{2}\right)^{\frac{1}{4}+\frac{a}{2\hbar}} D_{-\frac12+\frac{a}{\hbar}}\left(\frac{\sqrt{2}x}{\sqrt{\hbar}} \right), \quad \psi_n(x,\theta) = \psi_0(x,\theta + i \pi n) \label{eq:psix2-2a}
\end{equation}
with Wronskians $i(\psi_n,\psi_{n+1}) = 1$ and
\begin{equation}
i(\psi_{-1},\psi_{1}) = \frac{\left(\frac{\hbar}{2}\right)^{-a/\hbar}\sqrt{2\pi}}{\Gamma\left( \frac{1}{2} + \frac{a}{\hbar}\right)}, \quad i(\psi_{-1},\psi_2) = e^{-2i \pi \frac{a}{\hbar}}.
\end{equation}
The normalization in \eqref{eq:psix2-2a} is such that in its large $x$ asymptotic expansion $\frac{1}{\sqrt{\pm 2\partial_x S(x)}} e^{\mp S(x)}$, there is no constant term in the primitive $S(x)$.

In $\hbar$ asymptotics, $i(\psi_{-1},\psi_{1}) = (2e)^{1/\hbar}\left(1+ \frac{1}{24}\hbar + \frac{1}{1152} \hbar^2 - \frac{1003}{414720} \hbar^3 + O(\hbar^4)\right)$. Let us try to reproduce this in two other ways: contour integral of WKB momentum and using local coordinate systems. 

The WKB momentum is $p(x,\hbar) = \frac{\sqrt{x^2-2}}{\hbar} + p_1(x)\hbar + p_3(x)\hbar^3 + \dots$. Since
\begin{equation}
\int^L \sqrt{x'^2-2}dx' \overset{L\rightarrow \infty}{\longrightarrow} \frac{L^2}{2}+ const -\log L + O\Big(\frac{1}{L^2}\Big)
\end{equation}
we can regularize the integral at infinity by defining
\begin{equation}
\int^x_{\infty} \equiv  \lim\limits_{L \rightarrow \infty}\int^x_{L} + \left(\frac{L^2}{2} + A -\log L \right).
\end{equation}
We will choose the constant $A = 0$ in this article to normalize the WKB solutions. With this normalization, the large $x$ asymptotics will take the form of only power of $x$ without any constant\footnote{There are some other natural choices as well. For example, we chose $A = -\log \sqrt{2e}$ in \cite{KondoGLW} such that the regularized integral $\int^x_\infty \sqrt{x'^2-2}dx'$ coincides with $\int_{\sqrt{2}}^x \sqrt{x'^2-2}dx'$. With this, $\int_{-i \infty}^{i \infty}  \sqrt{x'^2-2}dx' = 0$, and consequently we don't have the prefactor $(2e)^{1/\hbar}$ in $i(\psi_{-1},\psi_{1})$}
\begin{equation}
\int_{\infty}^x  \sqrt{x'^2-2}dx' \sim \frac{x^2}{2} -\log x +\frac{1}{4}x^2 + \dots
\end{equation}
Therefore, the leading term of the integral of the WKB momentum is
\begin{equation}
\int_{-i \infty}^{i \infty}  \sqrt{x'^2-2}dx' =\log 2e, \quad \int_{-i \infty}^{i \infty} p_1(x) dx = -\frac{1}{24}, \quad \int_{-i \infty}^{i \infty} p_3(x) dx = \frac{7}{2880}.
\end{equation}
Therefore we get 
\begin{equation}
i(\psi_{-1},\psi_{1}) = e^{-\int_{-i \infty}^{i \infty}p(x,\hbar)} = (2e)^{1/\hbar}\left(1+ \frac{1}{24}\hbar + \frac{1}{1152} \hbar^2 - \frac{1003}{414720} \hbar ^3+ O(\hbar^4)\right).
\end{equation}

We can also find this result via going to the local coordinate system. Essentially one needs to figure out the constant of proportionality in $\left(\partial_x y(x,\hbar)\right)^{-1/2}A(y(x)) \propto \psi(x)$. To this end, we look at the large $x$ asymptotics of both sides. By definition, the large $x$ asymptotics of $\psi(x)$ is given by $\frac{1}{\sqrt{\pm 2\partial_x S(x)}} e^{\mp S(x)}$, where
\begin{align}
S(x) &= \frac{x^2}{2\hbar} - \frac{a}{\hbar}\log x +\left(\frac{a^2}{4\hbar} + \frac{3\hbar}{16} \right) \frac{1}{x^2} + \left(\frac{a^3}{8\hbar} + \frac{19a\hbar}{32} \right)\frac{1}{x^4} \notag\\
&\qquad\qquad\qquad\qquad\qquad + \left(\frac{5a^4}{48\hbar} +\frac{145a^2\hbar}{96} + \frac{99\hbar^3}{256} \right)\frac{1}{x^6}+\dots \label{eq:x2-2aS(x)}
\end{align}
which has no constant term according to our definition. On the other hand, the large $x$ asymptotics of $\left(\partial_x y(x,\hbar)\right)^{-1/2}A(y(x))$ exactly matches with \eqref{eq:x2-2aS(x)}, modulo a possible constant term in $S(x)$. With $y(x) = \hbar^{-2/3}(y_0+\hbar^2 y_2 + \hbar^4 y_4 +\dots)$ and \eqref{eq:localS(y)1st}, we have
\begin{align}
\tilde{S}(y) &= \frac{2y_0(x)^{3/2}}{3\hbar} + \left(\frac{5}{48y_0(x)^{3/2}} + y_0(x)^{1/2}y_2(x) \right)\hbar +\dots\\
& = \left(\frac{x^2}{2}-\log \sqrt{2e}-\log x+\frac{1}{4}\frac{1}{x^2} +\dots \right)\frac{1}{\hbar}  + \left(-\frac{1}{48} + \frac{3}{16x^2} +\dots \right) \hbar.
\end{align}
Therefore $S(x) = \tilde{S}(y(x)) + \frac12\left(\frac{1}{\hbar}\log 2e +\frac{1}{24}\hbar +\dots \right)$, and 
\begin{equation}
\psi_1(x) =  e^{\frac{1}{2}\left(\frac{1}{\hbar}\log 2e +\frac{1}{24}\hbar +\dots  \right)}\left(\partial_x y(x,\hbar)\right)^{-1/2}A_1(y(x))
\end{equation}
hence
\begin{equation}
i(\psi_{-1},\psi_{1}) = e^{ \frac{1}{\hbar}\log 2e +\frac{1}{24}\hbar +\dots  } (A_{-1}(y),A_{1}(y)) =e^{ \frac{1}{\hbar}\log 2e +\frac{1}{24}\hbar +\dots  }.
\end{equation}
There is one cross ratio defined by
\begin{equation}
\chi = \frac{(\psi_{0},\psi_{1})(\psi_{-1},\psi_{2})}{(\psi_{-1},\psi_{0})(\psi_{1},\psi_{2})} = e^{-2 \pi i \frac{a}{\hbar}}.
\end{equation}
Furthermore, the exact solutions are solved by parabolic cylinder functions, which we can use to test our numerical calculation.
\begin{figure}
	\centering
	\includegraphics[width=0.48\linewidth]{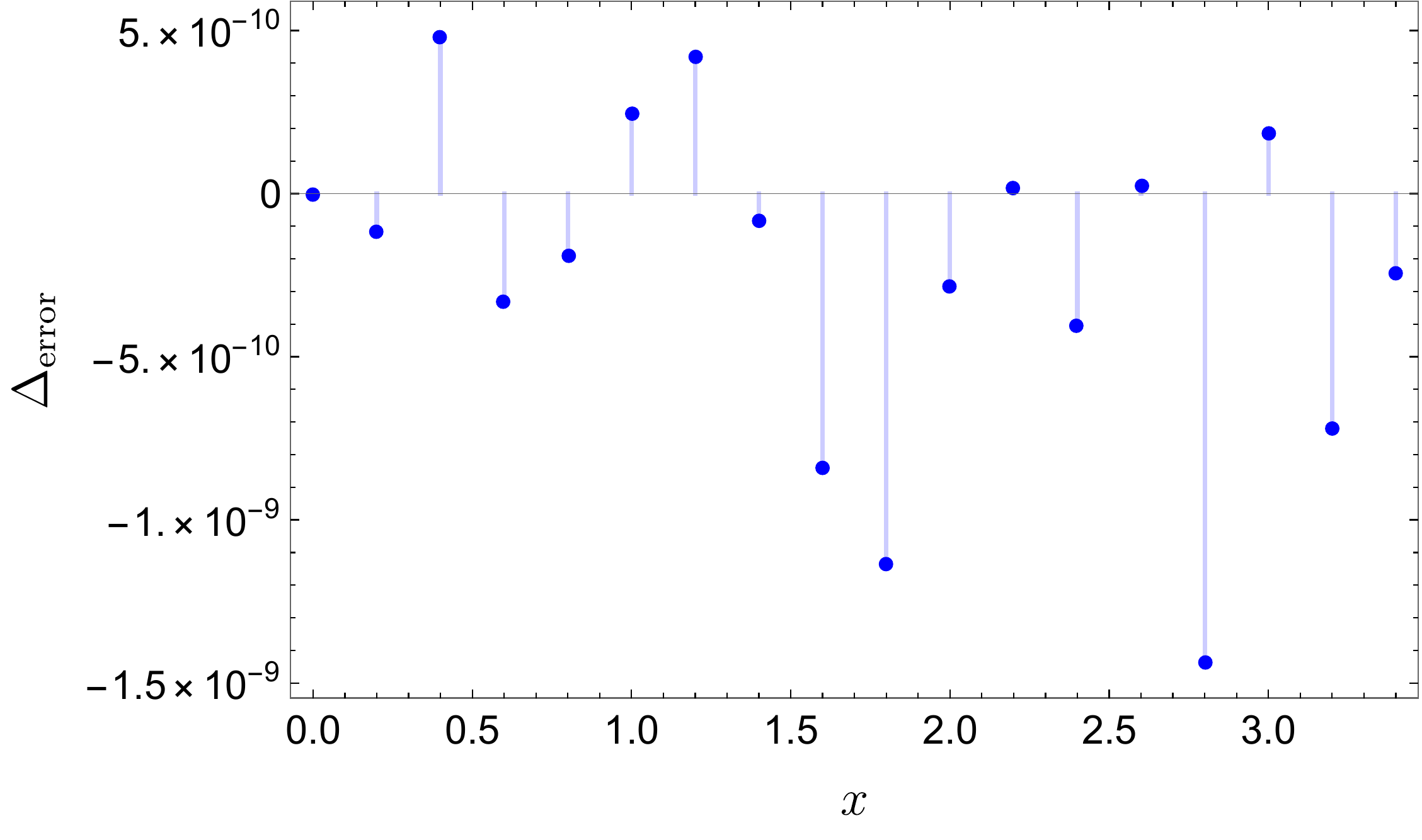}
	\includegraphics[width=0.48\linewidth]{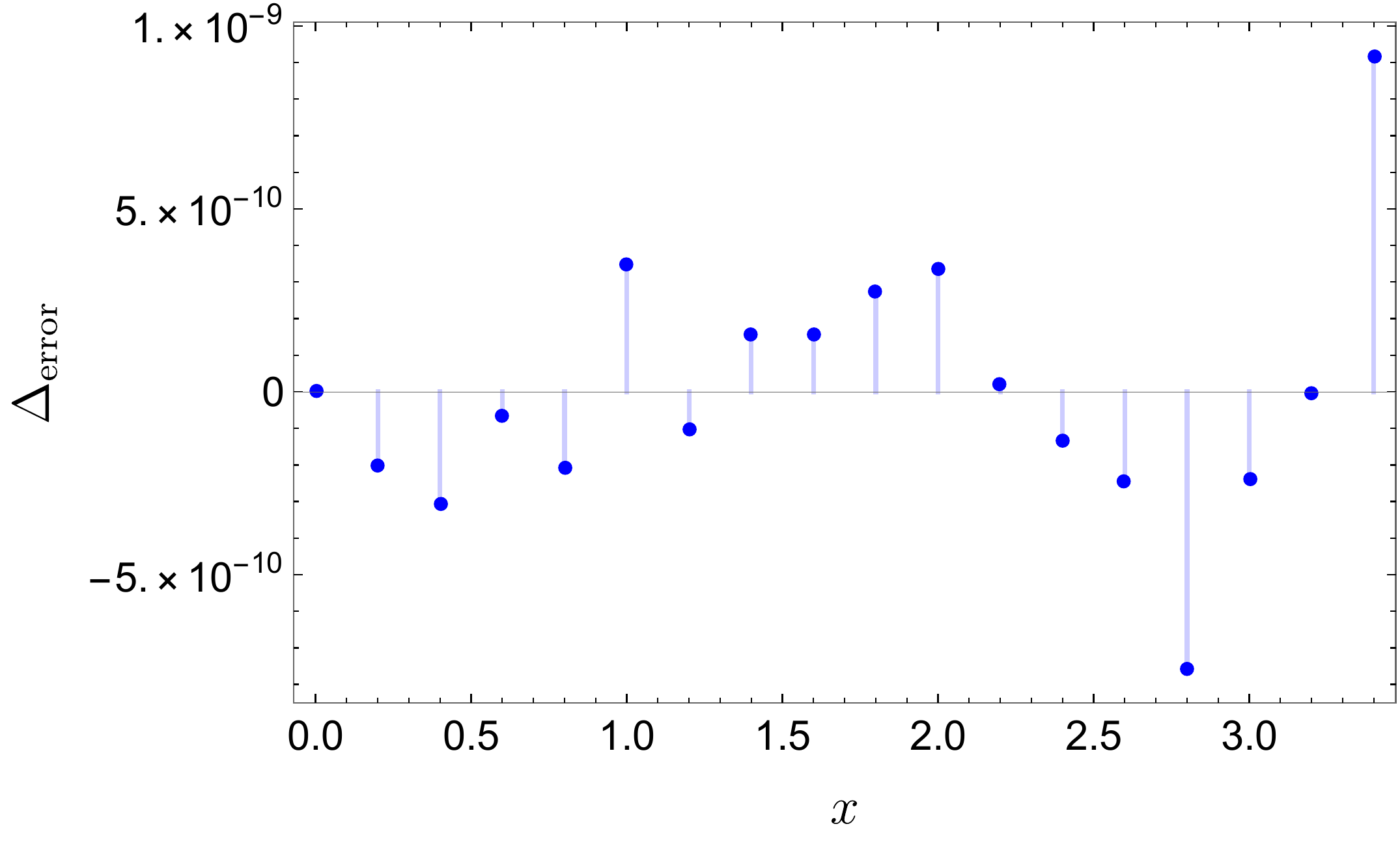}
	\caption{Numerical error $\Delta_{\mathrm{error}} \equiv \frac{\partial_x^2\psi(x)}{\psi(x)}-\frac{1}{\hbar^2}P(x)$. (Left) $P(x) = x^2 -2$ and $\hbar = 1$ (Right) $P(x) = x^3-x^2$ and $\hbar = 1$. This is just to illustrate numerical error is indeed very small.}
	\label{fig:numericerror2ndorder}
\end{figure}

\subsubsection{Ex: $P(x) = x^3-x^2$}
\begin{figure}
	\centering
	\includegraphics[width=0.4\linewidth]{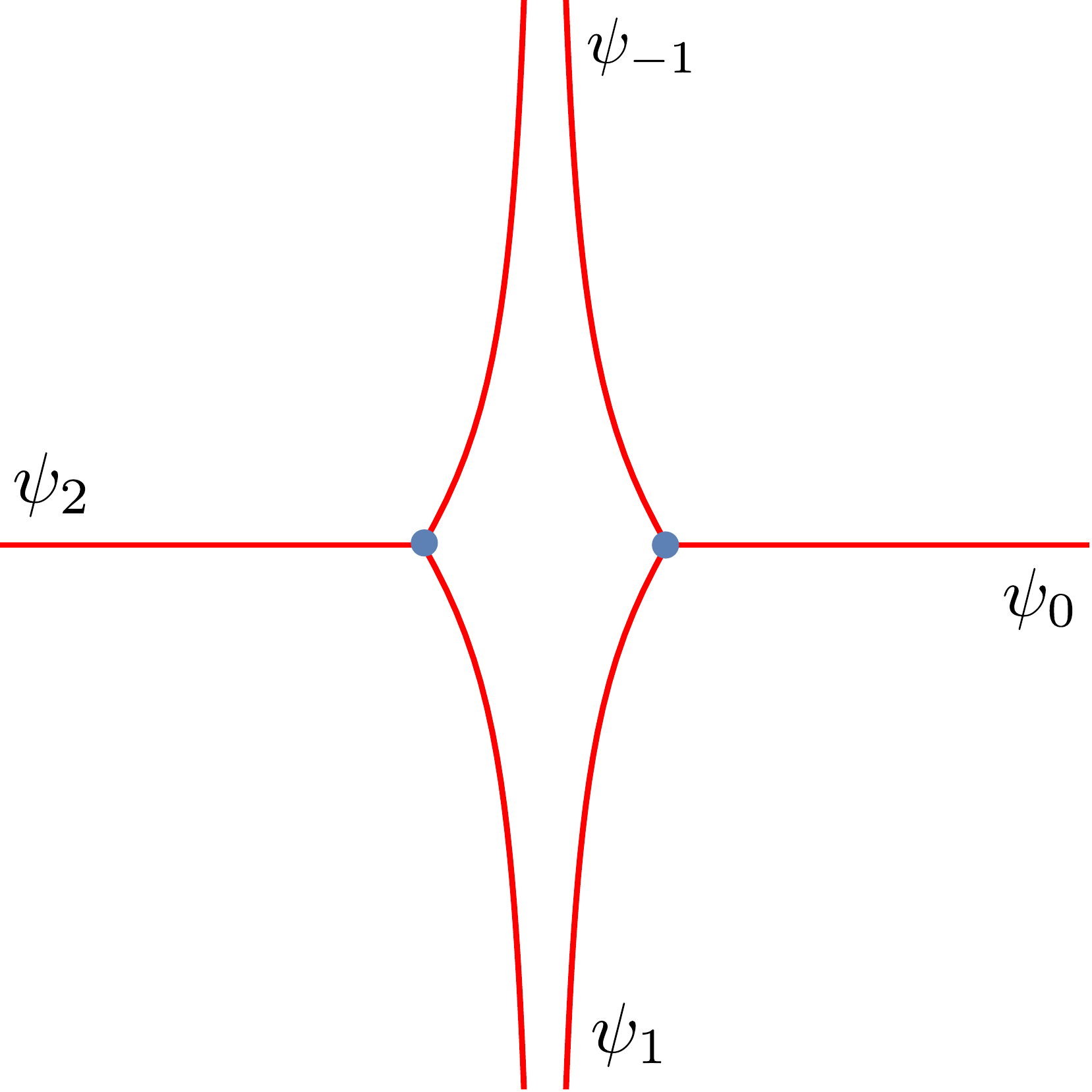}
	\includegraphics[width=0.4\linewidth]{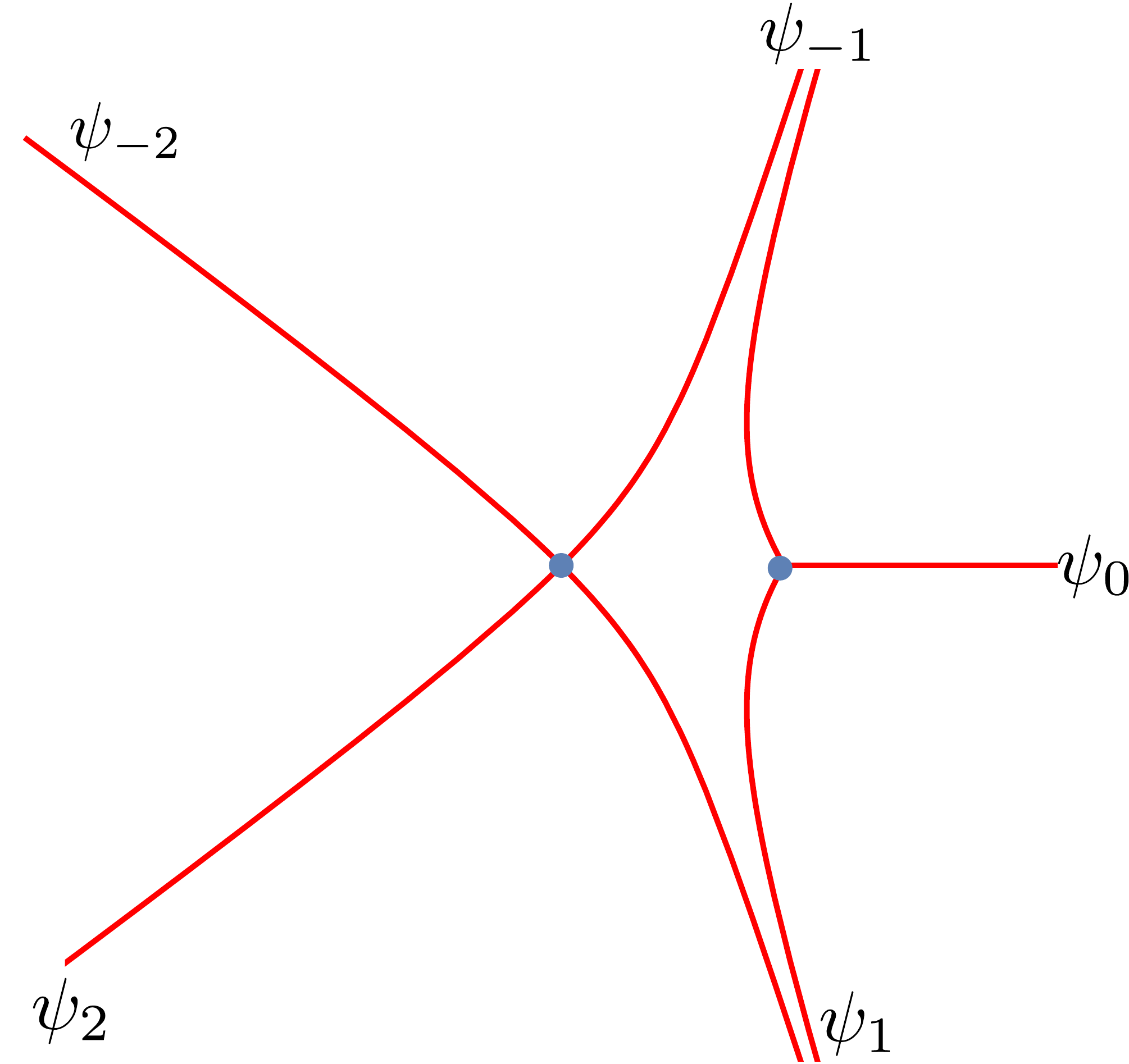}
	\caption{Stokes diagram for (Left) $P(x) = x^2-2a$, (Right) $P(x) = x^3-x^2$.}
	\label{fig:wkbx3mx2}
\end{figure}
There is one simple zero, one double zero and five asymptotic directions. The Stokes diagram is shown in Fig. \ref{fig:wkbx3mx2}.

In the local coordinate system around the double zero,
\begin{equation}
\tilde{T}(y,\hbar) = y^2 -2a(\hbar), \quad a(\hbar) = \frac{7 i}{64}\hbar -\frac{119119 i \hbar ^3}{131072}+  \frac{10775385621 i \hbar ^5}{268435456} + \dots
\end{equation}
Recall that 
\begin{equation}
y_2(x)  = \frac{1}{y_0(x)} \int_{x_0}^x y_0(x') \frac{8 a^{(0)}_0 (y_0')^4 -3 (y_0'')^2 + 2y_0'y_0'''}{8 y_0^2(y_0')^3} dx' 
\end{equation}
the integrand around $x = 0$ is
\begin{equation}
\frac{a^{(0)}-\frac{7i}{64}}{x} + \frac{-1687i-960 a^{(0)}}{5760} +\dots
\end{equation}
therefore we have to choose $a^{(0)} = \frac{7i}{64}$. And since typically $y_0(x) \sim \alpha x +\dots$, $y_2(x)$ will be nonzero at the zero $x=0$.

With a suitably chosen branch cut, namely from $1$ to positive infinity along the real axis, the large $x$ asymptotics is given by
\begin{equation}
\int_{1}^{x}\ i x'\sqrt{1-x'} dx' \sim c_1 x^{5/2} + c_2 x^{3/2} + c_3 x^{1/2} +0 + c_4 x^{-1/2} +\dots
\end{equation}
and $\int_{1}^{0}\ i x'\sqrt{1-x'}dx' = -\frac{4i}{15}$. Recall that we choose to regularize the integral at infinity by removing the powers of divergence without adding any constant term. As a result, the regularized contour integral of the WKB momentum
\begin{equation}
\int_{e^{-\frac{2\pi i}{5}}\infty}^{e^{\frac{2\pi i}{5}}\infty}\ i x'\sqrt{1-x'} dx' = 0.
\end{equation}
So the leading order of $i(\psi_{-1},\psi_1)$ is $1$.

Let's try to reproduce $i(\psi_{-1},\psi_1)$ using two local coordinate systems separately. We use $y_{\pm}(x)$ to denote the local coordinate system around the simple/double zero, respectively. At the leading order
\begin{align}
S_+(y_+(x)) = \frac{2y_{+,0}(x)^{3/2}}{3\hbar} + \dots = \frac{1}{\hbar} \int_1^x\ ix'\sqrt{1-x'} dx' +\dots \\
S_-(y_-(x)) = \frac{y_{-,0}(x)^{2}}{2\hbar} + \dots = \frac{1}{\hbar} \int_0^x\ ix'\sqrt{1-x'} dx' +\dots
\end{align}
One might find this puzzling: since the constant term in the large $x$ asymptotics of $S_+$ is zero but is nonzero in $S_-$. This means that 
\begin{align}
\left(\partial_x y_+(x)\right)^{-1/2} A_+(y(x)) = \psi(x) = e^{\pm \frac{4i}{15\hbar} + \dots} \left(\partial_x y_-(x)\right)^{-1/2} A_-(y(x)).
\end{align}
But this is not problematic because the nontrivial factor actually cancels in the Wronskian. Therefore, they give the same $i(\psi_{-1},\psi_1) = 1 + \dots$, as expected. 
This Wronskian actually has nontrivial $\hbar$ corrections, which we will present below.

First, of course, we can integrate the WKB momentum along the generic WKB line 
\begin{equation}
i(\psi_{-1},\psi_1)= \exp \int_{e^{-\frac{2\pi i}{5}}\infty}^{e^{\frac{2\pi i}{5}}\infty} p(x,\hbar)dx = \exp\left(0 + \frac{7\pi}{32}\hbar - \frac{119119 \pi}{65536}\hbar^3\dots \right) 
\end{equation}
\begin{equation}
\tilde{S}(y) = \frac{2y_0(x)^{3/2}}{3\hbar} + \left(\frac{5}{48y_0(x)^{3/2}} + y_0(x)^{1/2}y_2(x) \right)\hbar +\dots
\end{equation}
One can obtain the same result in local coordinate system around the simple zero and the double zero.

As an example of Wronskians that cannot be evaluated by the contour integral of the WKB momentum, let's try to calculate $i(\psi_{-1},\psi_2)$. Again we go to the local coordiante system around the double zero $x_- = 0$ and we have 
\begin{align}
\left(\frac{\partial y_-(x)}{\partial x}\right)^{-1/2} A_{-;a}(y_-(x)) \propto \psi_{-1}(x),\\
\left(\frac{\partial y_-(x)}{\partial x}\right)^{-1/2} A_{-;a+3}(y_-(x)) \propto \psi_{2}(x).
\end{align}
To find the proportionality constant, we can just look at the large $x$ asymptotics along the corresponding direction and compare both side term by term. The leading term is $\exp\left(-\frac{4i}{15\hbar}\right)$. The subleading terms can be found numerically.
\begin{figure}[ht]
	\centering
	\includegraphics[width=0.7\linewidth]{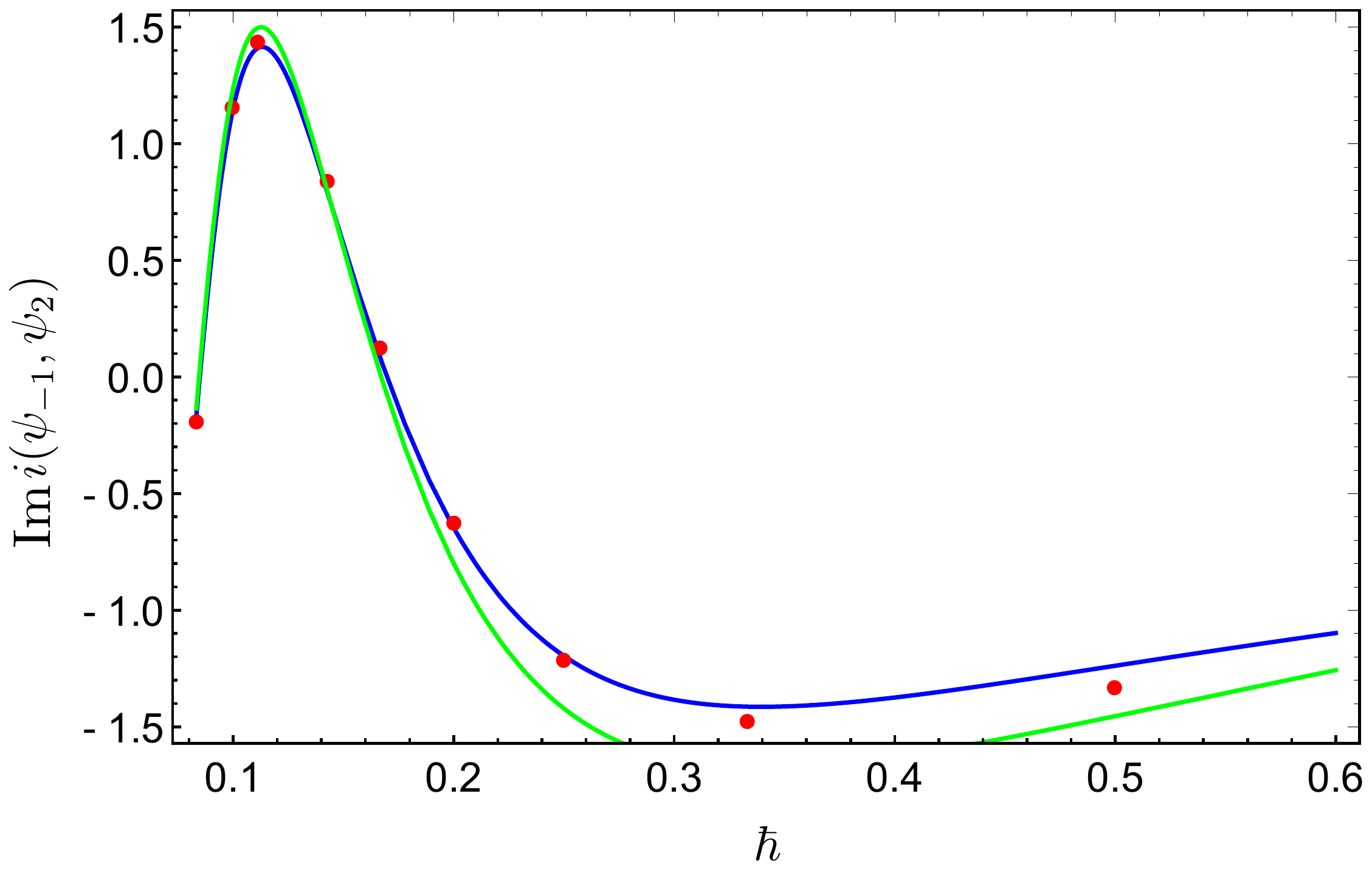}
	\caption{Evaluations of the Wronskian $i(\psi_{-1},\psi_2)$. The red dots are the numerical result. The blue line and the red line are the analytic prediction from the local coordinate system around the double zero up to $\hbar^{-1}$ and $\hbar$ order respectively given in \eqref{app:eq:chi2}. }
	\label{fig:wkbx3mx2wrm12}
\end{figure}

We are now ready to calculate the cross ratios. There are two independent cross ratios defined as follows
\begin{equation}
\chi_{1} \equiv \frac{\left(\psi_{-1}, \psi_{-2}\right)\left(\psi_{1}, \psi_{2}\right)}{\left(\psi_{-1}, \psi_{1}\right)\left(\psi_{2}, \psi_{-2}\right)}, \quad \chi_{2} \equiv \frac{\left(\psi_{0}, \psi_{1}\right)\left(\psi_{-1}, \psi_{2}\right)}{\left(\psi_{0}, \psi_{-1}\right)\left(\psi_{1}, \psi_{2}\right)}.
\end{equation}
$\chi_1$ is easily evaluated in the local coordinate system around the double zero. It coincides with the contour integral of the WKB momentum along a small circle around the double zero. From the end of the Subsection \ref{app:sec:localcoorddoublezero}, this evaluates to be 
\begin{equation}
\chi_1 = e^{2\pi i a(\hbar)}.
\end{equation}
The second cross ratio boils down to $-i (\psi_{-1},\psi_2)$, namely
\begin{equation}
-\chi_2 = i (\psi_{-1},\psi_2) = \sqrt{2} \exp\Big[-\frac{8 i}{15\hbar} - \hbar (\lim_{R \rightarrow +\infty}S_{\hbar}(e^{\frac{2\pi i}{5}} R) + S_{\hbar}(e^{-\frac{4\pi i}{5}} R)) \Big]\label{app:eq:chi2}
\end{equation}
where
\begin{align}
S_{\hbar}(x) = \frac{3}{16}\frac{1}{y_0^2} + y_0y_2 - a^{(0)} \log y_0  
\end{align}
and we also parametrize the coordinate transformation as usual
\begin{equation}
y(x) = y_0(x) + \hbar^2 y_2(x) + \dots
\end{equation}
This agrees quite well with the numerical evaluation shown in Fig. \ref{fig:wkbx3mx2wrm12}.

\subsection{Example: chiral WZW} \label{app:sec:WZW}
\subsubsection{trivial theory}
\begin{figure}[ht]
	\centering
	\includegraphics[width=0.7\linewidth]{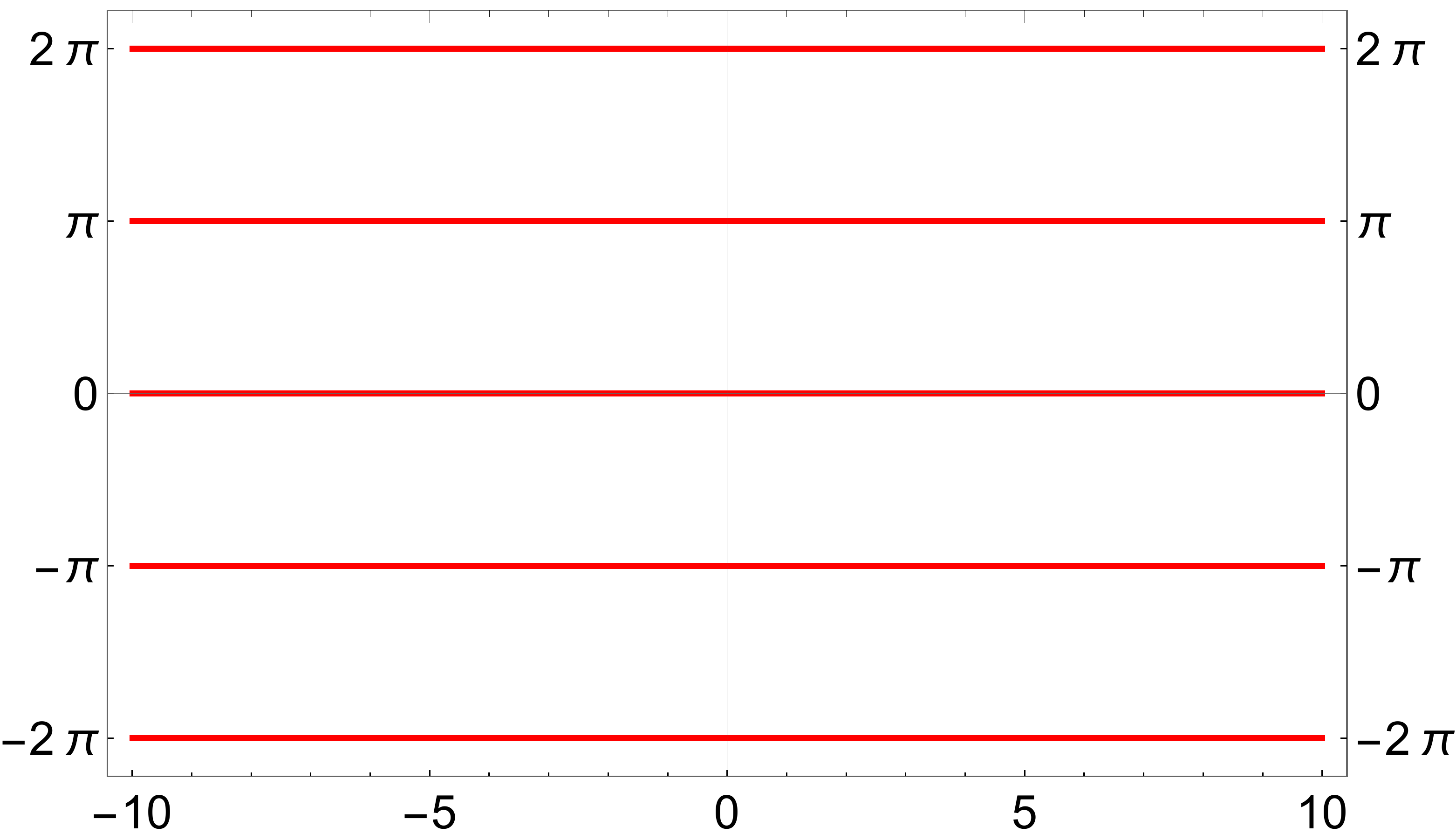}
	\caption{Stokes diagram for $P(x) = e^{2x}$, which corresponds to $\mathrm{SU}(2)_0$ trivial theory. There are infinite number of special WKB lines depicted as red paralell lines.}
	\label{fig:levelzerostokes}
\end{figure}

When the level $k=0$, we have a trivial theory with central charge $c= 0$ and the only primary operator being the vacuum
\begin{equation}
\partial_x^2 \psi(x) = e^{2\theta}e^{2x}\psi(x).
\end{equation}
The WKB diagram takes the form of Fig. \ref{fig:levelzerostokes}.
By use of the asymptotics of the Bessel function for large real positive argument
\begin{equation}
\frac{1}{\sqrt{\pi}} K_\nu (z) \sim \frac{1}{\sqrt{2z}} e^{-z}.
\end{equation}
Small solutions are given by
\begin{equation}
\begin{split}
\psi_0 &= \frac{1}{\sqrt{\pi}}K_0(e^{\theta+x}),\\
\psi_n &= \psi_0(x;\theta+n i \pi) = \frac{1}{\sqrt{\pi}} \Big(K_0(e^{\theta+x}) -i\pi n I_0(e^{\theta+x})\Big)
\end{split}\label{eq:smallsolexpx}
\end{equation}
with the Wronskians given by $i(\psi_n,\psi_{n'}) = n'-n$ from $(K_0(e^{\theta+x}),I_0(e^{\theta+x})) = 1$.

\subsubsection{Matching around the zero}
When $t(x) = 0$, after shifting the coordinate, the stress tensor looks like
\begin{equation}
\frac{1}{\hbar^2}e^{2x} x^k.
\end{equation}
In the local coordinate around the zero,
\begin{equation}
y^k +a_{k-2} \gamma^{k} y^{k-2}+ \dots +  a_{j} \gamma^{2+j} y^{j} +\dots+ a_0 \gamma^{2}
\end{equation}
where $\gamma = \hbar^{\frac{2}{k+2}}$. When $k=1$, it just equals $y$ without corrections in $\gamma$. When $k\geq 2$, the stress tensor in the local coordinate system generically have nonzero coefficients $a_i$. For example, when $k=2$, we found that
\begin{equation}
a= -\frac{1}{8} + \frac{40911}{1024} \hbar^2 + O(\hbar^4).
\end{equation}

\subsubsection{Matching around the negative infinity}\label{app:sec:matchingaround_negative_inf}
Suppose $\delta = x-x_{-\infty}$ is a local coordinate around $x_{-\infty}$, which has a large negative real part. Then
\begin{align}
P(x)& = e^{2\theta+2x}(1+gx)^k \\
& = e^{2\delta} e^{2\theta +2 x_{-\infty} +k\log(1+gx_{-\infty})} \left(1+\frac{\delta}{x_{-\infty} + \frac{1}{g}}\right)^k.
\end{align}
We would like to find $x_{-\infty}$ such that the exponent $2\theta +2 x_0 +k\log(1+gx_{-\infty}) = 0$. And hopefully in the IR limit $\theta \rightarrow \infty$, the denominator in the parenthesis $x_{-\infty} + \frac{1}{g}$ is large so we can perform perturbation theory. This indeed can be done. 

We can solve the equation  $2\theta +2 x_{-\infty} +k\log(1+gx_{-\infty}) = 0$ by\footnote{Note that the naive solution 
	\begin{equation}
	x_{-\infty} = -\frac{1}{g} + \frac{k}{2} W\left(\frac{2}{kg}e^{\frac{2}{k}(\frac{1}{g}-\theta)} \right) \sim -\frac{\theta}{1+\frac{gk}{2}} + \frac{2k}{(2+gk)^3} (g\theta)^2 - \frac{4k(4-gk)}{3(2+gk)^5}(g\theta)^3 + \dots
	\end{equation}
	is not the one we want, since $0> x_{-\infty} > - \frac{1}{g}$, and $ 0 \ll \theta <\frac{1}{g}$
}
\begin{equation}
x_{-\infty} \sim -\theta -\frac{1}{2} k \log (-g\theta) -\frac{k^2}{4} \frac{\log(-g\theta)}{\theta}+ O\Big(\frac{1}{\theta}\Big). \label{eq:xinfasymp}
\end{equation}
Apparently the imaginary part of $x_{-\infty}$ is neither arbitrary nor unique, and depends on the parity of $k$ and the imaginary part of $\theta$. However, importantly, it turns out that we can always choose $x_{-\infty}$ to lie on one of the special WKB lines, though the precise choice doesn't matter.  

On the other hand, we can also see this from a different perspective. Recall that our system only depends on the $g_{\mathrm{eff}}$, a particular combination of $g$ and $\theta$, given by
\begin{equation}
e^{-\frac{1}{g_{\mathrm{eff}}(\theta)}} g_{\mathrm{eff}}(\theta)^{\frac{k}{2}} \equiv e^{-\frac{1}{g}} g^{\frac{k}{2}} e^{\theta}.
\end{equation}
It is not hard to see that this is indeed the same equation as the one for $x_{-\infty}$ once we identify
\begin{equation}
x_{-\infty}(\theta) = \frac{1}{g_{\mathrm{eff}}(\theta)} - \frac{1}{g} \label{eq:xinfgeff}
\end{equation}
that satisfy
\begin{equation}
\frac{d x_{-\infty}(\theta)}{d\theta} = - \frac{1}{1+\frac{k}{2} \frac{1}{x_{-\infty}+\frac{1}{g}}}, \quad \left.\frac{d x_{-\infty}(\theta)}{d\theta}\right|_{\theta = \infty} = -1, \quad x_{-\infty}(\infty) = -\infty.
\end{equation}
\begin{figure}[t]
	\centering
	\includegraphics[width=0.6\linewidth]{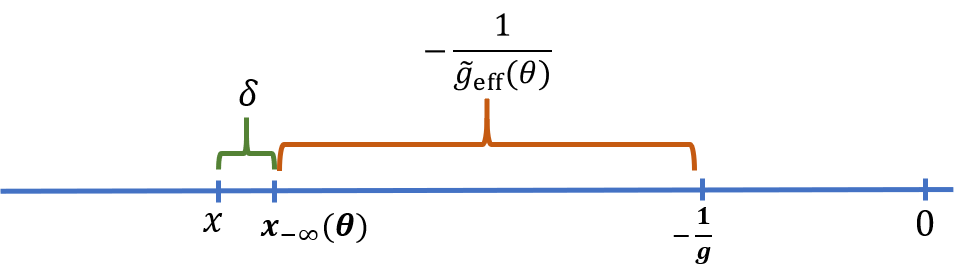}
	\includegraphics[width=0.4\linewidth]{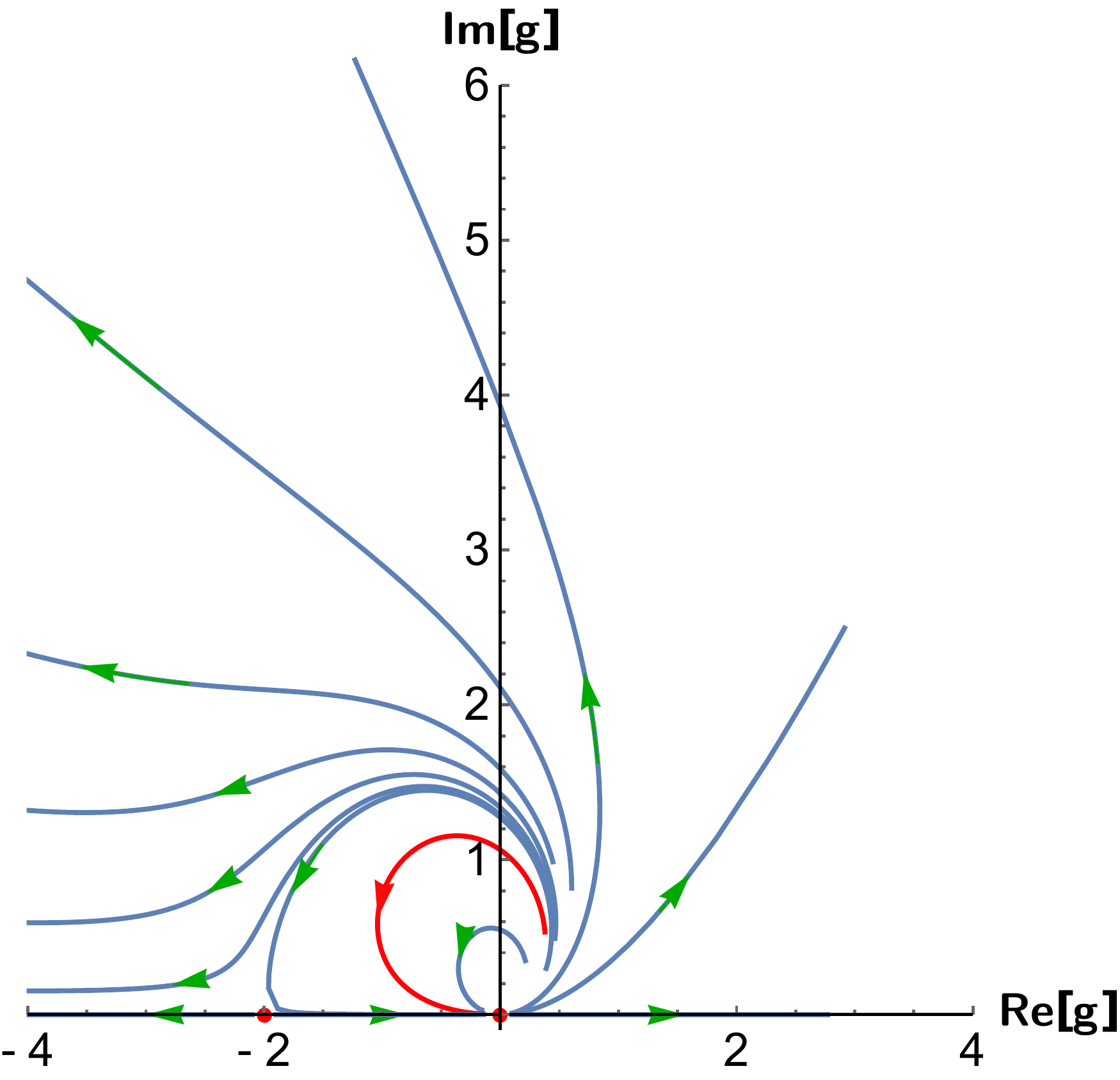}
	\includegraphics[width=0.4\linewidth]{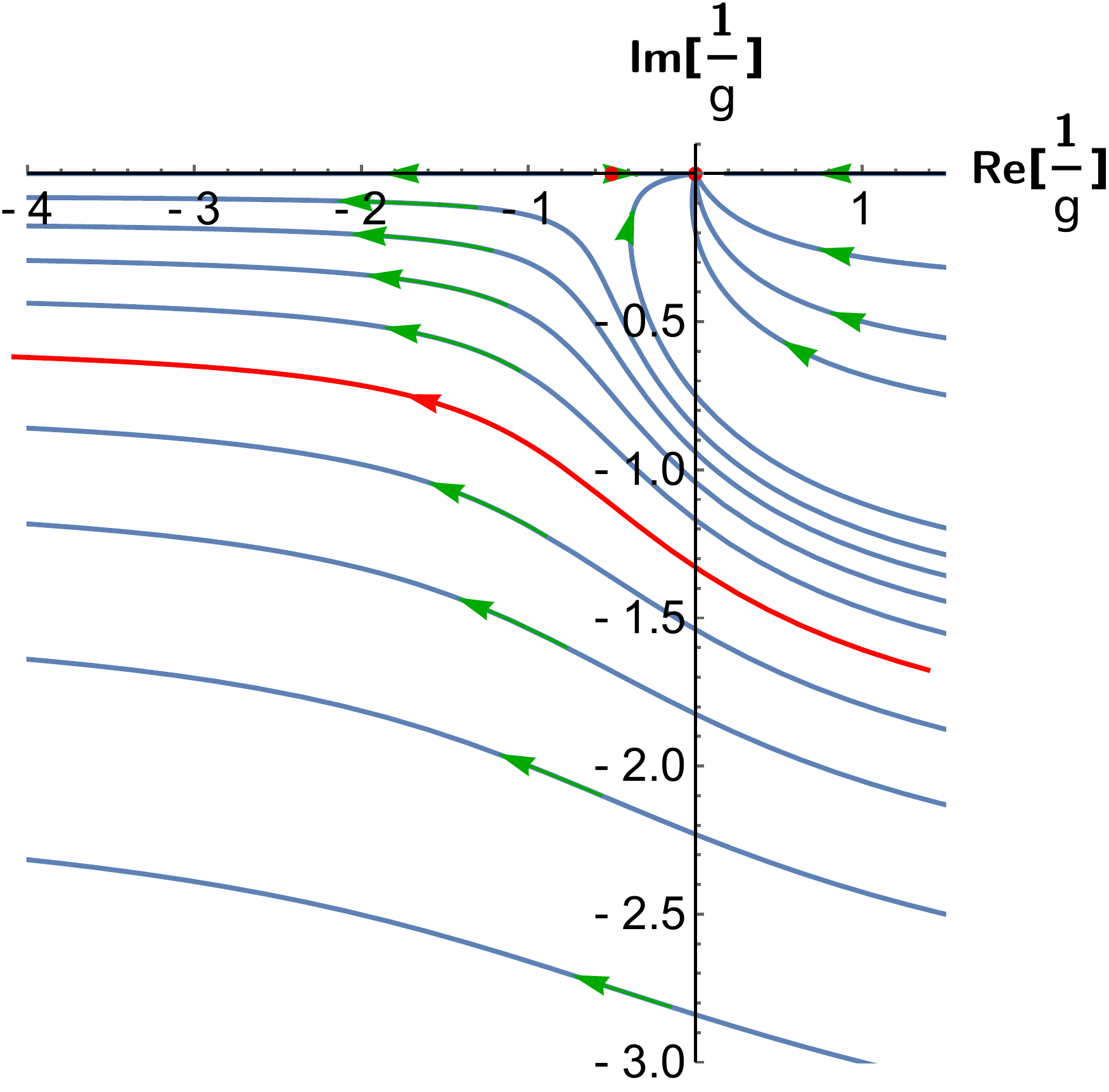}
	\caption{In the top figure, $g$ is assumed to be a some order $1$ constant, independent of $\theta$. $x_{-\infty}(\theta)$ is farther away from $-\frac{1}{g}$ as $\theta\rightarrow \infty$. $\delta$ is the local variation around $x_{-\infty}(\theta)$ that is complex. So it doesn't have to be on the real axis. In the bottom figures, the red line is an example of $g_{\mathrm{eff}}$ discussed in this section, namely an example of circular RG flow.}
	\label{fig:xinf}
\end{figure}
Or in terms of $g_{\mathrm{eff}}(\theta)$, it starts with $g = g_{\mathrm{eff}}(\theta = 0)$ that has some small imaginary part, and circles around in the complex $g_{\mathrm{eff}}(\theta)$ plane and goes back to the zero $g_{\mathrm{eff}}(\theta\rightarrow +\infty) \rightarrow 0^-$. Therefore the careful solution we found in \eqref{eq:xinfasymp}, especially the imaginary part of $x_{-\infty}$ is just to make sure we choose this circular type of RG flow, depicted as red lines in Fig. \ref{fig:xinf}.

Using the definition \eqref{eq:xinfgeff}, our quadratic differential is actually just
\begin{align}
P(x) = e^{2\delta} \left(1+ g_{\mathrm{eff}}(\theta){\delta}\right)^k.
\end{align}
And with the above solution of $x_{-\infty}$ and $g_{\mathrm{eff}}(\theta)$, we have $g_{\mathrm{eff}}(\theta\rightarrow +\infty) \rightarrow 0^-$, therefore the perturbation in $g_{\mathrm{eff}}$ is valid in the IR.\footnote{Of course we also have to make sure
	\begin{equation}
	|\delta| \ll -\frac{1}{g_{\mathrm{eff}}}.
	\end{equation}
} Note that $g_{\mathrm{eff}}(\theta)$ expands in large $\theta$ as
\begin{equation}
-\frac{1}{\theta} + \frac{-2 + gk\log(-g\theta)}{2g\theta^2} + \dots
\end{equation}

\bibliographystyle{JHEP}

\bibliography{mono}

\end{document}